\documentclass[preprint,10pt,sort,pdftex]{aastex}

\begin{document}
\title{AKARI/IRC Near-Infrared Spectral Atlas of Galactic Planetary Nebulae}
\author{Ryou Ohsawa,}
\email{ohsawa@ioa.s.u-tokyo.ac.jp}
\affil{Institute of Astronomy, Graduate School of Science, The University of Tokyo}
\affil{2-21-1 Osawa, Mitaka, Tokyo 181-0015, Japan}
\author{Takashi Onaka, Itsuki Sakon,}
\affil{Department of Astronomy, Graduate School of Science, The University of Tokyo}
\affil{7-3-1 Hongo, Bunkyo-ku, Tokyo 113-0033, Japan}
\author{Mikako Matsuura,}
\affil{School of Physics and Astronomy, Cardiff University}
\affil{Queen's Buildings, 5 The Parade, Roath, Cardiff CF24 3AA, United Kingdom}
\author{Hidehiro Kaneda}
\affil{Division of Particle and Astrophysical Science, Graduate School of Science, Nagoya University}
\affil{Furo-cho, Chikusa-ku, Nagoya, Aichi 464-8602, Japan}

\begin{abstract}
Near-infrared (2.5--5.0$\,\mu$m) low-resolution ($\lambda/\Delta\lambda{\sim}100$) spectra of 72 Galactic planetary nebulae (PNe) were obtained with the Infrared Camera (IRC) in the post-helium phase. The IRC, equipped with a $1'{\times}1'$ window for spectroscopy of a point source, was capable of obtaining near-infrared spectra in a slit-less mode without any flux loss due to a slit. The spectra show emission features including hydrogen recombination lines and the 3.3--3.5$\,\mu$m hydrocarbon features. The intensity and equivalent width of the emission features were measured by spectral fitting. We made a catalog\footnote{Available at \url{http://www.ir.isas.jaxa.jp/AKARI/Archive/Catalogues/IRC\_PNSPC/}} providing unique information on the investigation of the near-infrared emission of PNe. In this paper, details of the observations and characteristics of the catalog are described.
\end{abstract}
\keywords{Catalogs --- planetary nebulae: general}

\section{Introduction}\label{sec:introduction}
Planetary nebulae (PNe) are a late evolutionary stage of low- and intermediate-mass stars \citep[e.g.,][]{schonberner_structure_1996,blocker_stellar_1995}. They are surrounded by rich circumstellar material (CSM) ejected during the Asymptotic Giant Branch (AGB) phase \citep{schonberner_evolution_2005-1}. The CSM is eventually incorporated into the interstellar medium and consists of the ingredients for next-generation star-forming activity.

Infrared spectra are rich in CSM emission features. These features can be used to investigate mass, temperature and chemical properties of CSMs \citep[e.g.,][]{pottasch_planetary_2006,pottasch_iras_1984,bernard-salas_physical_2005,phillips_mid-_2011}. In some PNe, far-infrared emission is dominated by that from dust grains, providing information about the temperature and the total mass of dust grains. Spectroscopic observations in the mid-infrared (${\sim}5$--$40\,\mu$m) have been used to identify dust species \citep[e.g.,][]{volk_iras_1990,stanghellini_nature_2012}. Emission from stochastically heated dust grains also appears in the mid-infrared \citep{dwek_detection_1997,draine_infrared_2001}. The dust features commonly found are PAHs, MgS, silicate, and some potential cases of SiC. Near-infrared (${\sim}1$--$5\,\mu$m) continuum emission in PNe involves free-free emission, stellar continuum, and hot dust emission. Hydrogen recombination lines are prominent in the near-infrared, such as Brackett-$\alpha$ at 4.051$\,\mu$m. The PAH emission feature appears at 3.3$\,\mu$m \citep{beintema_rich_1996,roche_investigation_1996,tokunaga_high-resolution_1991}, while there is also emission from aliphatic C{\sbond}H bonds at around 3.4--3.5$\,\mu$m \citep{joblin_spatial_1996,sloan_variations_1997}.

Infrared space telescopes have been instrumental for PNe observations despite limited spatial resolution. Eliminating the effects of heavy terrestrial atmospheric absorption provides the critical advantage of continuous infrared spectroscopic coverage. \citet{volk_iras_1990} investigated 7--23$\,\mu$m low-resolution spectra of 170 Galactic PNe with the \textit{Infrared Astronomical Satellite (IRAS)}. The Short-Wave Spectrometer (SWS) and the Long-Wave Spectrometer (LWS) on-board the \textit{Infrared Space Observatory (ISO)} cover 2.5--45 and 45--197$\,\mu$m. A number of emission lines have been identified and the physical conditions of circumstellar nebulae have been investigated \citep[e.g.,][]{pottasch_planetary_2006,bernard-salas_physical_2005}. The Infrared Spectrograph (IRS) on-board the \textit{Spitzer} space telescope covers 5.5--37$\,\mu$m. \citet{stanghellini_nature_2012} obtained 157 spectra of compact Galactic PNe and investigated their dust emission. Spectra of PNe in the Magellanic Clouds were also obtained with the IRS \citep{stanghellini_galactic_2010,bernard-salas_unusual_2009,bernard-salas_neon_2008} thanks to its high sensitivity. The \textit{ISO}/SWS had obtained 2.5--5.0$\,\mu$m spectra of Galactic PNe. About 85 spectra of Galactic PNe are available in \citet{sloan_uniform_2003}. However, only a few PNe have spectra with a sufficient signal-to-noise ratio at 2.5--5.0$\,\mu$m to investigate weak features such as the 3.4--3.5$\,\mu$m aliphatic emission. The Infrared Camera (IRC) on-board \textit{AKARI} satellite, for the first time, provides near-infrared (2.5--5.0$\,\mu$m) spectroscopy with a sensitivity of a few mJy \citep{onaka_infrared_2007}.

The present paper reports near-infrared spectroscopy with the \textit{AKARI}/IRC. Observations were carried out as part of an Open Time Program for the post-helium phase, ``Near-Infrared Spectroscopy of Planetary Nebulae'' (PNSPC). Near-infrared spectra of 72 PNe were obtained. The data were compiled as a near-infrared spectral catalog. In this paper, we describe details of the observations and the characteristics of the catalog. Detailed information about the observations is given in Section \ref{sec:observations}. Section \ref{sec:data-reduction} describes the reduction procedures and the method to extract the spectra. The characteristics of the retrieved spectra and the  format of the compiled catalog are presented in Section \ref{sec:results}. The quality of the IRC spectra and the statistical properties of the catalog are discussed in comparison with the observations by other facilities in Section \ref{sec:discussion}. We summarize the paper in Section \ref{sec:summary}.

\section{Observations}\label{sec:observations}
\subsection{Target Selection}\label{sec:obs:targets}
\begin{figure}[!tp]
  \centering
  \includegraphics[width=1.0\linewidth]{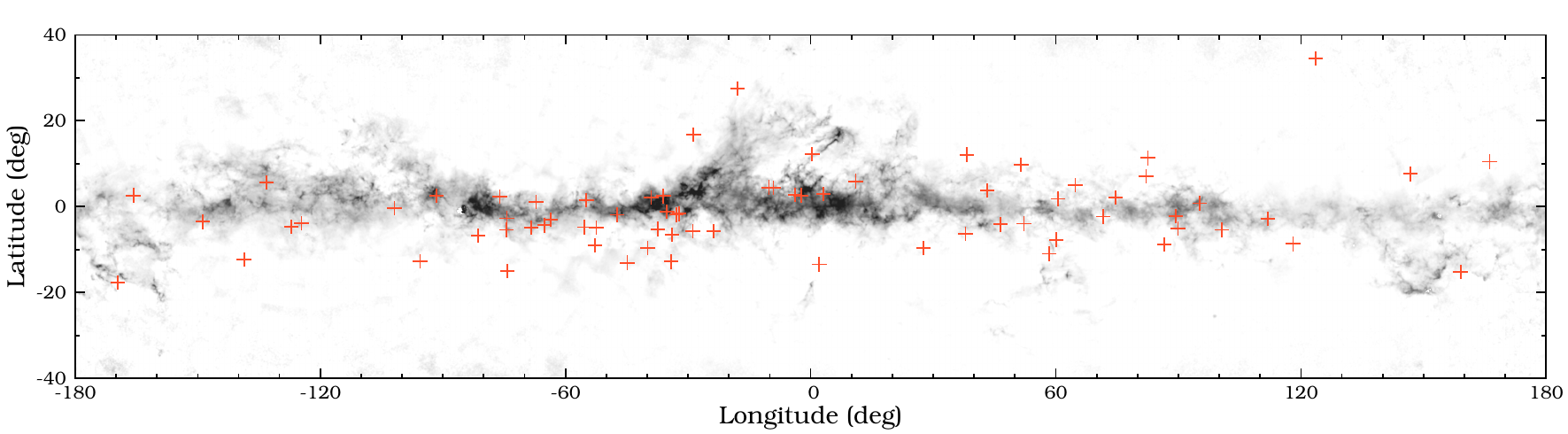}
  \caption{Distribution of the observed planetary nebulae (the crosses) overlaid on an extinction map in the Galactic coordinates created from the DSS \citep{dobashi_atlas_2005}. The gray scale indicates extinction at the $V$-band from 0.1 to 5\,mag.}
  \label{fig:pndistribution}
\end{figure}

The observed targets were selected from the Strasbourg-ESO Catalogue of Galactic Planetary Nebulae \citep{acker_strasbourg-eso_1992}, taking into account their NIR magnitude and angular size. The spectroscopic sensitivity of the IRC is a few$\,$mJy in the range of 2--5$\,\mu$m, while the saturation limit is about 10$\,$Jy. With these limits taken into account, objects whose $K$-band magnitude is between 5 and 13 were selected as targets for this program. The observations were carried out with the $1\arcmin{\times}1\arcmin$ Np-window \citep[\textit{see} Section~\ref{sec:obs:instruments} and][]{onaka_infrared_2007} in the slit-less mode. Therefore, compact objects were preferred to avoid degrading the spectral resolution. The candidate list was then narrowed to objects whose radius in the optical was smaller than $5\arcsec$, comparable to the full width half maximum of the IRC point spread function $({\sim}3\arcsec)$, to exclude apparently large objects. Finally, 72 objects were selected and their spectra were successfully obtained. Figure~\ref{fig:pndistribution} shows the distribution of the targets in the Galactic coordinates. The PNG\,ID, the AKARI observation ID, and the coordinates of the PNSPC samples are listed in Table~\ref{tab:basic_info}. Table~\ref{tab:basic_info_misc} summarizes miscellaneous information of the objects, including the optical (the $V$-band) and infrared (the 2MASS $K_s$-band) magnitudes.

The target selection method was independent of the chemistry and dust compositions of the circumstellar nebula. The catalog is not considered to be biased, in terms of chemical abundance and the chemistry of dust. However, the objects were selected based on their sizes. Since PNe expand as they evolve, the selection could be biased toward young PNe. The potential biases of the PNSPC samples are discussed in detail in Section~\ref{sec:discussion}.

\subsection{Instruments}\label{sec:obs:instruments}
Although the \textit{AKARI}/IRC has three channels \citep{onaka_infrared_2007}, only the NIR channel was available during the post-helium phase. The field of view of the NIR channel consists of four parts: the N/Nc-window, the Ns-slit, the Np-window, and the Nh-slit \citep[\textit{see}, Figure 3 in][]{onaka_infrared_2007}. The observations were performed with the Np-window, which was designed for point source, slit-less spectroscopy. Spectroscopy with the Np-window allows us to collect all of the flux from an object. The present spectroscopy was carried out with the grism, providing 2.5--5.0$\,\mu$m spectra with the spectral resolution of $\lambda/\Delta\lambda{\sim}100$, for point sources (${\lesssim}3''$). The spectral resolution was degraded for extended objects.

The observations were carried out with the observation template IRCZ4. This mode consists of three sequences: the first set is four spectroscopic images, the second set is a broad band image, and the final set is another sequence of four spectroscopic images. Thus, each pointing observation involves eight frames for spectroscopy and one frame for broad-band imaging. Every frame consists of short (4.58$\,$sec) and long (44.41$\,$sec) exposure images. Due to the short integration time, the typical signal-to-noise ratio of the short exposure images is worse than that of the long exposure images. In general, we used only the long exposure images. The short exposure images were used only for cases when the long exposure image showed saturated pixels.

Each object was observed either once, or twice. Thus, there were eight or sixteen frames for each object. However, some frames were not usable due to cosmic rays or unstable pointing during the exposure. The net integration time thus depends on the target, listed in the sixth column of Table~\ref{tab:basic_info}.

\begin{deluxetable}{ccccccccccc}
\tabletypesize{\small}
\tablewidth{0pt}

\tablecaption{Summary of Observations \label{tab:basic_info}}
\tablehead{
  \colhead{PN\,G} &
  \colhead{Obs. ID} &
  \colhead{RA (J2000)} &
  \colhead{DEC (J2000)} &
  \colhead{$N$$^{(a)}$} &
  \colhead{$t_{\rm int}${$^{(b)}$}} &
  \colhead{FWHM} &
  \multicolumn{3}{c}{Flags} \\
  \colhead{} & \colhead{} &
  \colhead{$hh{:}mm{:}ss$} &
  \colhead{$dd{:}mm{:}ss$} &
  \colhead{} &
  \colhead{${\rm sec.}$} &
  \colhead{${\rm \mu m}$} &
  \colhead{~$\mathcal{S}${$^{(c)}$}} &
  \colhead{~$\mathcal{D}${$^{(d)}$}} &
  \colhead{~$\mathcal{C}${$^{(e)}$}}
}
\startdata
000.3${+}$12.2 & $3460107$ & $17:01:33.6$ & $-21:49:33.5$ & $17$ & $754.97$ & $0.045$ & N & N & N  \\
002.0${-}$13.4 & $3460003$ & $18:45:50.7$ & $-33:20:35.0$ & $16$ & $710.56$ & $0.040$ & N & Y & N  \\
003.1${+}$02.9 & $3460005$ & $17:41:52.8$ & $-24:42:07.7$ & $9$ & $399.69$ & $0.045$ & N & Y & Y  \\
011.0${+}$05.8 & $3460012$ & $17:48:19.8$ & $-16:28:44.0$ & $18$ & $799.38$ & $0.035$ & N & N & N  \\
027.6${-}$09.6 & $3460016$ & $19:16:28.3$ & $-09:02:37.0$ & $9$ & $399.69$ & $0.040$ & N & Y & N  \\
037.8${-}$06.3 & $3460019$ & $19:22:56.9$ & $+01:30:48.0$ & $17$ & $754.97$ & $0.035$ & N & Y & N  \\
038.2${+}$12.0 & $3460020$ & $18:17:34.0$ & $+10:09:05.0$ & $9$ & $399.69$ & $0.040$ & N & N & N  \\
043.1${+}$03.8 & $3460093$ & $18:56:33.6$ & $+10:52:12.0$ & $9$ & $399.69$ & $0.035$ & N & Y & Y  \\
046.4${-}$04.1 & $3460021$ & $19:31:16.4$ & $+10:03:21.7$ & $9$ & $399.69$ & $0.040$ & N & Y & Y  \\
051.4${+}$09.6 & $3460022$ & $18:49:47.6$ & $+20:50:39.5$ & $17$ & $754.97$ & $0.040$ & N & N & N  \\
052.2${-}$04.0 & $3460094$ & $19:42:18.7$ & $+15:09:09.0$ & $18$ & $799.38$ & $0.035$ & N & Y & N  \\
058.3${-}$10.9 & $3460023$ & $20:20:08.8$ & $+16:43:54.0$ & $18$ & $799.38$ & $0.040$ & N & N & N  \\
060.1${-}$07.7 & $3460024$ & $20:12:42.9$ & $+19:59:23.0$ & $14$ & $621.74$ & $0.048$ & N & Y & N  \\
060.5${+}$01.8 & $3460025$ & $19:38:08.4$ & $+25:15:42.0$ & $18$ & $799.38$ & $0.035$ & N & Y & N  \\
064.7${+}$05.0 & $3460026$ & $19:34:45.3$ & $+30:30:59.2$ & $17$ & $\phn77.86$ & $0.048$ & Y & N & N  \\
071.6${-}$02.3 & $3460027$ & $20:21:03.8$ & $+32:29:24.0$ & $18$ & $799.38$ & $0.035$ & N & N & N  \\
074.5${+}$02.1 & $3460028$ & $20:10:52.5$ & $+37:24:41.0$ & $18$ & $799.38$ & $0.040$ & N & N & N  \\
082.1${+}$07.0 & $3460029$ & $20:10:23.7$ & $+46:27:39.0$ & $18$ & $799.38$ & $0.045$ & N & Y & N  \\
082.5${+}$11.3 & $3460030$ & $19:49:46.6$ & $+48:57:40.0$ & $9$ & $399.69$ & $0.035$ & N & Y & N  \\
086.5${-}$08.8 & $3460114$ & $21:33:08.3$ & $+39:38:09.7$ & $18$ & $799.38$ & $0.045$ & N & Y & N  \\
089.3${-}$02.2 & $3460031$ & $21:19:07.2$ & $+46:18:48.0$ & $18$ & $799.38$ & $0.038$ & N & Y & N  \\
089.8${-}$05.1 & $3460032$ & $21:32:31.0$ & $+44:35:47.7$ & $18$ & $799.38$ & $0.035$ & N & N & N  \\
095.2${+}$00.7 & $3460033$ & $21:31:50.2$ & $+52:33:52.0$ & $18$ & $799.38$ & $0.042$ & N & Y & N  \\
100.6${-}$05.4 & $3460034$ & $22:23:55.7$ & $+50:58:00.0$ & $18$ & $799.38$ & $0.035$ & N & N & N  \\
111.8${-}$02.8 & $3460035$ & $23:26:14.9$ & $+58:10:53.0$ & $8$ & $355.28$ & $0.038$ & N & N & N  \\
118.0${-}$08.6 & $3460096$ & $00:18:42.2$ & $+53:52:20.0$ & $16$ & $710.56$ & $0.035$ & N & Y & N  \\
123.6${+}$34.5 & $3460115$ & $12:33:06.8$ & $+82:33:50.1$ & $18$ & $799.38$ & $0.065$ & N & Y & N  \\
146.7${+}$07.6 & $3460097$ & $04:25:50.8$ & $+60:07:12.7$ & $18$ & $799.38$ & $0.035$ & N & N & N  \\
159.0${-}$15.1 & $3460098$ & $03:47:33.0$ & $+35:02:48.9$ & $9$ & $399.69$ & $0.045$ & N & Y & N  \\
166.1${+}$10.4 & $3460116$ & $05:56:23.9$ & $+46:06:17.2$ & $18$ & $799.38$ & $0.050$ & N & N & N  \\
190.3${-}$17.7 & $3460099$ & $05:05:34.3$ & $+10:42:23.8$ & $9$ & $399.69$ & $0.045$ & N & N & N  \\
194.2${+}$02.5 & $3460117$ & $06:25:57.2$ & $+17:47:27.3$ & $18$ & $799.38$ & $0.045$ & N & N & N  \\
211.2${-}$03.5 & $3460036$ & $06:35:45.1$ & $+00:05:38.0$ & $16$ & $710.56$ & $0.035$ & N & N & N  \\
221.3${-}$12.3 & $3460118$ & $06:21:42.7$ & $-12:59:14.0$ & $15$ & $666.15$ & $0.046$ & N & N & N  \\
226.7${+}$05.6 & $3460100$ & $07:37:18.9$ & $-09:38:50.0$ & $17$ & $754.97$ & $0.040$ & N & Y & N  \\
232.8${-}$04.7 & $3460037$ & $07:11:16.7$ & $-19:51:04.0$ & $14$ & $621.74$ & $0.035$ & N & N & N  \\
235.3${-}$03.9 & $3460038$ & $07:19:21.5$ & $-21:43:54.3$ & $17$ & $754.97$ & $0.035$ & N & N & N  \\
258.1${-}$00.3 & $3460039$ & $08:28:28.0$ & $-39:23:40.0$ & $17$ & $754.97$ & $0.042$ & N & N & N  \\
264.4${-}$12.7 & $3460040$ & $07:47:20.0$ & $-51:15:03.4$ & $17$ & $754.97$ & $0.035$ & N & Y & N  \\
268.4${+}$02.4 & $3460042$ & $09:16:09.6$ & $-45:28:42.8$ & $18$ & $799.38$ & $0.038$ & N & N & N  \\
278.6${-}$06.7 & $3460101$ & $09:19:27.5$ & $-59:12:00.3$ & $18$ & $799.38$ & $0.040$ & N & Y & N  \\
283.8${+}$02.2 & $3460044$ & $10:31:33.4$ & $-55:20:50.5$ & $16$ & $710.56$ & $0.045$ & N & Y & N  \\
285.4${-}$05.3 & $3460120$ & $10:09:20.8$ & $-62:36:48.5$ & $18$ & $799.38$ & $0.050$ & N & Y & N  \\
285.6${-}$02.7 & $3460045$ & $10:23:09.1$ & $-60:32:42.3$ & $17$ & $754.97$ & $0.042$ & N & Y & N  \\
285.7${-}$14.9 & $3460121$ & $09:07:06.3$ & $-69:56:30.6$ & $17$ & $754.97$ & $0.060$ & N & N & N  \\
291.6${-}$04.8 & $3460046$ & $11:00:20.0$ & $-65:14:57.8$ & $18$ & $799.38$ & $0.038$ & N & Y & N  \\
292.8${+}$01.1 & $3460102$ & $11:28:47.4$ & $-60:06:37.3$ & $18$ & $799.38$ & $0.042$ & N & Y & N  \\
294.9${-}$04.3 & $3460047$ & $11:31:45.4$ & $-65:58:13.7$ & $18$ & $799.38$ & $0.035$ & N & N & N  \\
296.3${-}$03.0 & $3460048$ & $11:48:38.2$ & $-65:08:37.3$ & $15$ & $666.15$ & $0.035$ & N & N & N  \\
304.5${-}$04.8 & $3460051$ & $13:08:47.3$ & $-67:38:37.6$ & $16$ & $710.56$ & $0.038$ & N & Y & N  \\
305.1${+}$01.4 & $3460122$ & $13:09:36.4$ & $-61:19:35.6$ & $18$ & $\phn82.44$ & $0.050$ & Y & N & N  \\
307.2${-}$09.0 & $3460052$ & $13:45:22.4$ & $-71:28:55.7$ & $18$ & $799.38$ & $0.035$ & N & Y & Y  \\
307.5${-}$04.9 & $3460053$ & $13:39:35.1$ & $-67:22:51.7$ & $18$ & $799.38$ & $0.040$ & N & Y & Y  \\
312.6${-}$01.8 & $3460123$ & $14:18:43.3$ & $-63:07:10.1$ & $16$ & $710.56$ & $0.056$ & N & Y & N  \\
315.1${-}$13.0 & $3460054$ & $15:37:11.2$ & $-71:54:52.9$ & $18$ & $\phn82.44$ & $0.042$ & Y & N & N  \\
320.1${-}$09.6 & $3460103$ & $15:56:01.7$ & $-66:09:09.2$ & $16$ & $710.56$ & $0.045$ & N & Y & N  \\
320.9${+}$02.0 & $3460056$ & $15:05:59.2$ & $-55:59:16.5$ & $9$ & $399.69$ & $0.042$ & N & Y & N  \\
322.5${-}$05.2 & $3460060$ & $15:47:41.2$ & $-61:13:05.6$ & $16$ & $710.56$ & $0.092$ & N & Y & Y  \\
323.9${+}$02.4 & $3460061$ & $15:22:19.4$ & $-54:08:13.1$ & $9$ & $399.69$ & $0.045$ & N & Y & N  \\
324.8${-}$01.1 & $3460062$ & $15:41:58.8$ & $-56:36:25.6$ & $8$ & $355.28$ & $0.036$ & N & Y & Y  \\
325.8${-}$12.8 & $3460063$ & $16:54:35.2$ & $-64:14:28.4$ & $16$ & $710.56$ & $0.035$ & N & Y & N  \\
326.0${-}$06.5 & $3460064$ & $16:15:42.3$ & $-59:54:01.0$ & $14$ & $621.74$ & $0.035$ & N & Y & Y  \\
327.1${-}$01.8 & $3460065$ & $15:58:08.1$ & $-55:41:50.3$ & $9$ & $399.69$ & $0.035$ & N & Y & Y  \\
327.8${-}$01.6 & $3460066$ & $16:00:59.1$ & $-55:05:39.7$ & $8$ & $355.28$ & $0.038$ & N & Y & N  \\
331.1${-}$05.7 & $3460067$ & $16:37:42.7$ & $-55:42:26.5$ & $17$ & $754.97$ & $0.035$ & N & Y & N  \\
331.3${+}$16.8 & $3460068$ & $15:12:50.8$ & $-38:07:32.0$ & $16$ & $710.56$ & $0.048$ & N & Y & N  \\
336.3${-}$05.6 & $3460069$ & $16:59:36.1$ & $-51:42:08.4$ & $18$ & $799.38$ & $0.035$ & N & Y & N  \\
342.1${+}$27.5 & $3460104$ & $15:22:19.3$ & $-23:37:32.0$ & $18$ & $799.38$ & $0.050$ & N & Y & N  \\
349.8${+}$04.4 & $3460105$ & $17:01:06.2$ & $-34:49:38.0$ & $9$ & $399.69$ & $0.040$ & N & Y & N  \\
350.9${+}$04.4 & $3460070$ & $17:04:36.2$ & $-33:59:18.0$ & $18$ & $799.38$ & $0.038$ & N & Y & N  \\
356.1${+}$02.7 & $3460076$ & $17:25:19.2$ & $-30:40:41.0$ & $18$ & $799.38$ & $0.035$ & N & Y & Y  \\
357.6${+}$02.6 & $3460082$ & $17:29:42.7$ & $-29:32:50.0$ & $15$ & $666.15$ & $0.035$ & N & Y & N  \\
\enddata

\tablecomments{$^{(a)}$ the number of exposures; $^{(b)}$ the net integration time; $^{(c)}$ the saturation flag, Y: the spectrum is (partly) saturated in long-exposure spectral images; N: the spectrum is not saturated; $^{(d)}$ the decomposition flag, Y: the spectral decomposition procedure was used; N: the spectrum was not affected by any neighbor object; $^{(e)}$ the contamination flag, Y: the decontamination procedure was used; N: the spectrum was not affected by any overlapping object.}

\end{deluxetable}

\begin{deluxetable}{ccccccc}
\tabletypesize{\small}
\tablewidth{0pt}

\tablecaption{Miscellaneous Information of Objects \label{tab:basic_info_misc_s}}
\tablehead{
  \colhead{PN\,G} &
  \colhead{$V^{(a)}$} &
  \colhead{$K_s^{(b)}$} &
  \colhead{$A_{V}({\rm H\beta})$} &
  \colhead{$A_{V}({\rm fit.})$} &
  \colhead{$T_{\rm eff}$} & \colhead{Refs. $T_{\rm eff}$} \\
  \colhead{} &
  \colhead{mag.} & \colhead{mag.} &
  \colhead{mag.} & \colhead{mag.} &
  \colhead{$10^3\,$K} & \colhead{}
}
\startdata
000.3${+}$12.2 & $\phn13.9$ & $\phn11.4$ & $\phn\phn0.90_{{-}0.03}^{{+}0.03}$ & $\phn\phn0.98_{{-}0.05}^{{+}0.15}$ & $\phn47$ & Ka76,Ph03 \\
002.0${-}$13.4 & $\phn14.1$ & $\phn11.1$ & $\phn\phn0.31_{{-}0.01}^{{+}0.02}$ & $\phn\phn0.38_{{-}0.04}^{{+}0.02}$ & $\phn60$ & PM89,PM91,Ph03 \\
003.1${+}$02.9 & ${>}17.0$ & $\phn11.6$ & $\phn\phn3.31_{{-}0.03}^{{+}0.03}$ & $\phn\phn\phn3.4_{{-}0.1}^{{+}0.6}$ & $\phn87$ & PM89,Ph03 \\
011.0${+}$05.8 & \nodata & $\phn11.8$ & $\phn\phn1.81_{{-}0.04}^{{+}0.04}$ & $\phn\phn1.79_{{-}0.06}^{{+}0.06}$ & $\phn93$ & PM91,Ph03 \\
027.6${-}$09.6 & $\phn15.2$ & $\phn11.9$ & $\phn\phn1.04_{{-}0.05}^{{+}0.06}$ & $\phn\phn\phn1.0_{{-}0.2}^{{+}0.9}$ & $\phn58$ & PM91,Ka76,KJ91,Ph03 \\
037.8${-}$06.3 & \nodata & $\phn9.47$ & $\phn\phn\phn1.4_{{-}0.1}^{{+}0.1}$ & $\phn\phn1.41_{{-}0.02}^{{+}0.08}$ & $\phn74$ & Ph03 \\
038.2${+}$12.0 & $\phn12.5$ & $\phn11.1$ & $\phn\phn0.82_{{-}0.05}^{{+}0.05}$ & $\phn\phn0.93_{{-}0.07}^{{+}0.07}$ & $\phn30$ & Ka78,Ph03 \\
043.1${+}$03.8 & $\phn14.9$ & $\phn12.1$ & $\phn\phn\phn1.9_{{-}0.1}^{{+}0.1}$ & $\phn\phn2.05_{{-}0.05}^{{+}0.46}$ & $\phn28$ & Ph03 \\
046.4${-}$04.1 & $\phn15.2$ & $\phn11.1$ & $\phn\phn1.42_{{-}0.02}^{{+}0.02}$ & $\phn\phn1.39_{{-}0.01}^{{+}0.01}$ & $\phn68$ & Ka76,KJ91,Ph03 \\
051.4${+}$09.6 & $\phn13.3$ & $\phn10.4$ & $\phn\phn0.88_{{-}0.04}^{{+}0.05}$ & $\phn\phn0.98_{{-}0.08}^{{+}0.11}$ & $\phn37$ & Ka76,Ka78,Ph03 \\
\enddata

\tablecomments{Table~\ref{tab:basic_info_misc_s} is published in its entirety in the electronic edition of the Astrophysical Journal. A portion is shown here for guidance regarding its form and content.}

\tablerefs{$^{(a)}$\citet{acker_strasbourg-eso_1992}; $^{(b)}$ \citet{skrutskie_two_2006}; HF83: \citet{harrington_planetary_1983}, Ka76: \citet{kaler_exciting_1976}, Ka78: \citet{kaler_forbidden_1978}, KJ91: \citet{kaler_central_1991}, Lu01: \citet{lumsden_infrared_2001}, Me88: \citet{mendez_high_1988}, Ph03: \citet{phillips_relation_2003}, PM89: \citet{preite-martinez_energy-balance_1989}, PM91: \citet{preite-martinez_energy-balance_1991}.}

\end{deluxetable}

\section{Data Reduction}\label{sec:data-reduction}
\subsection{Bias Subtraction and Flux Calibration}\label{sec:red:basic}
The basic data reduction procedure used here, includes subtraction of dark frames, linearity correction, saturation masking, flat fielding, and subtraction of background emission. All steps were carried out with the IRC Spectroscopy Toolkit for Phase 3 (Version 20111121)\footnote{Available at \url{http://www.ir.isas.jaxa.jp/ASTRO-F/Observation/DataReduction/IRC/}}. Pixels affected by cosmic rays were masked by hand. Some pixels were saturated by bright objects in the long exposure images. The saturated pixels were replaced by the corresponding pixels from the short exposure images. Objects whose spectra were saturated in the long exposure images are marked in the eighth column of Table~\ref{tab:basic_info}. Spectroscopic images were aligned by maximizing the correlation among the images. Then, the images were combined by median stacking, producing a two-dimensional spectrum. Frames that were affected by unstable pointing during the exposure were excluded from stacking. The toolkit also produced a two-dimensional noise map, which was given as a standard deviation of the stacked frames.

\subsection{Extraction of One-Dimensional Spectrum}\label{sec:red:extraction}
Some objects were extended with respect to the point spread function (PSF) of the IRC. To make a homogeneous data reduction, we assumed that all the sources were extended. We used a sufficiently wide aperture (${>}10\,$pixels) to extract a one-dimensional spectrum that included all of the fluxes from the object, and did not apply any aperture correction. The standard error in the spectrum was calculated using the noise map produced by the toolkit.

\begin{figure}[!t]
  \centering
  \includegraphics[width=1.0\linewidth]{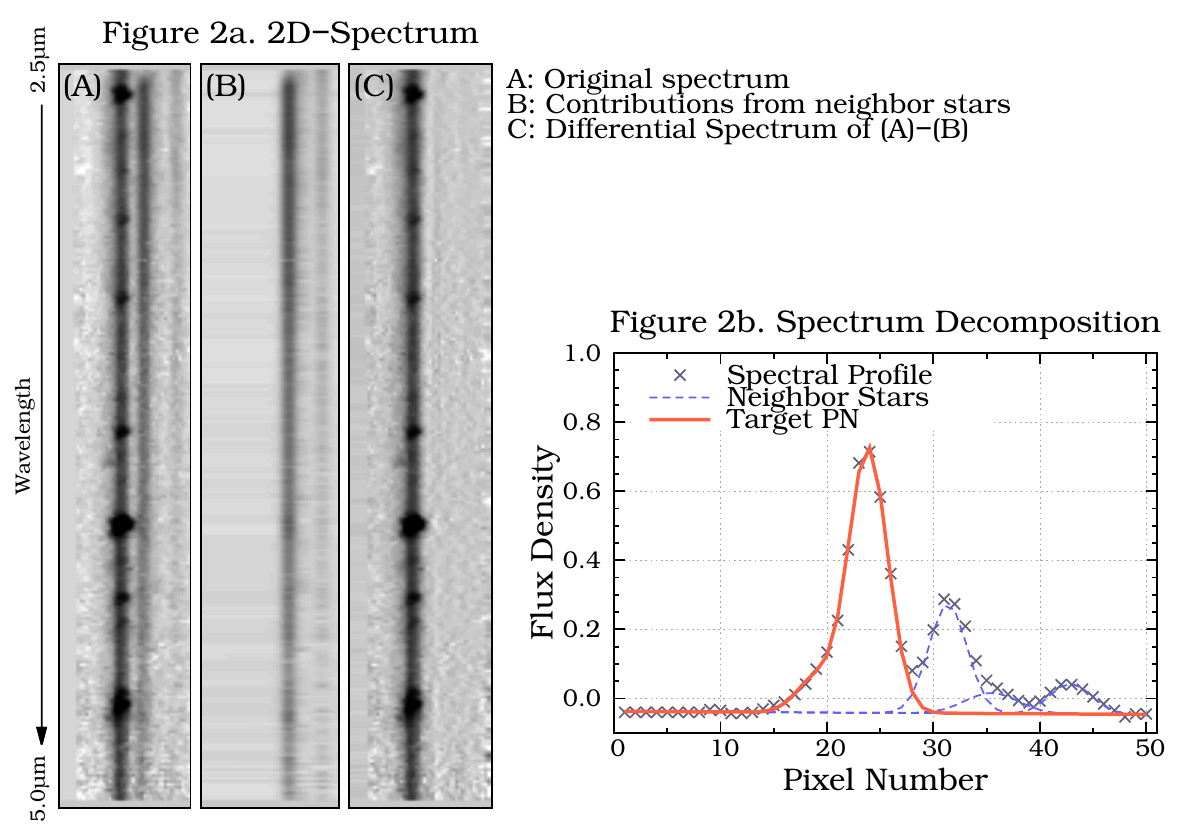}
  \caption{Schematic view of the spectral decomposition procedure. (\textit{Left}) Panels A, B, and C represent an original two-dimensional spectrum, estimated contributions from the neighboring stars, and an isolated target spectrum. (\textit{Right}): Spatial profiles of the two-dimensional spectrum (Panel A of Figure~\ref{fig:decompose}a). The profiles are decomposed by fitting (\textit{see}, text). The contributions from neighboring stars are shown by the blue dashed lines. The component of the target is shown by the red solid line.}
  \label{fig:decompose}
\end{figure}

It was difficult to use a wide aperture for objects which were located in crowded regions. \citet{noble_survey_2013} invented a novel method to extract the spectrum of a certain object in a crowded region by profile fitting. We used a similar method to remove the contribution from neighboring objects. Figure~\ref{fig:decompose} shows a schematic view of the method (hereafter, the spectral decomposition procedure). Panel A of Figure~\ref{fig:decompose}a shows an example two-dimensional spectrum of a target in a crowded region. The horizontal and vertical axes correspond to the position and wavelength. The spectrum of the target PN runs through the center of the panel. The spectra of neighboring objects are located to the right side of the PN spectrum. Figure~\ref{fig:decompose}b shows the spatial profiles of the spectra (Panel A). The spectra were decomposed by fitting with a combination of Gaussian functions. The red solid line shows the PN profile, and the blue dashed lines show those of the neighboring objects. The fitting was performed for every spectral element. The instrumental spatial profile of a spectrum was not a Gaussian, rather a combination of Gaussian profiles provided the best fit to the profile. Panel B of Figure~\ref{fig:decompose}a shows the estimated contribution from the neighboring objects. By subtracting Panel B from Panel A, the two-dimensional spectrum of the target PN was extracted (Panel C). Finally, the one-dimensional spectrum of the target PN was obtained from Panel C with a wide aperture, as mentioned above. The uncertainty in the contamination-corrected spectrum was estimated by taking into account errors in the spectral decomposition procedure. The targets which are in crowded regions are denoted in the ninth column of Table~\ref{tab:basic_info}.

\subsubsection{Subtraction of Contamination}\label{sec:red:decont}
\begin{figure}[!t]
  \centering
  \includegraphics[width=1.0\linewidth]{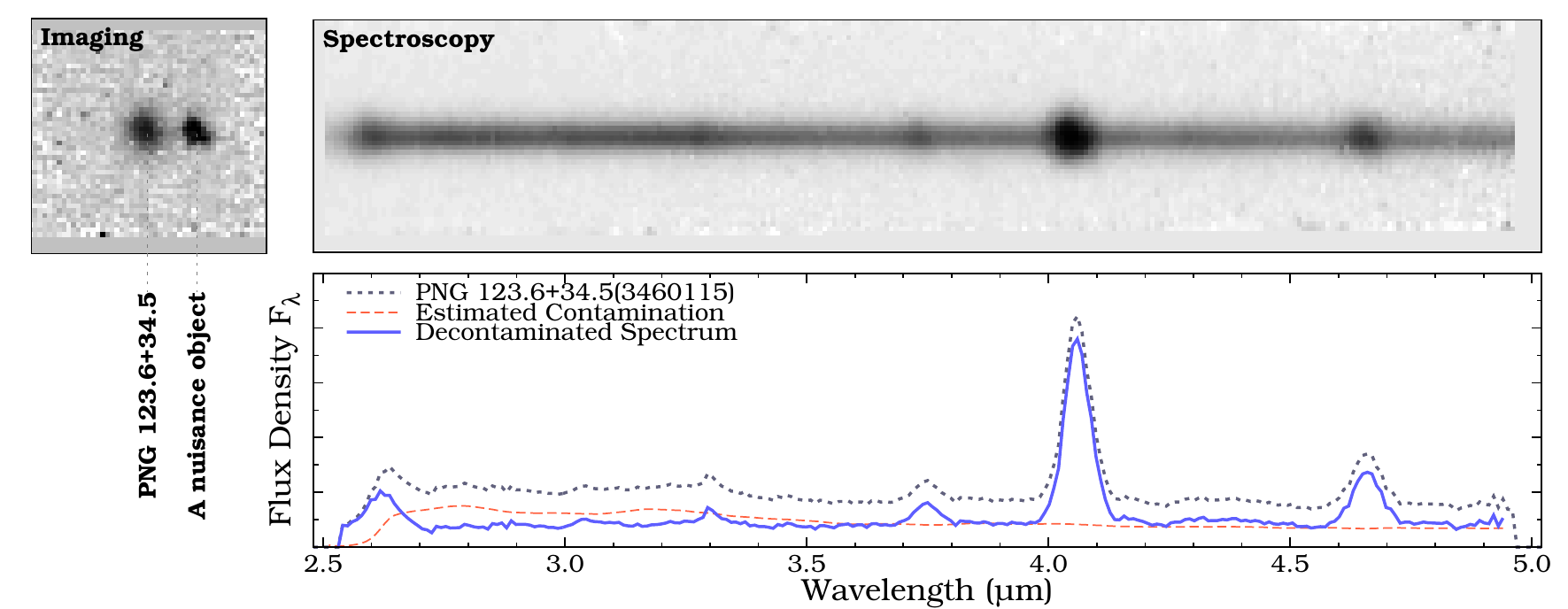}
  \caption{Schematic view of the decontamination procedure. The upper left panel shows the locations of a PN and an overlapping star, while the upper right panel shows the obtained spectrum with contamination from the overlapping star. The contaminated spectrum is shown by the gray dotted line in the bottom panel. The red dashed line shows the estimated contamination from the overlapping star. The decontaminated spectrum of the PN is shown by the blue solid line.}
  \label{fig:decontami}
\end{figure}
For some observations, the targets were found to be aligned with neighboring objects, along the direction of dispersion. In such cases, the spectra of these objects fully overlapped with each other. This made retrieval of the target profile spectrum, by profile fitting alone, impossible (\textit{see} the upper panels in Figure \ref{fig:decontami}). To remove the contamination from the overlapping objects, we developed a method to estimate the amount of the contamination (hereafter, the decontamination procedure). The method basically follows the procedure used in \citet{sakon_data_2008}. At first, a one-dimensional spectrum was extracted with a wide aperture. We assumed that overlapping objects were stars and their spectra did not show any particular spectral features in the spectral range in question. Their 2.5--5.0$\,\mu$m spectra ($\tilde{f}_{\nu}$) were estimated from the 2MASS, \textit{AKARI}, and \textit{WISE} photometric data by fitting with a blackbody function. The calibration of the overlapping objects required correction since the overlapping object was shifted along the direction of dispersion. The amount of contamination $f^{c}_{\nu}(\lambda)$ was converted from $\tilde{f}_{\nu}$ using the distance between the target and the overlapping objects in the direction of the dispersion, $d$, in units of pixel and the spectral response function of the IRC, $R(\lambda)$, in units of ${\rm ADU}\,{\rm mJy^{-1}}$. We defined $\delta$ as the pixel scale of the spectrum in units of $\mu$m$\,{\rm pixel}^{-1}$. The amount of the contamination was estimated by
\begin{equation}
  \label{eq:contamination}
  f^{c}_{\nu}(\lambda) =
  \frac{
    \tilde{f}_{\nu}(\lambda{-}\delta d)R(\lambda{-}\delta d)
  } {
    R(\lambda)
  }.
\end{equation}
We defined $f^{\rm obs}_{\nu}$ as the obtained spectrum, including the contamination. The decontaminated spectrum was defined by $f^{\rm obj}_{\nu} = f^{\rm obs}_{\nu} - f^{c}_{\nu}$. A schematic view of the method is given in Figure \ref{fig:decontami}. The upper left panel shows the location of the target (PN\,G123.6+34.5) and overlapping object. The direction of dispersion is horizontal. The spectrum with the contamination is shown in the upper right panel. The contaminated one-dimensional spectrum is shown by the gray dotted line in the bottom panel. The estimated contamination and decontaminated spectra are shown by the red dashed and blue solid lines. We assume that the spectrum of the overlapping star does not have any significant features and that the decontamination procedure does not affect the estimate of emission feature intensities (although continuum flux could be somewhat affected). The targets with overlapping objects are flagged as contaminated objects in the tenth column of Table~\ref{tab:basic_info}.

\section{Results}\label{sec:results}
\subsection{Near-Infrared Spectroscopy of Planetary Nebulae Spectral Atlas}\label{sec:res:atlas}
Details of the observations are summarized in Table~\ref{tab:basic_info}, which includes the PNG\,ID, the observation ID, the location, the number of exposures, the total integration time, the spectral PSF size, and the flags to denote data quality. The PNSPC catalog contains extracted one-dimensional spectra shown in Figure~\ref{fig:allspectrum} (plots for all the sources are available in the electric edition). Most of the spectra cover $2.5$--$5.0\,\mu$m. Some spectra were affected by scattered light inside the instrument or column-pull-down effects of the detector, which were not correctable at present. The part of the spectrum severely affected by such issues has been excluded from the catalog.

\begin{figure}[!tp]
  \centering
  \includegraphics[width=1.0\linewidth]{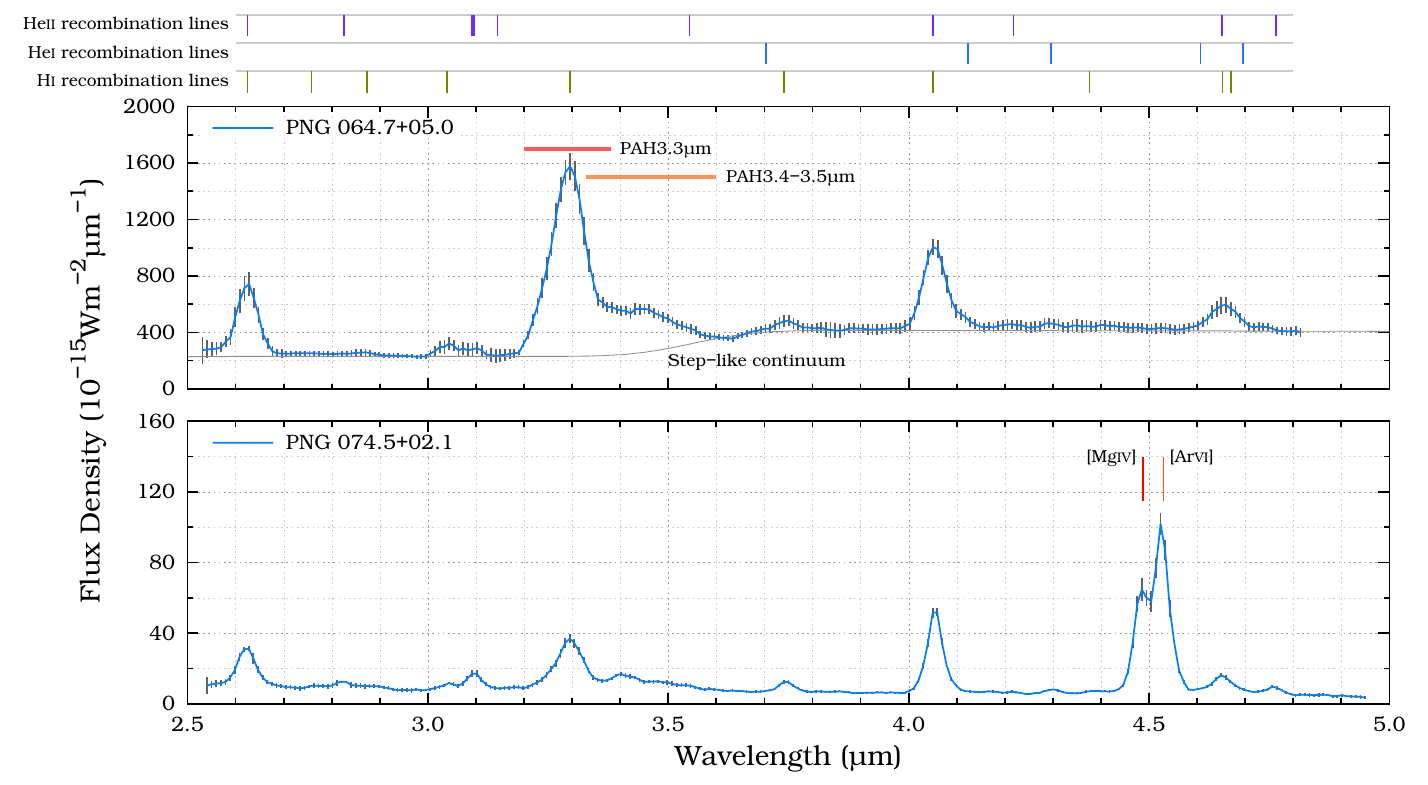}
  \caption{Representative spectra in the PNSPC catalog. The top and bottom panels show the spectra of PNG\,064.7$+$05.0 and PNG\,074.5$+$02.1 by the blue solid lines with errors. Series of recombination lines (\ion{H}{1}, \ion{He}{1}, and \ion{He}{2}) are indicated by the vertical lines. The both spectra show the PAH feature in 3.3$\,\mu$m and the aliphatic feature complex in 3.4--3.5$\,\mu$m. The spectrum of PNG\,074.5$+$02.1 shows fine-structure lines of [\ion{Mg}{4}] at 4.487$\,\mu$m and [\ion{Ar}{6}] at 4.530$\,\mu$m. Step-like continuum emission is seen in the spectra of PNG\,064.7$+$05.0 (\textit{see} Section~\ref{sec:res:fitting}).}
  \label{fig:samplespc}
\end{figure}

Two representative spectra in the PNSPC catalog are shown in Figure~\ref{fig:samplespc}. The top panel is the spectrum of PNG\,064.7$+$05.0, and the bottom panel is that of PNG\,074.5$+$02.1. The spectra showed several emission features: \ion{H}{1} and \ion{He}{2} recombination lines, the PAH feature at $3.3\,\mu$m, and the aliphatic feature complex in $3.4$--$3.5\,\mu$m. Some spectra also showed \ion{He}{2} recombination lines and [\ion{Mg}{4}] at ${\rm 4.49\,\mu m}$ and [\ion{Ar}{6}] at ${\rm 4.53\,\mu m}$ fine-structure lines, indicating that the excitation of the object was high. Some objects showed molecular hydrogen lines, but they were weak and not frequently detected. No prominent absorption features were detected in any of the objects.

\begin{figure}[!tp]
  \centering
  \includegraphics[width=1.0\linewidth]{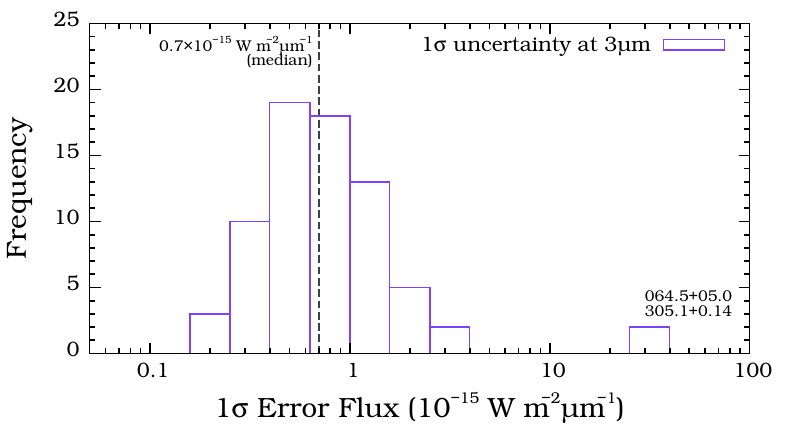}
  \includegraphics[width=1.0\linewidth]{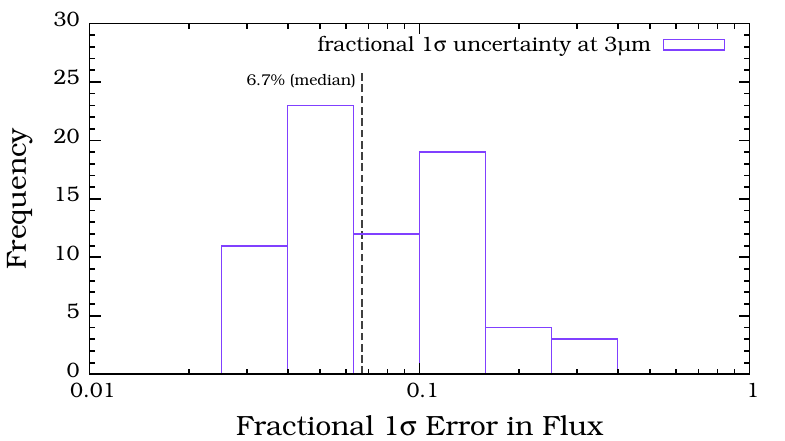}
  \caption{Histograms of the uncertainty in continuum emission at $3.0\,\mu$m. The top panel shows the histogram of absolute 1$\sigma$ errors. The gray dashed line indicates the median of the uncertainty ($0.7{\times}10^{-15}\,{\rm W\,m^{-2}\,\mu m^{-1}}$). The data around 30$\mathit{\times}10^{-15}\,{\rm W\,m^{-2}\,\mu m^{-1}}$ are very bright sources of PNG\,064,7${+}$0.5 and 305.1${+}$01.4, for which the spectra were entirely saturated in the long exposure images. The bottom panel shows the histogram of fractional 1$\sigma$ errors. The median value (${\sim}$6.7$\%$) is shown by the gray dashed line.}
  \label{fig:sigmahist}
\end{figure}

The top panel of Figure~\ref{fig:sigmahist} shows a histogram of the uncertainty in continuum emission at $3.0\,\mu$m. The gray dashed line indicates the median of the uncertainty, which is about $0.7{\times}10^{-15}\,{\rm W\,m^{-2}\,\mu m^{-1}}$ or $2\,$mJy. When the object is brighter than $2$--$5{\times}10^{-13}\,{\rm W\,m^{-2}\,\mu m^{-1}}$, the pixels become saturated in the long exposure image. For moderately bright objects, pixels around emission lines were saturated. The spectra of extremely bright objects were completely saturated in the long exposure images. The signal-to-noise ratio of the spectrum depends on the fraction of saturated pixels. PNG\,064.7${+}$05.0 and 305.1${+}$01.4 are extremely bright, with their spectrum completely saturated in the long exposure images. Figure~\ref{fig:sigmahist} (top) indicates that their uncertainties at $3.0\,\mu$m are about 50 times worse than the others. The fractional uncertainty at $3.0\,\mu$m is shown in the bottom panel of Figure~\ref{fig:sigmahist}. The median of the fractional uncertainty is about 6.4\%, indicated by the vertical dashed line. This means that, for half of the PNSPC samples, the continuum emission at $3.0\,\mu$m was detected with a signal-to-noise ratio larger than about 15.

The full width half maximum (FWHM) of the spectral PSF of the IRC, equivalent to the FWHM of an unresolved emission line, is about $0.035\,\mu$m. Thus, the spectral resolution is about $\lambda/\Delta\lambda \sim 100$. When the object is extended for the IRC, the spectral resolution is degraded. The FWHM of the spectral PSF, estimated from spectral fitting (\textit{see} Section~\ref{sec:res:fitting}), is listed in the seventh column of Table~\ref{tab:basic_info}. PN\,G322.5${-}$05.2 is the most extended object in the PNSPC catalog and its spectral resolution was estimated to be about $\lambda/\Delta\lambda \sim 40$ at 4$\,\mu$m.

\subsection{Extinction}\label{sec:res:extinction}
Foreground extinction towards the objects was estimated based on an intensity ratio of H$\beta$ to Brackett-$\alpha$ (Br$\alpha$). The intensities of H$\beta$ were obtained from literature \citep[and references therein]{acker_strasbourg-eso_1992}. We measured the intensities of Br$\alpha$ from the \textit{AKARI} spectra by fitting with a Gaussian function. We defined $I_{\rm H\beta}$ and $I_{\rm Br\alpha}$ as the intensities of H$\beta$ and Br$\alpha$. The observed intensity ratio follows the relation:
\begin{equation}
  \label{eq:lineextinction}
  \frac{I_{\rm H\beta}^{\rm obs}}{I_{\rm Br\alpha}^{\rm obs}} =
  \frac{I_{\rm H\beta}^{0}{\rm e}^{-\tau_{\rm H\beta}}}
  {I_{\rm Br\alpha}^{0}{\rm e}^{-\tau_{\rm Br\alpha}}},
\end{equation}
where $I^{\rm obs}_X$ and $I^{0}_{X}$ are the observed and intrinsic intensities of the line $X$, and $\tau_X$ is the extinction of the line $X$. We assumed that the nebula was totally opaque for Lyman photons (Case B in \citet{baker_physical_1938}). Then, the intrinsic intensity ratio $\left({I_{\rm H\beta}^{0}}/{I_{\rm Br\alpha}^{0}}\right)$ was assumed to be ${\sim}12.853$, given by the Case B line ratio calculated for the electron density of $10^4\,{\rm cm^{-3}}$ and the electron temperature of $10^4\,$K from \citet{storey_recombination_1995}.  We adopted the extinction curve given by \citet{mathis_interstellar_1990} and the extinction at the $V$-band was derived as
\begin{equation}
  \label{eq:av_lineratio}
  {\rm A}_V({\rm H\beta}) =
  2.39-0.95\ln\left(
    \frac{I_{\rm H\beta}^{\rm obs}}{I_{\rm Br\alpha}^{\rm obs}}
  \right).
\end{equation}
The coefficients in Equation~(\ref{eq:av_lineratio}) depends on the assumed electron temperature and density. When the assumed electron temperature increases to $2{\times}10^4\,$K, ${\rm A}_V$ will increase by about 0.3. The effect of the electron density on ${\rm A}_V$ is much smaller than 0.3 for the range from $10^2$ to $10^6\,{\rm cm^{-3}}$. The systematic uncertainty in ${\rm A}_V$, chiefly attributable to the electron temperature uncertainty, was estimated to be less than $0.3\,$mag. The estimated ${\rm A}_V({\rm H\beta})$ is listed in the fourth colunm of Table~\ref{tab:basic_info_misc}.

\subsection{Spectral Fitting}\label{sec:res:fitting}
The intensities of the spectral features were measured by spectral fitting. The fitted function was assumed to be a linear combination of the spectral features as
\begin{equation}
  \label{eq:fittedfunc}
  F_{\lambda}(\lambda) =
  \Bigl[
    f_{\lambda}^{\rm cont}(\lambda) +
    f_{\lambda}^{\rm line}(\lambda) +
    f_{\lambda}^{\rm dust}(\lambda)
  \Bigr]{\rm e}^{-\tau(\lambda)},
\end{equation}
where $f_{\lambda}^{\rm cont}$, $f_{\lambda}^{\rm line}$, and $f_{\lambda}^{\rm dust}$ were the components of continuum, emission lines, and dust features, and $\tau(\lambda)$ was the optical depth at $\lambda$. The functional shape of $\tau(\lambda)$ was given by a cubic spline function that interpolated the extinction curve of \citet{mathis_interstellar_1990} at every data point of the IRC spectrum. The amount of the extinction at the $V$-band, ${\rm A}_V = \tau(V)$, was included as a fitting parameter.

The continuum emission, $f_{\lambda}^{\rm cont}$, was given by a polynomial function, $f_{\lambda}^{\rm cont} = \sum_{l=0}^N c_l \lambda^l$, where $c_l$'s were the fitting parameters. The order $N$ was defined for each object, usually $N=2$ or $3$. This was sufficient to represent the contributions from free-free and hot thermal dust emission. Some PNe showed a sudden increase in the continuum emission around $\lambda \sim 3.6\,\mu$m, for instance, PN\,G064.7${+}$05.0 in Figure~\ref{fig:samplespc}. The existence of this emission was mentioned in \citet{boulanger_near-infrared_2011}, but the carrier of this emission has not been identified. It was difficult to fit this profile with such a low-order polynomial function. Thus, for those objects with such an increase, we added a step-like function in the components of the continuum emission as
\begin{equation}
  \label{eq:stepfunc}
  f_{\lambda}^{\rm cont}(\lambda) =
  \sum_{l=0}^N c_l \lambda^l +
  c\left[~
    \frac{1}{2} + \frac{1}{2}{\rm erf}\left(
      \frac{\lambda - 3.65{\,\rm \mu m}}{\Delta_{\rm step}}
    \right)\,
  \right],
\end{equation}
where $c$ was the fitting parameter, ${\rm erf}(x)$ was the error function, and $\Delta_{\rm step}$ was the width of the step function, which was tentatively fixed at 0.135$\,\mu$m. This component strongly appears in PNG\,037.8${-}$06.3, 064.7${+}$050, 089.8${-}$05.1, and 268.4${+}$02.4.

The emission line profile was approximated by a Gaussian function. The line emission component, $f_{\lambda}^{\rm line}$, was approximated by a combination of Gaussian functions:
\begin{equation}
  \label{eq:linefunc}
  f_{\lambda}^{\rm line}(\lambda) =
  \sum_m \frac{c_m}{\sqrt{2\pi}\sigma}
  \exp\left[\,
    -\frac{(\lambda-\lambda_{m0})^2}{2\sigma^2}
  \,\,\right],
\end{equation}
where $\lambda_{m0}$ was the central wavelengths of the $m$-th emission line and $c_m$'s were the parameters to indicate the strength of the lines. The central wavelengths were fixed in the fitting. The line width of the Br$\alpha$ line was measured by fitting with a Gaussian function prior to the fitting and the line widths of other emission lines were assumed to be the same as Br$\alpha$. The lines included in the fitting are listed in Table~\ref{tab:linelist}. The assignment and central wavelength are shown in the first and second column. In the fitting, we did not include the lines which were apparently not detected. Some of the \ion{H}{1} and \ion{He}{2} recombination lines appear at the same wavelength and it was impossible to isolate them. Thus, the relative intensities of the \ion{H}{1} and \ion{He}{2} recombination lines were fixed, assuming the Case B condition. The relative intensity was adopted from \citet{storey_recombination_1995} for the electron temperature and the electron density of $10^4\,$K and $10^4\,{\rm cm^{-2}}$. The variations in the relative intensity with the electron temperature and density were almost negligible compared to the uncertainty in the observed spectra.

\begin{table}[!t]
  \centering
  \caption{List of Line Features}\label{tab:linelist}
  \begin{tabular}{lc}
    \tableline\tableline
    Name & $\lambda_0$ ($\mu$m) \\
    \tableline
    Brakett-$\alpha^\dagger$ & 4.051 \\
    Brakett-$\beta^\dagger$  & 2.625 \\
    Pfund-$\beta^\dagger$    & 4.653 \\
    Pfund-$\gamma^\dagger$   & 3.740 \\
    Pfund-$\delta^\dagger$   & 3.296 \\
    Pfund-$\epsilon^\dagger$ & 3.039 \\
    Pfund-$\zeta^\dagger$    & 2.873 \\
    Pfund-$\eta^\dagger$     & 2.758 \\
    Humpleys-$\epsilon^\dagger$ & 4.671 \\
    Humpleys-$\zeta^\dagger$    & 4.376 \\
    \ion{He}{1}(3P$^{\rm o}$-3D)  & 3.704 \\
    \ion{He}{1}(1P$^{\rm o}$-1D)  & 4.123 \\
    \ion{He}{1}(3S-3P$^{\rm o}$)  & 4.296 \\
    \ion{He}{1}(1P$^{\rm o}$-1S)  & 4.607 \\
    \ion{He}{1}(3P$^{\rm o}$-3S)  & 4.695 \\
    & \\
    & \\
    \tableline\tableline
  \end{tabular}
  \begin{tabular}{lc}
    \tableline\tableline
    Name & $\lambda_0$ ($\mu$m) \\
    \tableline
    \ion{He}{2}(7-6)$^\ddagger$    & 3.092 \\
    \ion{He}{2}(9-7)$^\ddagger$    & 2.826 \\
    \ion{He}{2}(8-7)$^\ddagger$    & 4.764 \\
    \ion{He}{2}(12-8)$^\ddagger$   & 2.625 \\
    \ion{He}{2}(11-8)$^\ddagger$   & 3.096 \\
    \ion{He}{2}(10-8)$^\ddagger$   & 4.051 \\
    \ion{He}{2}(14-9)$^\ddagger$   & 3.145 \\
    \ion{He}{2}(13-9)$^\ddagger$   & 3.544 \\
    \ion{He}{2}(12-9)$^\ddagger$   & 4.218 \\
    \ion{He}{2}(14-10)$^\ddagger$  & 4.652 \\
    H$_2$(1-0)  & 2.802 \\
    H$_2$(1-0)  & 3.003 \\
    H$_2$(1-0)  & 3.234 \\
    H$_2$(0-0)  & 3.836 \\
    H$_2$(0-0)  & 4.181 \\
    $[$\ion{Mg}{4}$]$  & 4.487 \\
    $[$\ion{Ar}{6}$]$  & 4.530 \\
    \tableline\tableline
  \end{tabular}
  \tablecomments{$^{\dagger,\ddagger}$: relative intensities of these lines are fixed assuming the Case B condition.}
\end{table}

The intrinsic spectral profiles of the PAH emission and the aliphatic feature complex were assumed to be given by a combination of Lorentzian functions which need to be convoluted with the resolution given by the instrument. Thus, their spectral profiles were approximated by the Lorentzian function convoluted with the spectral PSF of the IRC, which was approximated by a Gaussian. The component from dust emission, $f_{\lambda}^{\rm dust}$, was defined by
\begin{equation}
  \label{eq:dustfunc}
  f_{\lambda}^{\rm dust}(\lambda) =
  \sum_n \frac{c_n}{\pi\sqrt{2\pi}\sigma}
  \int{\rm d}\lambda' {\rm e}^{-\frac{(\lambda-\lambda')^2}{2\sigma^2}}
  \frac{\Delta_n}{\Delta_n^2 + (\lambda' - \lambda_{n0})^2},
\end{equation}
where $\Delta_n$ and $\lambda_{n0}$ were the width and peak wavelength of the $n$-th dust emission and $c_n$'s were the fitting parameters. The adopted values of $\Delta_n$'s and $\lambda_{n0}$'s were listed in Table~\ref{tab:dustlist}. The PAH feature at 3.3$\,\mu$m was mainly composed of the 3.30$\,\mu$m feature, but the 3.25$\,\mu$m feature was added to account for a slight change in the spectral profile of the 3.3$\,\mu$m feature. For instance the 3.3$\,\mu$m feature of PNG\,320.9${+}$02.0 show an extended blue component, which cannot be well-fitted without the 3.25$\,\mu$m component. \citet{mori_observations_2012} used a combination of four Lorentzian functions to describe the 3.4--3.5$\,\mu$m feature complex ({C{\sbond}H{$_{al}$}}) in spectra obtained with the IRC. We used a combination of the four components to describe the {C{\sbond}H{$_{al}$}} feature. The central wavelengths and widths were adopted from \citet{mori_observations_2012}.

\begin{table}[!t]
  \centering
  \caption{List of Dust Features}\label{tab:dustlist}
  \begin{tabular}{lcc}
    \tableline\tableline
    Name & $\lambda_0$ ($\mu$m) & $\Delta$ ($\mu$m)\\
    \tableline
    PAH\,3.25$^\dagger$  & 3.248 & 0.005 \\
    PAH\,3.30$^\dagger$  & 3.296 & 0.021 \\
    C{\sbond}H{$_{al}$}\,3.41$^\ddagger$ & 3.410 & 0.030 \\
    C{\sbond}H{$_{al}$}\,3.46$^\ddagger$ & 3.460 & 0.013 \\
    C{\sbond}H{$_{al}$}\,3.51$^\ddagger$ & 3.512 & 0.013 \\
    C{\sbond}H{$_{al}$}\,3.56$^\ddagger$ & 3.560 & 0.013 \\
    \tableline\tableline
  \end{tabular}
  \tablecomments{$^{\dagger}$: they belong to the PAH feature at 3.3$\,\mu$m. $^{\ddagger}$: they belong to the aliphatic feature complex in 3.4--3.5$\,\mu$m.}
\end{table}

Define $F_{\lambda}^{\rm obj}(\lambda)$ and $\sigma_{\lambda}(\lambda)$ as the observed flux and its uncertainty at $\lambda$. We assumed that the residual $\Delta_{\lambda}(\lambda) = F_{\lambda}^{\rm obj}(\lambda) - F_{\lambda}(\lambda)$ followed a normal distribution with the variance of $\sigma^2_\lambda(\lambda)$. The likelihood function was defined as
\begin{equation}
  \label{eq:likelihood}
  \mathcal{L} \propto \prod_i 
  \exp\left[
    -\frac{1}{2}\left(
      \frac{\Delta_{\lambda}(\lambda_i)}{\sigma_{\lambda}(\lambda_i)}
    \right)^2
  \right],
\end{equation}
where $\lambda_i$ was the wavelength of the $i$-th spectral element. The fitting parameters could be defined by maximizing Equation (\ref{eq:likelihood}). However, it was difficult to accurately derive the amount of the extinction, ${\rm A}_V$, only from the near-infrared spectrum (\textit{see}, Appendix \ref{sec:app:uncertainAv}). To make the fitting stable, we used the extinction estimated in Section \ref{sec:res:extinction} to provide a prior distribution of the extinction
\begin{equation}
  \label{eq:avprior}
  P({\rm A}_V) \propto
  \exp\left[
    - \frac{1}{2}\left(
      \frac{{\rm A}_V {-} {\rm A}_V({\rm H\beta})}{\sigma''}
    \right)^2
  \right],
\end{equation}
where ${\rm A}_V({\rm H\beta})$ was the extinction estimated from the H$\beta$ to Br$\alpha$ intensity ratio and $\sigma''$ was tentatively fixed at $0.3\,$mag, estimated from the systematic error. The fitting parameters were derived by maximizing the product of Equations (\ref{eq:likelihood}) and (\ref{eq:avprior}) with the constraints that the parameters of ${\rm A}_V$, $c$, $c_m$ and $c_n$ should be non-negative. The uncertainties of the fitting parameters were estimated based on a Monte Carlo simulation: The spectrum of the target, $\tilde{F}(\lambda)$ was simulated assuming a normal distribution of flux with a mean of $F^{\rm obj}_\lambda(\lambda)$ and a variance of $\sigma_\lambda(\lambda)$. The fitting parameters were derived for $\tilde{F}(\lambda)$. This procedure was repeated 1\,000 times and the probability distributions of the fitting parameters were obtained. The confidence interval of the fitting parameters was defined to account for 68\% of the trials.

\begin{figure}
  \centering
  \includegraphics[width=1.0\linewidth]{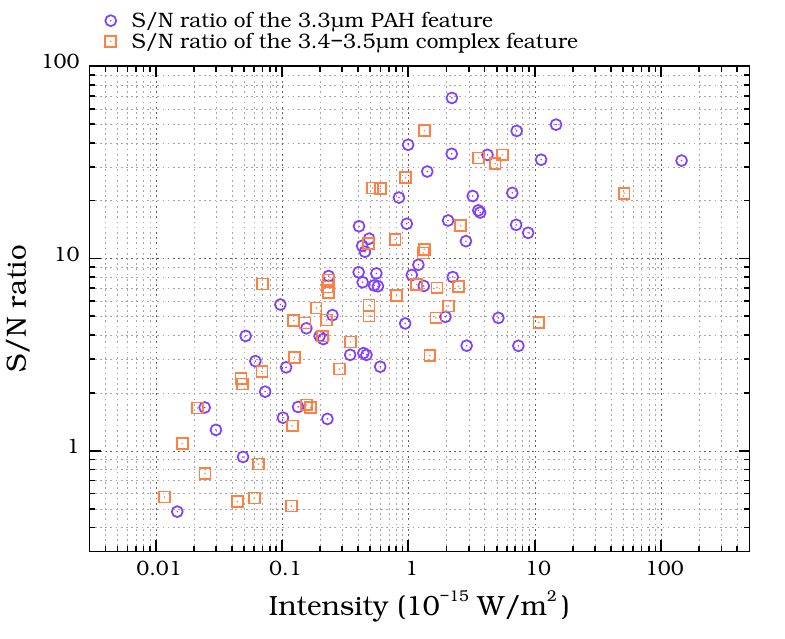}
  \caption{The signal-to-noise ratio (${\rm S/N}$) is plotted against the intensity of the emission features, indicating a typical detection limit of emission features. The blue circles represent the ${\rm S/N}$ ratio of the 3.3$\,\mu$m PAH feature, while the red squares show that of the 3.4--3.5$\,\mu$m aliphatic feature complex.}
  \label{fig:detectionlimit}
\end{figure}

The extinction obtained by the fitting (${\rm A}_V(\rm fit.)$) is listed in the fifth column of Table~\ref{tab:basic_info_misc}. Table~\ref{tab:intensity} lists the extinction-corrected intensities of Br$\alpha$ at $4.05\,\mu$m, \ion{He}{1} at $4.30\,\mu$m, \ion{He}{2} at $3.09\,\mu$m, the PAH feature at $3.3\,\mu$m, the aliphatic feature complex in $3.4$--$3.5\,\mu$m, [\ion{Mg}{4}] at $4.49\,\mu$m, and [\ion{Ar}{6}] at $4.53\,\mu$m. These emission features were strong and widely detected among the PNSPC samples. Other line features were weak or not detected in the majority of the targets and thus were not included in the table. The equivalent widths of those are listed in Table \ref{tab:equivw}. Note that [\ion{Mg}{4}] and [\ion{Ar}{6}] are so close that their spectral profiles are in part overlapping. Their uncertainties are relatively large compared to other lines, because the effect of overlapping was taken into account. Figure~\ref{fig:detectionlimit} shows the signal-to-noise ratio against the intensity of the $3.3\,\mu$m PAH feature and the 3.4--3.5$\,\mu$m aliphatic feature complex by the blue circles and the red squares. Figure~\ref{fig:detectionlimit} indicates that the emission feature stronger than ${\sim}2{\times}10^{-16}\,{\rm W\,m^{-2}}$ was detected with a 3$\sigma$-confidence level.

\begin{deluxetable}{cccccccc}
\rotate
\tabletypesize{\small}
\tablewidth{0pt}

\tablecaption{Extinction Corrected Line Intensity \label{tab:intensity_s}}
\tablehead{
  \colhead{PN\,G} &
  \colhead{Bracket-$\alpha$} &
  \colhead{\ion{He}{2}{\footnotesize$(3.09{\rm\mu m})$}} &
  \colhead{\ion{He}{1}{\footnotesize$(4.30{\rm\mu m})$}} &
  \colhead{{PAH}{\footnotesize$(3.3{\rm\mu m})$}} &
  \colhead{{C{\sbond}H{$_{al}$}}{\footnotesize$(3.4$-$3.5{\rm\mu m})$}} &
  \colhead{[\ion{Mg}{4}]{\footnotesize$(4.49{\rm\mu m})$}} &
  \colhead{[\ion{Ar}{6}]{\footnotesize$(4.53{\rm\mu m})$}} \\
  \colhead{} &
  \colhead{$10^{-15}\,{\rm W\,m^{-2}}$} &
  \colhead{$10^{-15}\,{\rm W\,m^{-2}}$} &
  \colhead{$10^{-15}\,{\rm W\,m^{-2}}$} &
  \colhead{$10^{-15}\,{\rm W\,m^{-2}}$} &
  \colhead{$10^{-15}\,{\rm W\,m^{-2}}$} &
  \colhead{$10^{-15}\,{\rm W\,m^{-2}}$} &
  \colhead{$10^{-15}\,{\rm W\,m^{-2}}$}
}
\startdata
000.3${+}$12.2 & $\phn\phn\phn2.74_{{-}0.04}^{{+}0.05}$ & \nodata & $\phn\phn\phn0.14_{{-}0.03}^{{+}0.03}$ & ${<}   0.11$ & \nodata & \nodata & \nodata \\
002.0${-}$13.4 & $\phn\phn\phn2.05_{{-}0.03}^{{+}0.04}$ & \nodata & $\phn\phn\phn0.31_{{-}0.02}^{{+}0.03}$ & $\phn\phn\phn0.25_{{-}0.05}^{{+}0.05}$ & ${<}   0.10$ & \nodata & \nodata \\
003.1${+}$02.9 & $\phn\phn\phn3.16_{{-}0.04}^{{+}0.04}$ & ${<}   0.53$ & $\phn\phn\phn0.43_{{-}0.04}^{{+}0.03}$ & $\phn\phn\phn0.94_{{-}0.16}^{{+}0.25}$ & $\phn\phn\phn0.49_{{-}0.08}^{{+}0.09}$ & $\phn\phn\phn0.35_{{-}0.04}^{{+}0.05}$ & \nodata \\
011.0${+}$05.8 & $\phn\phn\phn1.09_{{-}0.04}^{{+}0.03}$ & $\phn\phn\phn0.13_{{-}0.03}^{{+}0.03}$ & ${<}   0.13$ & \nodata & \nodata & $\phn\phn\phn1.11_{{-}0.08}^{{+}0.05}$ & $\phn\phn\phn0.20_{{-}0.05}^{{+}0.04}$ \\
027.6${-}$09.6 & $\phn\phn\phn1.12_{{-}0.02}^{{+}0.04}$ & \nodata & $\phn\phn\phn0.10_{{-}0.01}^{{+}0.01}$ & \nodata & \nodata & \nodata & \nodata \\
037.8${-}$06.3 & $\phn\phn\phn4.61_{{-}0.10}^{{+}0.24}$ & \nodata & $\phn\phn\phn1.80_{{-}0.09}^{{+}0.13}$ & $\phn\phn\phn6.64_{{-}0.21}^{{+}0.40}$ & $\phn\phn\phn2.60_{{-}0.14}^{{+}0.21}$ & \nodata & \nodata \\
038.2${+}$12.0 & $\phn\phn\phn2.18_{{-}0.06}^{{+}0.04}$ & \nodata & ${<}   0.13$ & $\phn\phn\phn0.45_{{-}0.05}^{{+}0.04}$ & $\phn\phn\phn0.12_{{-}0.02}^{{+}0.03}$ & \nodata & \nodata \\
043.1${+}$03.8 & $\phn\phn\phn0.45_{{-}0.01}^{{+}0.01}$ & \nodata & \nodata & ${<}   0.06$ & ${<}   0.01$ & \nodata & \nodata \\
046.4${-}$04.1 & $\phn\phn\phn2.43_{{-}0.01}^{{+}0.02}$ & $\phn\phn\phn0.11_{{-}0.02}^{{+}0.02}$ & $\phn\phn\phn0.32_{{-}0.01}^{{+}0.02}$ & $\phn\phn\phn0.40_{{-}0.05}^{{+}0.04}$ & ${<}   0.16$ & \nodata & \nodata \\
051.4${+}$09.6 & $\phn\phn\phn3.25_{{-}0.02}^{{+}0.06}$ & \nodata & $\phn\phn\phn0.52_{{-}0.02}^{{+}0.02}$ & $\phn\phn\phn0.97_{{-}0.07}^{{+}0.05}$ & $\phn\phn\phn0.23_{{-}0.05}^{{+}0.04}$ & \nodata & \nodata \\
\enddata

\tablecomments{Table~\ref{tab:intensity_s} is published in its entirety in the electronic edition of the Astrophysical Journal. A portion is shown here for guidance regarding its form and content.}
\end{deluxetable}

\begin{deluxetable}{cccccccc}
\rotate
\tabletypesize{\small}
\tablewidth{0pt}

\tablecaption{Extinction Corrected Line Equivalent Width \label{tab:equivw_s}}
\tablehead{
  \colhead{PN\,G} &
  \colhead{Bracket-$\alpha$} &
  \colhead{\ion{He}{2}{\footnotesize$(3.09{\rm\mu m})$}} &
  \colhead{\ion{He}{1}{\footnotesize$(4.30{\rm\mu m})$}} &
  \colhead{{PAH}{\footnotesize$(3.3{\rm\mu m})$}} &
  \colhead{{C{\sbond}H{$_{al}$}}{\footnotesize$(3.4$-$3.5{\rm\mu m})$}} &
  \colhead{[\ion{Mg}{4}]{\footnotesize$(4.49{\rm\mu m})$}} &
  \colhead{[\ion{Ar}{6}]{\footnotesize$(4.53{\rm\mu m})$}} \\
  \colhead{} &
  \colhead{{\AA}} &
  \colhead{{\AA}} &
  \colhead{{\AA}} &
  \colhead{{\AA}} &
  \colhead{{\AA}} &
  \colhead{{\AA}} &
  \colhead{{\AA}}
}
\startdata
000.3${+}$12.2 & $\phn2825.6_{{-}38.2}^{{+}47.3}$ & \nodata & $\phn\phn154.5_{{-}32.7}^{{+}35.5}$ & ${<}  130.7$ & \nodata & \nodata & \nodata \\
002.0${-}$13.4 & $\phn2938.4_{{-}39.9}^{{+}64.0}$ & \nodata & $\phn\phn481.9_{{-}27.1}^{{+}42.1}$ & $\phn\phn435.3_{{-}84.6}^{{+}86.8}$ & ${<}  181.4$ & \nodata & \nodata \\
003.1${+}$02.9 & $\phn2105.7_{{-}25.5}^{{+}26.5}$ & ${<}  236.7$ & $\phn\phn343.7_{{-}30.2}^{{+}25.6}$ & $\phn\phn665.7_{{-}113.7}^{{+}176.0}$ & $\phn\phn365.1_{{-}59.5}^{{+}68.0}$ & $\phn\phn324.3_{{-}42.0}^{{+}51.4}$ & \nodata \\
011.0${+}$05.8 & $\phn2005.1_{{-}68.8}^{{+}61.4}$ & $\phn\phn144.8_{{-}34.2}^{{+}32.1}$ & ${<}  266.9$ & \nodata & \nodata & $\phn2605.4_{{-}179.7}^{{+}127.9}$ & $\phn\phn486.2_{{-}111.4}^{{+}101.8}$ \\
027.6${-}$09.6 & $\phn3229.6_{{-}44.8}^{{+}128.8}$ & \nodata & $\phn\phn328.8_{{-}27.6}^{{+}35.8}$ & \nodata & \nodata & \nodata & \nodata \\
037.8${-}$06.3 & $\phn\phn482.1_{{-}10.0}^{{+}25.4}$ & \nodata & $\phn\phn182.2_{{-}\phn8.6}^{{+}13.6}$ & $\phn1342.1_{{-}42.1}^{{+}80.2}$ & $\phn\phn531.1_{{-}29.1}^{{+}42.6}$ & \nodata & \nodata \\
038.2${+}$12.0 & $\phn2863.3_{{-}84.4}^{{+}54.5}$ & \nodata & ${<}  192.5$ & $\phn\phn709.0_{{-}75.8}^{{+}55.0}$ & $\phn\phn204.3_{{-}38.7}^{{+}47.2}$ & \nodata & \nodata \\
043.1${+}$03.8 & $\phn1316.2_{{-}37.5}^{{+}42.9}$ & \nodata & \nodata & ${<}  230.0$ & \nodata & \nodata & \nodata \\
046.4${-}$04.1 & $\phn3171.2_{{-}\phn5.0}^{{+}29.3}$ & $\phn\phn\phn99.6_{{-}18.0}^{{+}17.8}$ & $\phn\phn475.1_{{-}\phn9.3}^{{+}29.6}$ & $\phn\phn595.7_{{-}75.1}^{{+}65.3}$ & ${<}  250.8$ & \nodata & \nodata \\
051.4${+}$09.6 & $\phn1882.4_{{-}13.5}^{{+}36.4}$ & \nodata & $\phn\phn315.8_{{-}12.1}^{{+}15.0}$ & $\phn\phn740.7_{{-}56.1}^{{+}41.7}$ & $\phn\phn180.1_{{-}42.6}^{{+}32.7}$ & \nodata & \nodata \\
\enddata

\tablecomments{Table~\ref{tab:equivw_s} is published in its entirety in the electronic edition of the Astrophysical Journal. A portion is shown here for guidance regarding its form and content.}

\end{deluxetable}

\begin{figure*}[p]
  \centering
  \includegraphics[width=1.0\linewidth]{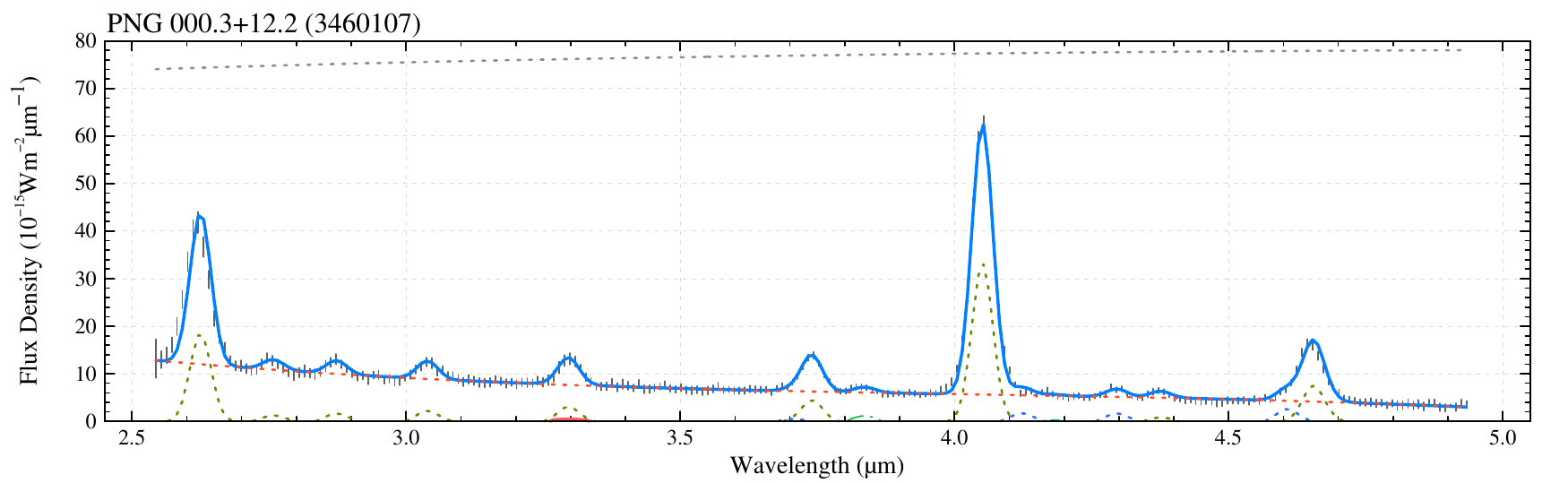}
  \includegraphics[width=1.0\linewidth]{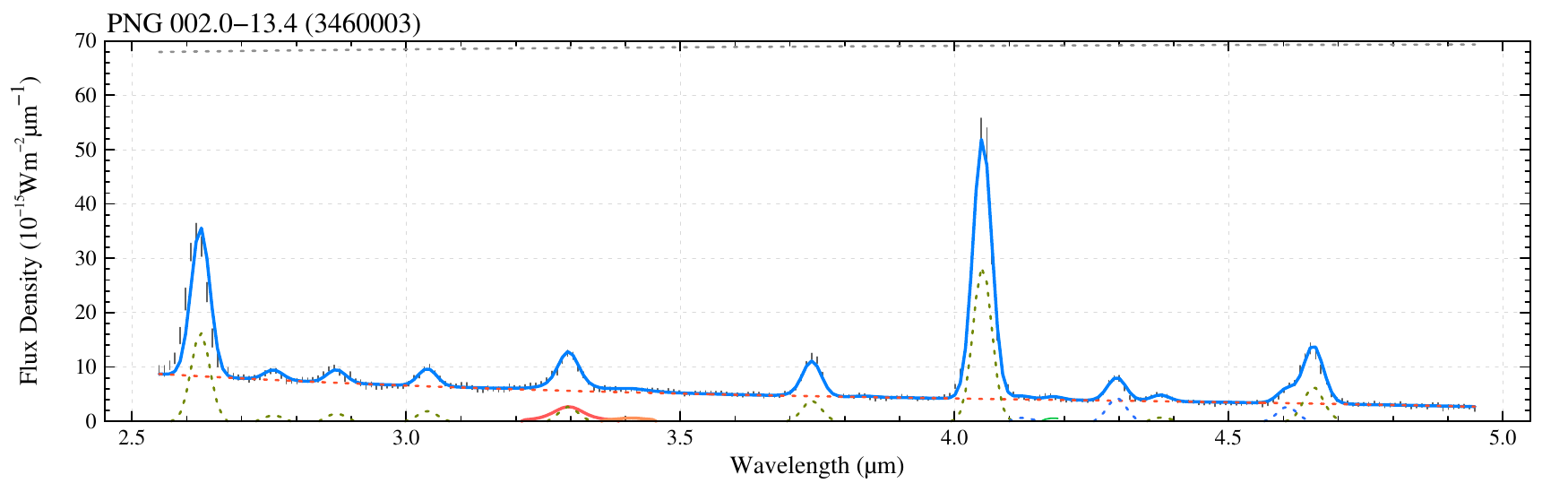}
  \includegraphics[width=1.0\linewidth]{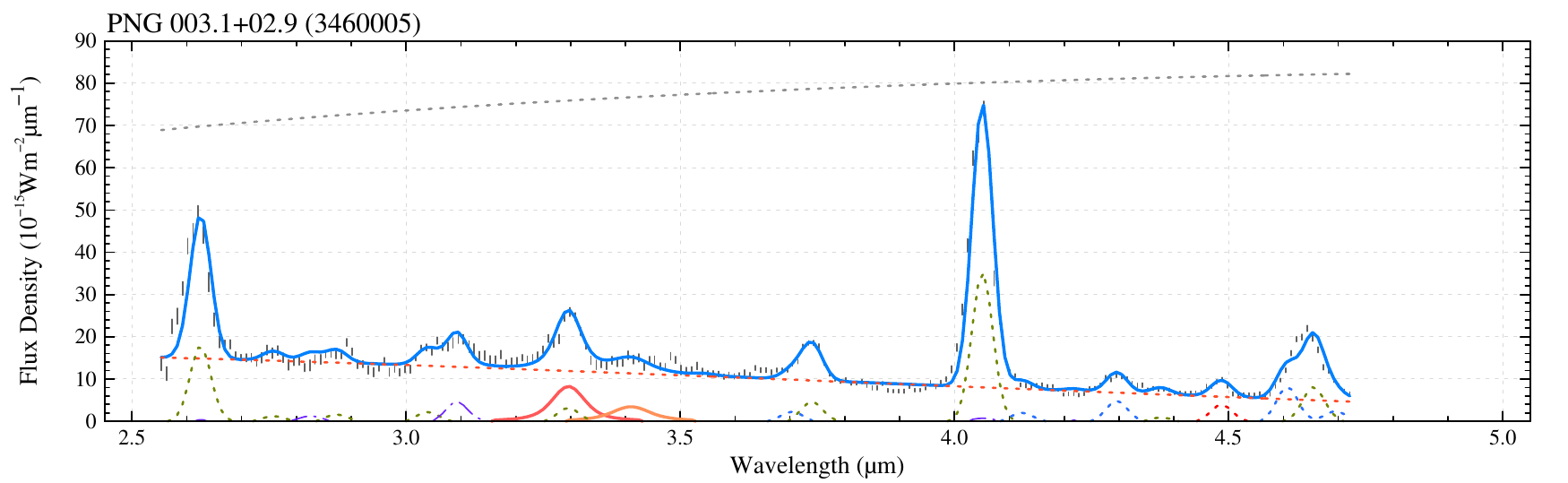}
  \includegraphics[width=1.0\linewidth]{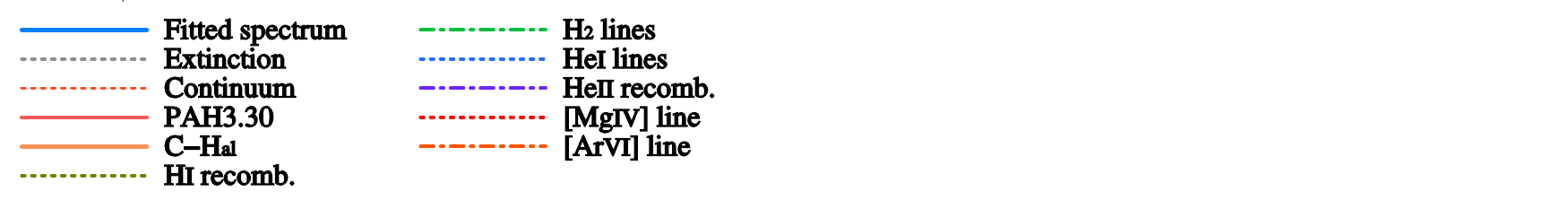}
  \caption{The \textit{AKARI}/IRC 2.5--5.0$\,\mu$m spectra. Explanations of the lines are shown in the bottom of the figures. Plots for all sources are available in the electronic edition of the journal.}
  \label{fig:allspectrum_s}
\end{figure*}

\section{Discussion}\label{sec:discussion}
\subsection{Biases in the PNSPC samples}\label{sec:dis:biases}

\subsubsection{Galactic Coordinates}\label{sec:dis:coordinates}
\begin{figure}[!tp]
  \centering
  \includegraphics[width=0.45\linewidth]{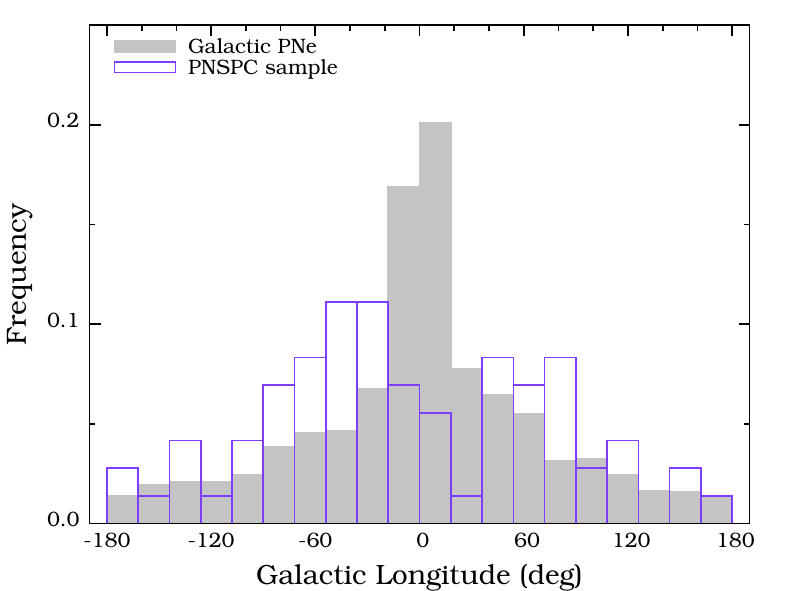}
  \includegraphics[width=0.45\linewidth]{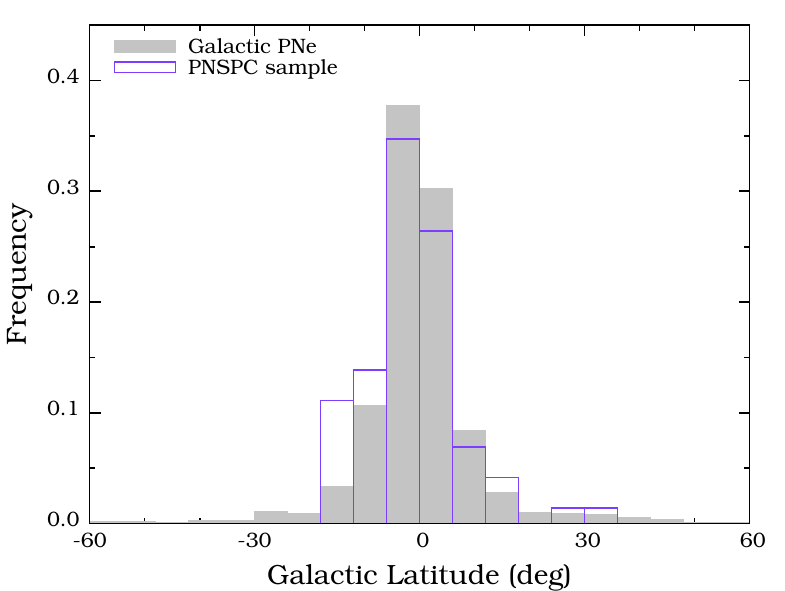}
  \caption{Histograms of PNe in the Galactic coordinates. The left panel shows the histogram of the Galactic longitude, while the right panel shows that of the Galactic latitude. The histograms for the PNSPC objects are shown by the blue boxes. The gray filled boxes represent the distribution of all the Galactic PNe in \citet{acker_strasbourg-eso_1992}.}
  \label{fig:coordhist}
\end{figure}
\begin{figure}[!tp]
  \centering
  \includegraphics[width=1.0\linewidth]{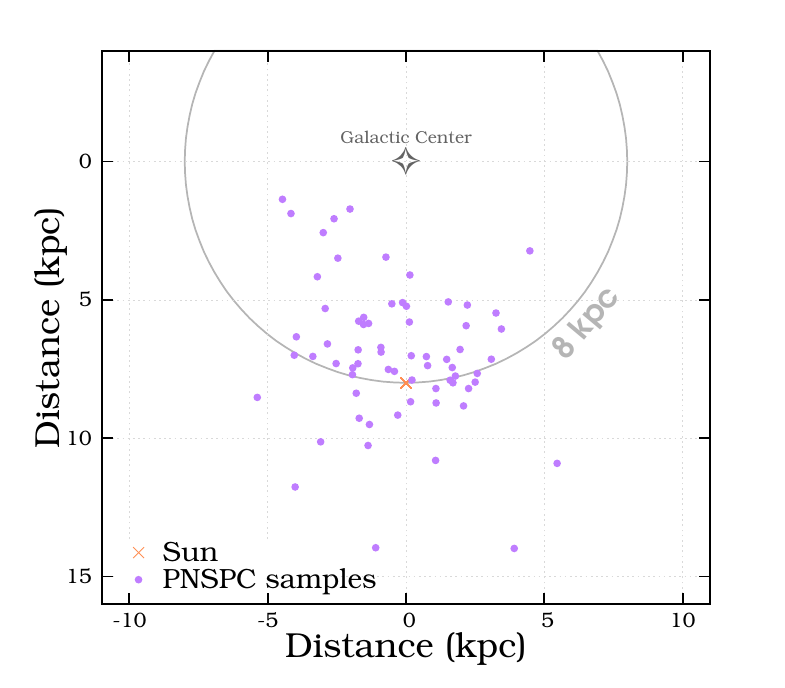}
  \caption{Distribution of the PNSPC samples on the Galactic Plane. The red cross shows the locus of the Sun. The purple circles show the positions of the PNSPC samples.}
  \label{fig:galcoord}
\end{figure}

We compare the distribution of the PNSPC samples in the Galactic coordinates with that of the PNe in \citet{acker_strasbourg-eso_1992}, which are assumed to represent the distribution of all the PNe in the Milky Way. The histograms of the Galactic longitude and latitude are shown in the left and right panels of Figure~\ref{fig:coordhist}. The histograms of the PNSPC samples are shown by the blue lines, while the histograms of PNe in \citet{acker_strasbourg-eso_1992} are shown by the gray filled bars. The total number of the PNe in \citet{acker_strasbourg-eso_1992} is 1143. The longitude histogram of the PNSPC samples looks broader than that of the PNe in \citet{acker_strasbourg-eso_1992}, while the latitude histogram of the PNSPC samples does not show a large deviation from that of the PNe in \citet{acker_strasbourg-eso_1992}. We investigate the differences by Kolmogorov-Smirnov test. The values of $\chi^2_{\nu{=}2}$ are about 16 for the longitude and 4 for the latitude. The results suggest that the deviation of the longitude is highly significant and that the PNSPC samples are biased toward PNe in the Galactic disk rather than those in the bulge. Figure~\ref{fig:galcoord} shows the position of the PNSPC samples projected onto the Galactic plane. The distances to the objects are taken from literature \citep[][and references therein]{acker_new_1978, pottasch_distances_1983, maciel_catalogue_1984, amnuel_statistical_1984, gathier_distances_1986, gathier_distances_1986-1, sabbadin_planetary_1986, cahn_catalogue_1992, acker_strasbourg-eso_1992}. Figure~\ref{fig:galcoord} shows that most of the PNSPC samples are within $5$\,kpc of the Sun, even when accounting for uncertainties in the distance estimate. This also supports that the PNSPC samples mainly consist of PNe in the Galactic disk.

\subsubsection{Effective Temperature and Age}\label{sec:dis:teffbias}
The central star of a PN becomes hot as it evolves. The effective temperature of the PN can be used as an indicator of the age of the PN. The timescale of the evolution depends on the mass of the central star \citep[e.g.,][]{blocker_stellar_1995,schonberner_late_1983}. Although the individual age of the PN is difficult to accurately derive from the effective temperature, the distribution of the effective temperature reflects the bias in the age of the PNSPC samples. The effective temperatures were collected from literature. The temperature and its references are listed in the sixth and seventh columns of Table~\ref{tab:basic_info_misc}. The effective temperature is estimated with several different methods. The effective temperatures estimated based on a model atmosphere should be the most reliable. The temperatures estimated in \citet{mendez_high_1988} and \citet{harrington_planetary_1983} are preferentially used, if available. The typical error for temperatures measured by this method is about 10\% \citep{mendez_high_1988}. Otherwise, the effective temperatures estimated with the Zanstra or the Energy Balance methods are used. The Zanstra temperatures are estimated based on \ion{H}{1}, \ion{He}{1}, and \ion{He}{2} recombination lines. Since the radiation-bounded condition is assumed in the Zanstra method, the Zanstra temperature estimated using the line with the highest ionization potential is most reliable. The Zanstra temperatures are used with the preferential order of \ion{He}{2}, \ion{He}{1}, and \ion{H}{1}. A typical uncertainty in the Zanstra temperature is as large as 10,000$\,$K in the worst case \citep{phillips_relation_2003}. \citet{preite-martinez_energy-balance_1991} reported that the typical uncertainty in the temperature with the Energy Balance method is no more than 20\%. We assume that a typical uncertainty in these methods is about 20\%. Finally, the effective temperature was obtained for $67$ of $72$ objects in the catalog. When multiple references are available, the averaged value is adopted.

\begin{figure}[!tp]
  \centering
  \includegraphics[width=1.0\linewidth]{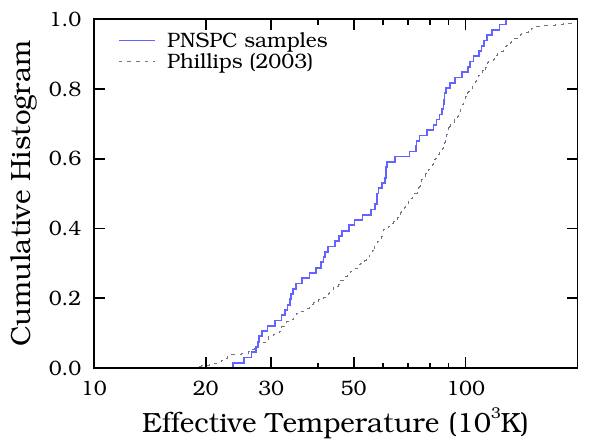}
  \caption{Cumulative histogram of the effective temperature of Galactic PNe. The blue solid line shows the histogram for the PNSPC samples. The gray dashed line represents the temperature distribution of all the Galactic PNe in \citet{phillips_relation_2003}.}
  \label{fig:teffbias}
\end{figure}

Figure~\ref{fig:teffbias} shows the distribution of the effective temperature in the PNSPC samples. \citet{phillips_relation_2003} collected the effective temperature of 373 Galactic PNe using the Zanstra method. Their samples were widely collected from literature without any limitation. Thus, we assume that their samples are not biased and that the temperature distribution of Phillips' samples represents that of the whole Galactic PNe (gray dashed line). The median of the effective temperature for the PNSPC samples is about $50\,000\,{\rm K}$, while that for Phillips' data is about $80\,000\,{\rm K}$. Figure \ref{fig:teffbias} indicates that the PNSPC samples are biased toward low-temperature or young PNe, possibly due to the target selection bias: the targets are limited by the apparent size (${\sim}3''$).

\subsection{Accuracy of the Absolute Intensity}\label{sec:dis:accuracy}
\begin{figure}[!tp]
  \centering
  \includegraphics[width=1.0\linewidth]{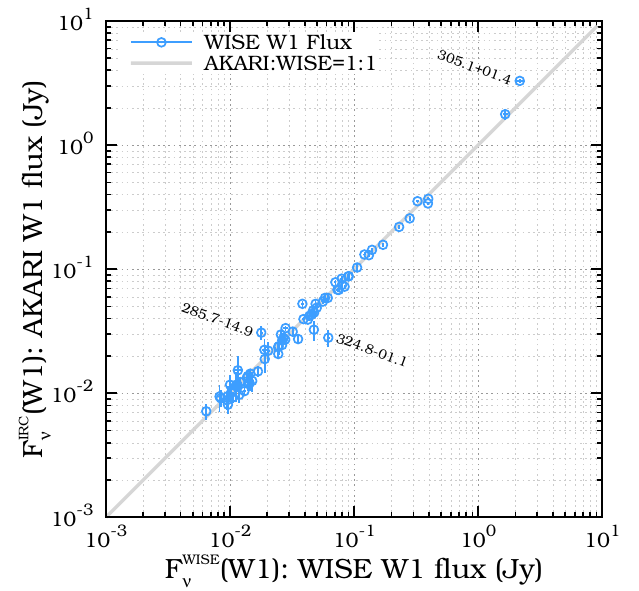}
  \caption{Flux comparison between the \textit{AKARI} and \textit{WISE} observations. The vertical axis shows the flux in the \textit{WISE} W1 filter estimated from the \textit{AKARI}/IRC spectra, while the horizontal axis shows the flux in the \textit{WISE} All-Sky Release Catalog \citep{cutri_wise_2012}. When the deviation between the \textit{AKARI} and \textit{WISE} W1 fluxes is larger than 50\%, the data point is labeled with the PNG\,ID.}
  \label{fig:comparewise}
\end{figure}

\begin{figure}[!tp]
  \centering
  \includegraphics[width=1.0\linewidth]{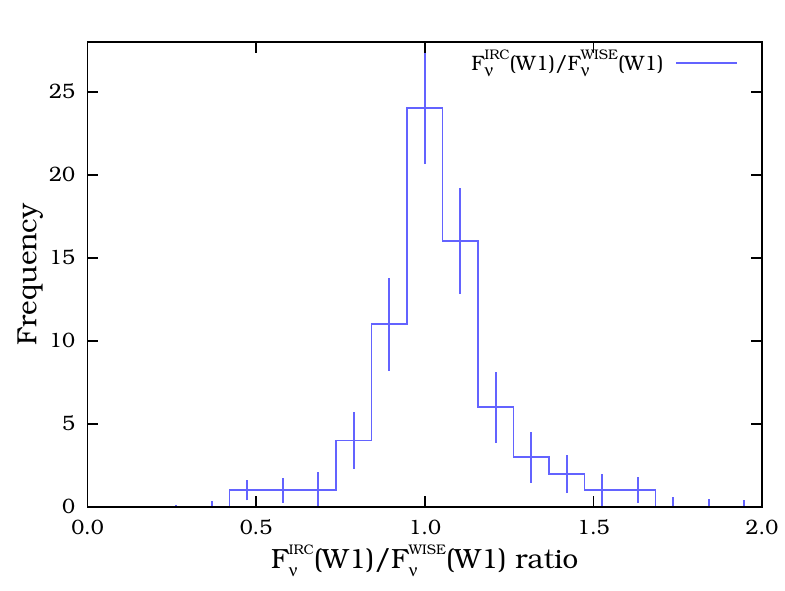}
  \caption{Distribution of the flux ratio: the flux in the \textit{WISE} W1 filter estimated from the \textit{AKARI}/IRC spectra to that from the \textit{WISE} All-Sky Release Catalog.}
  \label{fig:wisehist}
\end{figure}

One of the major characteristics of the PNSPC catalog is slit-less spectroscopy in the Np-window, which allows us to collect all of the flux from the target. Since the IRC spectrum totally covers the \textit{WISE} W1 band, the accuracy of the absolute flux density is evaluated by comparing the PNSPC spectra with the \textit{WISE} W1 photometry from WISE All-Sky Release Catalog \citep{cutri_wise_2012}. Define $R^{\rm W1}$ as the relative response curve of the \textit{WISE} W1 band \citep[per photon;][]{wright_wide-field_2010}. The flux in the W1 band estimated from the IRC spectrum is defined as
\begin{equation}
  \label{eq::akariW1}
  F^{\rm IRC}_{\nu}({\rm W1}) =
  \frac{\displaystyle \int F^{\rm obj}_{\nu}(\nu)
    R^{\rm W1}(\nu)\frac{{\rm d}\nu}{\nu}}
  {\displaystyle \int \left(\frac{\nu_{\rm iso}}{\nu}\right)^2
    R^{\rm W1}(\nu) \frac{{\rm d}\nu}{\nu}},
\end{equation}
where $\nu_{\rm iso}$ is the isophotal frequency of the W1 band (3.35$\,\mu$m) and $F^{\rm obj}_{\nu}$ is the observed flux density in units of Jy. Figure~\ref{fig:comparewise} shows the estimated flux density in the W1 band against the \textit{WISE} W1 catalog data. It confirms that the W1 flux density estimated from the IRC spectra, $F^{\rm IRC}_{\nu}({\rm W1})$, is consistent with the \textit{WISE} W1 data in a wide range from about $0.01$ to $5\,{\rm Jy}$. The histogram of the \textit{AKARI} to \textit{WISE} W1 flux ratio is shown in Figure~\ref{fig:wisehist}. Errors in Figure~\ref{fig:wisehist} are calculated by a Monte Carlo simulation with the uncertainty in the flux ratio. The ratios are distributed with the mean of $1.01{\pm}0.01$ and the standard deviation of $0.21{\pm}0.02$. Three PNe (PNG\,285.7$-$14.9, 324.8$-$01.1, and 305.1$+$01.4) show significant deviations larger than 50\%. For PNG\,285.7$-$14.9 and 305.1$+$01.4, their \textit{AKARI} W1 fluxes are larger than those of \textit{WISE}. Both objects are apparently extended in \textit{WISE} W1 images. The deviations may be attributed to the loss of the flux in \textit{WISE} photometry. Furthermore, PNG\,305.1$+$01.4 is the brightest object in the PNSPC sample. The \textit{WISE} detector starts to saturate in the W1 band at $\sim$0.17$\,{\rm Jy}$ \citep{wright_wide-field_2010}. This may also cause an error in \textit{WISE} photometry of PNG\,305.1$+$01.4. The \textit{AKARI} W1 flux of PNG\,324.8$-$01.1 is smaller than the \textit{WISE} W1 flux. PNG\,324.8$-$01.1 is located in a crowded region, so that the IRC field of view may not be wide enough to estimate background emission accurately. The loss of the \textit{AKARI} W1 flux is attributable to the error in the estimation of the background emission. Thus, although the absolute continuum intensity of PNG\,324.8$-$01.1 may be erroneous, the intensity of emission features is not affected by the estimation of background emission. The absolute flux of the present catalog is concluded to be accurate within $20$\%.

\subsection{Extinction within the Individual Nebula}\label{sec:dis:extinction}
\begin{figure}[!tp]
  \centering
  \includegraphics[width=1.0\linewidth]{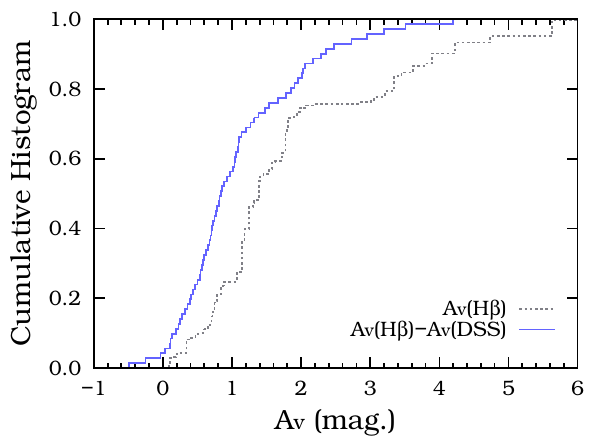}
  \caption{Cumulative histogram of the extinction at the $V$-band. The dotted line shows the histogram of the total extinction towards the PN. The solid line shows the histogram of the extinction without the contribution from the interstellar medium.}
  \label{fig:exthist}
\end{figure}

The PNSPC spectra include Brackett-$\alpha$ (Br$\alpha$), the hydrogen recombination line at 4.05$\,\mu$m, which enables us to estimate the extinction towards the objects using the relative intensity of H$\beta$ to Br$\alpha$ (the intensity of H$\beta$ is from \citet{acker_strasbourg-eso_1992}, \textit{see} Section~\ref{sec:res:extinction}). The cumulative histogram of the total extinction towards the PNSPC samples, $A_{V}({\rm H}\beta)$, is shown by the dotted line in Figure~\ref{fig:exthist}. The extinction toward the PNSPC samples are less than $6\,$mag. at the $V$-band. About 70\% of the PNSPC samples have $A_{\rm V}({\rm H}\beta) < 2\,$mag., suggesting that most of the PNSPC samples are not heavily obscured. This is consistent with the Galactic distribution of the PNSPC sample shown in Section~\ref{sec:dis:coordinates}.

\begin{figure}[!tp]
  \centering
  \includegraphics[width=1.0\linewidth]{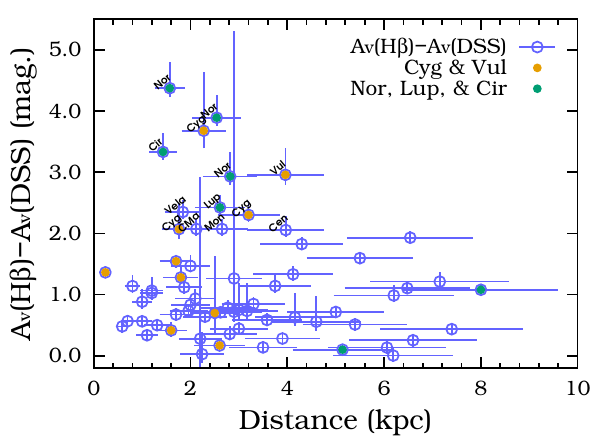}
  \caption{Net extinction at the $V$-band is plotted against a distance towards the PN. Errors in the distance are assumed to be about 20\%. The name of the constellation is added next to the symbols for the PN with the net extinction ${>}2.0\,$mag. The PNe in Cygnus and Vulpecula are indicated by the orange filled circles, while the PNe in Norma, Lupus and Circinus are indicated by the green filled circles.}
  \label{fig:distext}
\end{figure}

The extinction values calculated in Section~\ref{sec:res:extinction} include the contribution from their circumstellar envelope, $A_{V,\,{\rm PN}}$, and the interstellar medium to the object, $A_{V,\,{\rm ISM}}$. The contribution from the interstellar medium, $A_{V,\,{\rm ISM}}$, can be estimated using the extinction map based on the Digitized Sky Survey \citep{dobashi_atlas_2005,dobashi_erratum:_2005}. The extinction map is given in the resolution of $2\arcmin{\times}2\arcmin$. We calculate the averaged extinction values within a $6\arcmin$ radius around the object $A_{V}({\rm DSS})$ and assumed them as $A_{V,\,{\rm ISM}}$. \citet{dobashi_atlas_2005} showed that a systematic error in $A_{V}({\rm DSS})$ is typically less than 0.2$\,$mag for $A_{V}({\rm DSS}) < 5\,$mag. We estimate the uncertainty in $A_{V,\,{\rm ISM}}$ from the scatter of $A_{V}({\rm DSS})$ within the $6\arcmin$ circle taking into account the systematic error. Note that the extinction $A_{V,\,{\rm ISM}}$ estimated here may have a large uncertainty. The extinction from the circumstellar envelope, defined by $A_{V}({\rm H}\beta) - A_{V}({\rm DSS})$, is shown by the solid line in Figure~\ref{fig:exthist}. Some objects show negative extinction values (${\sim}0.5\,$mag.), possibly due to the overestimation of $A_{V,\,{\rm ISM}}$. Figure~\ref{fig:distext} shows the dependence of $A_{V}({\rm H}\beta) - A_{V}({\rm DSS})$ on the distance towards the PN. We find that PNe with extinction ${>}2\,$mag. frequently appear around 2\,kpc. The name of the constellation is added next to those PNe, indicating that most of them are in the regions around Cygnus and Norma as indicated by the orange and green symbols. The extinction $A_{V,\,{\rm ISM}}$ for these PNe may be underestimated or have large local fluctuations, but this affects only about 10\% of the PNSPC samples. Except for the objects in the Cygnus and Norma regions, Figure~\ref{fig:distext} does not show any trend with the distance, suggesting that the $A_{V}({\rm H}\beta) - A_{V}({\rm DSS})$ value is attributable to the extinction internal to the PN. Figure~\ref{fig:exthist} indicates that the median of the circumstellar extinction is about 0.8$\,$mag. and about $40\%$ of the PNe show extinction larger than $1\,$mag. suggesting that the circumstellar envelope of a PN is generally optically thick in the UV. \citet{ohsawa_unusual_2012} investigated the spatial distribution of mid-infrared dust emission of PN\,G095.2${+}$00.7, a young and UV-thick Galactic PN, and suggest that the evolution of a UV-thick circumstellar envelope may have a significant impact on the appearance of the mid-infrared spectrum of the PN. The present results suggest that PNe spectra should be investigated by taking into account the radiation transfer within the nebula, and the gradient in the dust temperature.


\subsection{Equivalent Width of Ionized Gas Emission Lines}\label{sec:dis:lines}
\begin{figure}[!tp]
  \centering
  \includegraphics[width=0.45\linewidth]{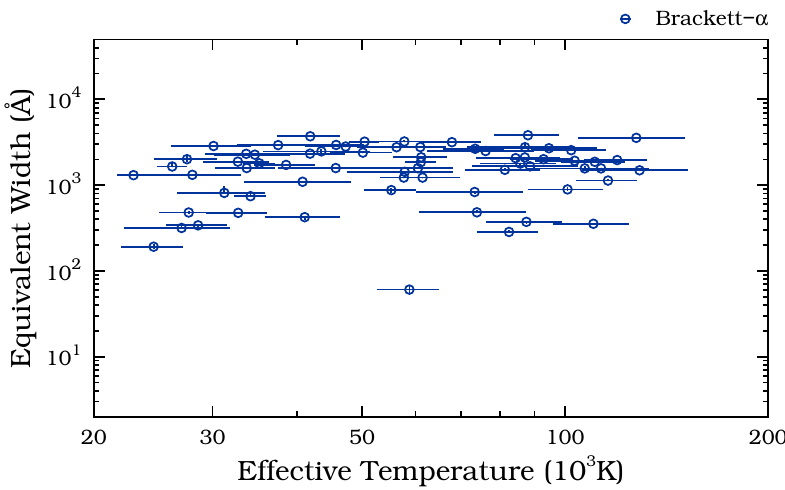}
  \includegraphics[width=0.45\linewidth]{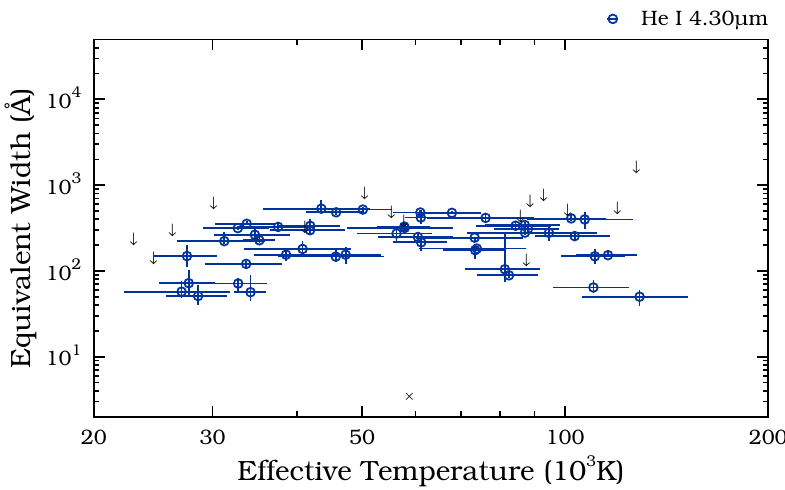}
  \includegraphics[width=0.45\linewidth]{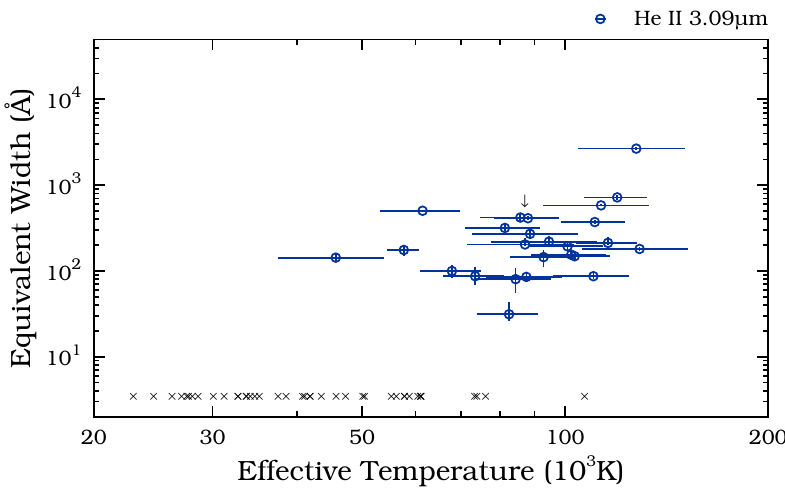}
  \includegraphics[width=0.45\linewidth]{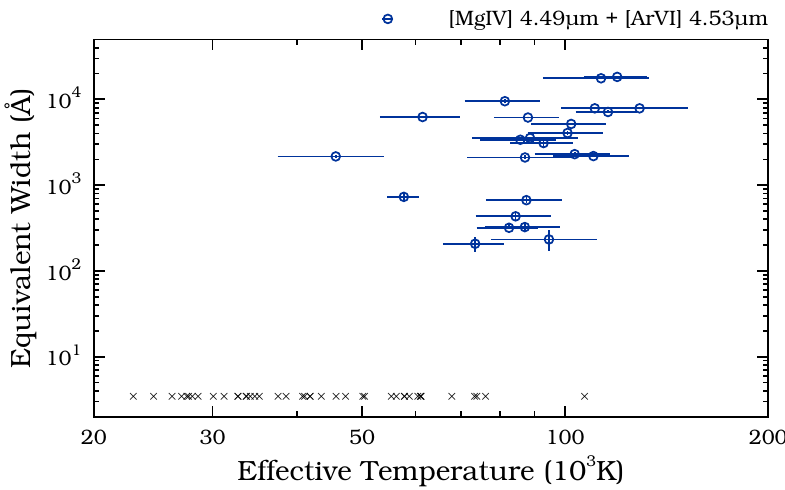}
  \caption{Equivalent widths of Brackett-$\alpha$ at 4.05$\,\mu$m (upper left), \ion{He}{1} at 4.30$\,\mu$m (upper right), \ion{He}{2} at 3.09$\,\mu$m (lower left), and [\ion{Mg}{4}] at 4.49$\,\mu$m ${+}$ [\ion{Ar}{6}] at 4.53$\,\mu$m (lower right) against the effective temperature. The downward arrows indicate a 3-$\sigma$ upper limit. When the emission is not detected, the effective temperature is indicated by the crosses.}
  \label{fig:gaslines}
\end{figure}

\begin{figure}
  \centering
  \includegraphics[width=1.0\linewidth]{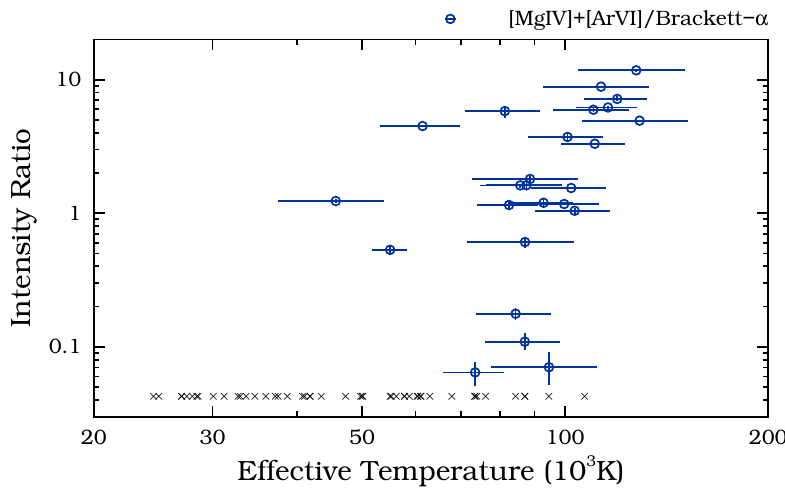}
  \caption{Intensity ratio of the summation of [\ion{Mg}{4}] and [\ion{Ar}{6}] to Brackett-$\alpha$ against the effective temperature. When the emission of [\ion{Mg}{4}] and [\ion{Ar}{6}] is not detected, the effective temperature is indicated by the crosses.}
  \label{fig:ratio-MgAr2BrA}
\end{figure}

The spectra in the PNSPC catalog contain several emission lines, including hydrogen and helium recombination lines and fine-structure lines of [\ion{Mg}{4}] at 4.49$\,\mu$m and [\ion{Ar}{6}] at 4.53$\,\mu$m. These emission lines should reflect the characteristics of the radiation field of the circumstellar envelope. Figure~\ref{fig:gaslines} shows the equivalent widths of Br$\alpha$, \ion{He}{1} at 4.30$\,\mu$m, \ion{He}{2} at 3.09$\,\mu$m, and the summation of [\ion{Mg}{4}] and [\ion{Ar}{6}] against the effective temperature of the central star. Since the [\ion{Mg}{4}] and [\ion{Ar}{6}] lines are close to each other, the equivalent width of the summation of these lines is used. The data whose signal-to-noise ratio of the equivalent width is less than three are replaced by 3-$\sigma$ upper limits. When the emission is not detected, the effective temperature is indicated by the crosses. The equivalent width of Br$\alpha$ is typically about $1$--$3{\times}10^3\,${\AA} and does not show a clear dependence on the effective temperature. The \ion{He}{1} line is seen from 20\,000 to 150\,000$\,$K. Its equivalent width is almost constant at about $2{\times}10^2\,${\AA}. It shows a small increase from 30\,000--50\,000$\,$K, possibly explained by the increase in photons with enough energy to ionize ${\rm He}$ with increasing effective temperature. When the effective temperature exceeds 50\,000$\,$K, the \ion{He}{1} equivalent width stops increasing and shows a small decrease. At the same time, the \ion{He}{2} lines start to be detected, suggesting that high-energy photons are in part used to ionize ${\rm He}^+$. The decrease in the \ion{He}{1} equivalent width is attributable to the ionization balance between ${\rm He}^+$ and ${\rm He}^{++}$. The \ion{He}{2}, [\ion{Mg}{4}], and [\ion{Ar}{6}] lines are not detected until the effective temperature exceeds about 50\,000$\,$K. The ionization potentials of \ion{He}{2}, \ion{Mg}{4}, and \ion{Ar}{6} are 54.41, 109.27, and 91.00$\,$eV. Although the ionization potential of \ion{He}{2} is about two times lower than those of \ion{Mg}{4}, and \ion{Ar}{6}, the \ion{He}{2}, [\ion{Mg}{4}], and [\ion{Ar}{6}] lines start to appear at almost the same temperature in Figure~\ref{fig:gaslines}. The equivalent width of the \ion{He}{2} line is typically about $2{\times}10^2\,${\AA}, not showing a clear trend with the effective temperature. The equivalent width of the [\ion{Mg}{4}]${+}$[\ion{Ar}{6}] becomes as large as $8{\times}10^3\,${\AA} at a high temperature of about 100\,000$\,$K. Figure~\ref{fig:ratio-MgAr2BrA} shows the intensity ratio of the [\ion{Mg}{4}]${+}$[\ion{Ar}{6}] to Br$\alpha$. The effective temperature of the PNe with the ratio larger than unity ranges about 500\,000--150\,000$\,$K. The [\ion{Mg}{4}] and [\ion{Ar}{6}] lines are an indicator of high-temperature PNe as expected. But note that some PNe with high effective temperature show a ratio less than unity. A ratio less than unity does not necessarily indicate a PN with a low effective temperature.

\subsection{PAH Features of Galactic PNe in the Near-Infrared}\label{sec:dis:pahfeatures}
The PAH features appear both in the near- and mid-infrared, and the PAH features in the mid-infrared are stronger than those in the near-infrared \citep[e.g.,][]{schutte_theoretical_1993,draine_infrared_2001}. The PAH features of PNe in the mid-infrared have been intensively investigated with \textit{Spitzer} \citep[e.g.,][]{bernard-salas_unusual_2009,stanghellini_spitzer_2007,stanghellini_nature_2012}. Surveys of the 3.3$\,\mu$m PAH feature in Galactic PNe were carried out by \citet{roche_investigation_1996} and \citet{smith_survey_2008}. Their samples were, however, biased towards carbon-rich objects. The PNSPC samples are not selected based on their chemistry. The PNSPC catalog provides a useful data set to investigate the near-infrared PAH features in Galactic PNe.

\begin{figure}[!tp]
  \centering
  \includegraphics[width=1.0\linewidth]{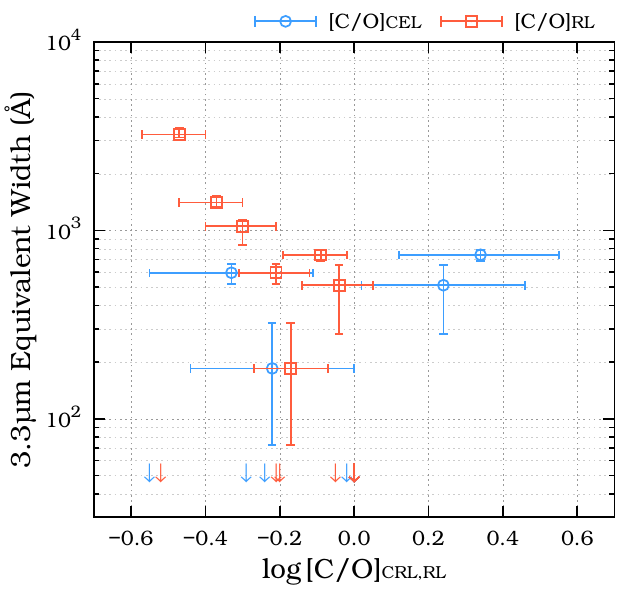}
  \caption{Equivalent width of the 3.3$\,\mu$m PAH feature against the carbon-to-oxygen (C/O) ratio. The C/O ratios are from \citet{delgado-inglada_c/o_2014}. The C/O ratio measured by collisionally excited lines (CEL) is shown by the blue circles, while the C/O ratio measured by recombination lines (RL) is by the red squares. The arrows indicate the non-detection of the 3.3$\,\mu$m PAH feature.}
  \label{fig:coratio_PAH330}
\end{figure}

The type of dust features (carbon-rich or oxygen-rich) is thought to be closely related to the abundance ratio of nebular gas. Several studies have suggested that the intensities of the PAH features increases with the carbon-to-oxygen abundance (C/O) ratio \citep[e.g.,][]{cohen_infrared_1986,roche_investigation_1996,cohen_polycyclic_2005,smith_survey_2008}. However, PAH formation both in carbon-rich and oxygen-rich environments has been suggested recently \citep{guzman-ramirez_pah_2014,guzman-ramirez_carbon_2011}. \citet{delgado-inglada_c/o_2014} measured the C/O ratio of 51 Galactic PNe and confirmed that the PAH emission can be seen even in PNe with ${\rm C/O} < 1$. The C/O ratios of 11 PNe in the PNSPC catalog were measured by \citet{delgado-inglada_c/o_2014}. Figure~\ref{fig:coratio_PAH330} shows the equivalent width of the 3.3$\,\mu$m PAH feature against the C/O ratio. The C/O ratio measured by collisionally excited lines (CEL) is indicated by the blue circles, while that measured by recombination lines (RL) is shown by the red squares. The downward arrows indicate non-detection of the 3.3$\,\mu$m PAH features. Figure~\ref{fig:coratio_PAH330} shows that the 3.3$\,\mu$m PAH emission is detected even in oxygen-rich environments, consistent with \citet{delgado-inglada_c/o_2014}. Due to the small number of samples, however, it is difficult to claim the correlation between the strength of the 3.3$\,\mu$m PAH feature and the C/O ratio. Extensive measurements of the C/O ratio are encouraged.

\begin{figure}[!tp]
  \centering
  \includegraphics[width=1.0\linewidth]{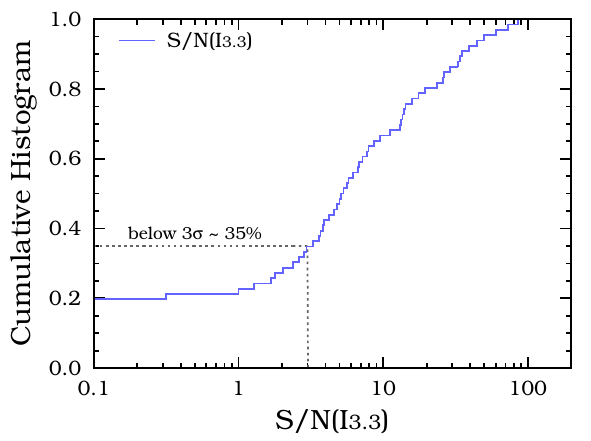}
  \caption{Cumulative histogram of the signal-to-noise ratio of the 3.3$\,\mu$m PAH feature $\left({\rm S/N}(I_{3.3})\right)$. The gray dotted-line shows the 35th percentile, which is the non-detection rate of the 3.3$\,\mu$m PAH feature when the detection limit is defined as ${\rm S/N}(I_{3.3}) \geq 3$.}
  \label{fig:pahdetect}
\end{figure}

Figure~\ref{fig:pahdetect} shows a cumulative histogram of the signal-to-noise ratio of the 3.3$\,\mu$m PAH feature (hereafter, ${\rm S/N}(I_{3.3})$). The gray dotted line shows the 35th percentile, which corresponds to ${\rm S/N}(I_{3.3}) = 3$. The result suggests that about 65\% of the PNSPC samples show band emission at 3.3$\,\mu$m. The 5--38$\,\mu$m spectra of Galactic PNe with \textit{Spitzer}/IRS are available in \citet{stanghellini_nature_2012}. We counted the number of PNe with PAH features, obtaining 40\% (60 of 150 PNe). Despite the intrinsic weakness of the near-infrared PAH features, the PAH detection rate is as high in the near-infrared as in the mid-infrared. The 3.3$\,\mu$m feature is thought to come from the smallest PAH populations, which are most fragile, and thus primarily reflect environment variations \citep{schutte_theoretical_1993,micelotta_polycyclic_2010,allain_photodestruction_1996-1,allain_photodestruction_1996}. The PNSPC catalog is useful to investigate the evolution of the PAH features and the processing of PAHs in circumstellar environments.

\section{Summary \& Conclusion}\label{sec:summary}
Near-infrared (2.5--5.0$\,\mu$m) spectra of 72 Galactic PNe were obtained at high sensitivity using the \textit{AKARI}/IRC. The PNSPC program provided a set of near-infrared spectra that collected all of the flux from the objects. The absolute flux of spectra provided by the PNSPC program is in agreement with the \textit{WISE} W1 photometry within ${\sim}20$\%. These spectra were compiled into the PNSPC catalog. The intensity and equivalent width of Brackett-$\alpha$, \ion{He}{1} at 4.30$\,\mu$m, \ion{He}{2} at 3.09$\,\mu$m, the $3.3\,\mu$m PAH feature, the 3.4--3.5$\,\mu$m aliphatic feature complex, [\ion{Mg}{4}] at $4.49\,\mu$m, and [\ion{Ar}{6}] at $4.53\,\mu$m were measured. The detection limit of the emission features was as faint as $2{\times}10^{-16}\,{\rm W\,m^{-2}}$. The PNSPC catalog is the largest data set of near-infrared spectra of Galactic PNe. It provides unique information for the investigation of 2.5--5.0$\,\mu$m emission of Galactic PNe.

The Galactic coordinates of the PNSPC samples suggest that the objects in the PNSPC catalog are biased toward PNe in the Galactic disk rather than those in the Galactic bulge. The median of the effective temperature of the PNSPC samples is lower than that of the whole Galactic PNe by about 30,000$\,$K. This indicates that the PNSPC catalog is biased toward young PNe. This bias possibly originates in the criterion on the apparent size in the target selection.

The PNSPC catalog also provides the extinction toward the objects. Even after the contribution from the interstellar medium is taken into account, Figure~\ref{fig:exthist} suggests that 40\% of PNe have extinction at the $V$-band larger than unity, which is attributed to the extinction in the circumstellar envelope. The present result suggests that for a large fraction of PNe a circumstellar envelope is optically thick in the UV.

The equivalent width of Brakett-$\alpha$ does not show a clear dependence on the effective temperature. The variations in the \ion{He}{1} and \ion{He}{2} equivalent widths can be attributed to the ionization balance between ${\rm He}^{+}$ and ${\rm He}^{++}$. The [\ion{Mg}{4}] at 4.49$\,\mu$m and [\ion{Ar}{6}] at 4.53$\,\mu$m lines are only detected when the effective temperature becomes higher than 50\,000$\,$K. The [\ion{Mg}{4}] and [\ion{Ar}{6}] lines are good indicators of PNe with a hot central star.

The 3.3$\,\mu$m PAH feature is detected in about 65\% of the PNSPC samples. This detection rate is comparable to that reported from mid-infrared observations \citep{stanghellini_nature_2012}. The present result suggests that the \textit{AKARI}/IRC is as sensitive as the \textit{Spitzer}/IRS in terms of detecting the PAH features. The PNSPC catalog provides a suitable data set to investigate PAHs in circumstellar environments. The near-infrared PAH features are attributed to small-sized PAHs, which are sensitive to harsh environments and processed faster than larger ones. The processing and evolution of PAHs during the PN phase will be discussed in the forth-coming paper.

\acknowledgements
The present results are based on observations with \textit{AKARI}, a JAXA project with the participation of ESA. We greatly appreciate all the people who worked in the operation and maintenance of those instruments. This publication makes use of data products from the \textit{Wide-field Infrared Survey Explorer}, which is a joint project of the University of California, Los Angeles, and the Jet Propulsion Laboratory/California Institute of Technology, funded by the National Aeronautics and Space Administration. This research has made use of the SIMBAD database, operated at CDS, Strasbourg, France. This work is supported in part by Grant-in-Aids for Scientific Research (25-8492, 23244021, and 26247074) by the Japan Society of Promotion of Science (JSPS).

\bibliographystyle{apj}

\setcounter{table}{1}
\begin{deluxetable}{ccccccc}
\tabletypesize{\small}
\tablewidth{0pt}

\tablecaption{Miscellaneous Information of Objects \label{tab:basic_info_misc}}
\tablehead{
  \colhead{PN\,G} &
  \colhead{$V^{(a)}$} &
  \colhead{$K_s^{(b)}$} &
  \colhead{$A_{V}({\rm H\beta})$} &
  \colhead{$A_{V}({\rm fit.})$} &
  \colhead{$T_{\rm eff}$} & \colhead{Refs. $T_{\rm eff}$} \\
  \colhead{} &
  \colhead{mag.} & \colhead{mag.} &
  \colhead{mag.} & \colhead{mag.} &
  \colhead{$10^3\,$K} & \colhead{}
}
\startdata
000.3${+}$12.2 & $\phn13.9$ & $\phn11.4$ & $\phn\phn0.90_{{-}0.03}^{{+}0.03}$ & $\phn\phn0.98_{{-}0.05}^{{+}0.15}$ & $\phn47$ & Ka76,Ph03 \\
002.0${-}$13.4 & $\phn14.1$ & $\phn11.1$ & $\phn\phn0.31_{{-}0.01}^{{+}0.02}$ & $\phn\phn0.38_{{-}0.04}^{{+}0.02}$ & $\phn60$ & PM89,PM91,Ph03 \\
003.1${+}$02.9 & ${>}17.0$ & $\phn11.6$ & $\phn\phn3.31_{{-}0.03}^{{+}0.03}$ & $\phn\phn\phn3.4_{{-}0.1}^{{+}0.6}$ & $\phn87$ & PM89,Ph03 \\
011.0${+}$05.8 & \nodata & $\phn11.8$ & $\phn\phn1.81_{{-}0.04}^{{+}0.04}$ & $\phn\phn1.79_{{-}0.06}^{{+}0.06}$ & $\phn93$ & PM91,Ph03 \\
027.6${-}$09.6 & $\phn15.2$ & $\phn11.9$ & $\phn\phn1.04_{{-}0.05}^{{+}0.06}$ & $\phn\phn\phn1.0_{{-}0.2}^{{+}0.9}$ & $\phn58$ & PM91,Ka76,KJ91,Ph03 \\
037.8${-}$06.3 & \nodata & $\phn9.47$ & $\phn\phn\phn1.4_{{-}0.1}^{{+}0.1}$ & $\phn\phn1.41_{{-}0.02}^{{+}0.08}$ & $\phn74$ & Ph03 \\
038.2${+}$12.0 & $\phn12.5$ & $\phn11.1$ & $\phn\phn0.82_{{-}0.05}^{{+}0.05}$ & $\phn\phn0.93_{{-}0.07}^{{+}0.07}$ & $\phn30$ & Ka78,Ph03 \\
043.1${+}$03.8 & $\phn14.9$ & $\phn12.1$ & $\phn\phn\phn1.9_{{-}0.1}^{{+}0.1}$ & $\phn\phn2.05_{{-}0.05}^{{+}0.46}$ & $\phn28$ & Ph03 \\
046.4${-}$04.1 & $\phn15.2$ & $\phn11.1$ & $\phn\phn1.42_{{-}0.02}^{{+}0.02}$ & $\phn\phn1.39_{{-}0.01}^{{+}0.01}$ & $\phn68$ & Ka76,KJ91,Ph03 \\
051.4${+}$09.6 & $\phn13.3$ & $\phn10.4$ & $\phn\phn0.88_{{-}0.04}^{{+}0.05}$ & $\phn\phn0.98_{{-}0.08}^{{+}0.11}$ & $\phn37$ & Ka76,Ka78,Ph03 \\
052.2${-}$04.0 & $\phn18.1$ & $\phn12.0$ & $\phn\phn1.68_{{-}0.04}^{{+}0.05}$ & $\phn\phn1.62_{{-}0.09}^{{+}0.09}$ & $\phn61$ & Lu01,PM89,KJ91,Ph03 \\
058.3${-}$10.9 & $\phn14.4$ & $\phn9.67$ & $\phn\phn0.66_{{-}0.05}^{{+}0.05}$ & $\phn\phn0.18_{{-}0.04}^{{+}0.08}$ & $\phn50$ & PM89,Ka76,Ph03 \\
060.1${-}$07.7 & $\phn18.0$ & $\phn11.5$ & $\phn\phn1.19_{{-}0.02}^{{+}0.02}$ & $\phn\phn1.28_{{-}0.02}^{{+}0.03}$ & $129$ & Ph03 \\
060.5${+}$01.8 & \nodata & $\phn11.8$ & $\phn\phn\phn4.1_{{-}0.2}^{{+}0.2}$ & $\phn\phn\phn3.5_{{-}0.4}^{{+}0.5}$ & \nodata & \nodata \\
064.7${+}$05.0 & $\phn12.5$ & $\phn8.11$ & $\phn\phn1.44_{{-}0.04}^{{+}0.04}$ & $\phn\phn1.43_{{-}0.04}^{{+}0.05}$ & $\phn28$ & Lu01,Ka76,Ka78,Ph03 \\
071.6${-}$02.3 & ${>}15.7$ & $\phn10.1$ & $\phn\phn\phn4.4_{{-}0.1}^{{+}0.1}$ & $\phn\phn\phn5.1_{{-}0.7}^{{+}1.3}$ & \nodata & \nodata \\
074.5${+}$02.1 & $\phn18.4$ & $\phn11.0$ & $\phn\phn3.43_{{-}0.07}^{{+}0.07}$ & $\phn\phn3.47_{{-}0.04}^{{+}1.32}$ & $\phn81$ & Ph03 \\
082.1${+}$07.0 & ${>}15.6$ & $\phn10.7$ & $\phn\phn1.71_{{-}0.08}^{{+}0.07}$ & $\phn\phn\phn1.7_{{-}0.1}^{{+}2.3}$ & $\phn87$ & Ph03 \\
082.5${+}$11.3 & $\phn14.5$ & $\phn12.2$ & $\phn\phn0.17_{{-}0.04}^{{+}0.04}$ & $\phn\phn0.16_{{-}0.05}^{{+}0.08}$ & $\phn63$ & Me88 \\
086.5${-}$08.8 & $\phn17.3$ & $\phn12.5$ & $\phn\phn0.60_{{-}0.03}^{{+}0.03}$ & $\phn\phn0.52_{{-}0.03}^{{+}0.03}$ & $\phn61$ & Lu01,Ph03 \\
089.3${-}$02.2 & $\phn12.1$ & $\phn9.20$ & $\phn\phn\phn2.0_{{-}0.5}^{{+}0.8}$ & $\phn\phn1.90_{{-}0.01}^{{+}5.40}$ & \nodata & \nodata \\
089.8${-}$05.1 & $\phn16.7$ & $\phn9.49$ & $\phn\phn1.99_{{-}0.04}^{{+}0.04}$ & $\phn\phn2.02_{{-}0.01}^{{+}0.02}$ & $\phn83$ & Ka78,KJ91,Ph03 \\
095.2${+}$00.7 & \nodata & $\phn11.2$ & $\phn\phn\phn5.5_{{-}0.2}^{{+}0.3}$ & $\phn\phn\phn5.4_{{-}0.1}^{{+}0.4}$ & $\phn58$ & Lu01 \\
100.6${-}$05.4 & $\phn15.5$ & $\phn12.0$ & $\phn\phn0.52_{{-}0.03}^{{+}0.03}$ & $\phn\phn0.47_{{-}0.07}^{{+}0.05}$ & $\phn74$ & Ka78,KJ91,Ph03 \\
111.8${-}$02.8 & $\phn13.8$ & $\phn8.81$ & $\phn\phn1.92_{{-}0.10}^{{+}0.10}$ & $\phn\phn2.12_{{-}0.76}^{{+}0.01}$ & $\phn41$ & Ka76,Ph03 \\
118.0${-}$08.6 & $\phn14.2$ & $\phn12.9$ & $\phn\phn0.41_{{-}0.10}^{{+}0.10}$ & $\phn\phn0.32_{{-}0.03}^{{+}0.02}$ & $\phn55$ & Lu01,Ka76,KJ91,Ph03 \\
123.6${+}$34.5 & $\phn13.4$ & $\phn11.9$ & $\phn\phn0.50_{{-}0.03}^{{+}0.03}$ & $\phn\phn0.60_{{-}0.03}^{{+}0.09}$ & $\phn50$ & HF83 \\
146.7${+}$07.6 & $\phn14.0$ & $\phn11.2$ & $\phn\phn1.55_{{-}0.09}^{{+}0.09}$ & $\phn\phn1.56_{{-}0.02}^{{+}0.29}$ & $\phn27$ & Ph03 \\
159.0${-}$15.1 & $\phn15.8$ & $\phn12.6$ & $\phn\phn0.44_{{-}0.04}^{{+}0.04}$ & $\phn\phn0.40_{{-}0.06}^{{+}0.02}$ & $\phn89$ & Ph03 \\
166.1${+}$10.4 & $\phn11.6$ & $\phn10.6$ & $\phn\phn0.63_{{-}0.02}^{{+}0.02}$ & $\phn\phn0.64_{{-}0.03}^{{+}0.02}$ & $\phn39$ & Ka76,Ka78,Ph03 \\
190.3${-}$17.7 & $\phn14.4$ & $\phn13.2$ & $\phn\phn0.34_{{-}0.03}^{{+}0.03}$ & $\phn\phn0.36_{{-}0.04}^{{+}0.10}$ & $\phn61$ & PM89,Ka76,KJ91,Ph03 \\
194.2${+}$02.5 & $\phn17.8$ & $\phn10.6$ & $\phn\phn1.40_{{-}0.06}^{{+}0.06}$ & $\phn\phn1.41_{{-}0.12}^{{+}0.04}$ & $\phn88$ & PM89,Ph03 \\
211.2${-}$03.5 & $\phn15.8$ & $\phn10.6$ & $\phn\phn\phn3.5_{{-}0.2}^{{+}0.2}$ & $\phn\phn3.49_{{-}0.02}^{{+}0.03}$ & $\phn33$ & Lu01,PM89,Ph03 \\
221.3${-}$12.3 & $\phn17.9$ & $\phn11.0$ & $\phn\phn0.91_{{-}0.01}^{{+}0.01}$ & $\phn\phn0.91_{{-}0.02}^{{+}0.10}$ & $111$ & PM91,Ph03 \\
226.7${+}$05.6 & $\phn16.9$ & $\phn12.8$ & $\phn\phn1.63_{{-}0.03}^{{+}0.02}$ & $\phn\phn1.61_{{-}0.05}^{{+}0.11}$ & $\phn46$ & Ph03 \\
232.8${-}$04.7 & $\phn13.9$ & $\phn9.48$ & $\phn\phn2.96_{{-}0.06}^{{+}0.06}$ & $\phn\phn2.99_{{-}0.09}^{{+}0.39}$ & $\phn25$ & Lu01,PM89,Ph03 \\
235.3${-}$03.9 & $\phn14.1$ & $\phn10.6$ & $\phn\phn\phn1.6_{{-}0.2}^{{+}0.2}$ & $\phn\phn1.62_{{-}0.01}^{{+}0.01}$ & $\phn29$ & Lu01,PM89,Ph03 \\
258.1${-}$00.3 & $\phn16.2$ & $\phn10.9$ & $\phn\phn4.02_{{-}0.03}^{{+}0.03}$ & $\phn\phn4.02_{{-}0.04}^{{+}0.05}$ & $\phn60$ & PM91,Ph03 \\
264.4${-}$12.7 & $\phn14.7$ & $\phn12.2$ & $\phn\phn0.72_{{-}0.07}^{{+}0.07}$ & $\phn\phn\phn0.6_{{-}0.1}^{{+}0.1}$ & $\phn56$ & PM91,KJ91 \\
268.4${+}$02.4 & $\phn18.7$ & $\phn10.6$ & $\phn\phn3.80_{{-}0.03}^{{+}0.03}$ & $\phn\phn3.91_{{-}0.05}^{{+}0.23}$ & $110$ & PM91,Ph03 \\
278.6${-}$06.7 & \nodata & $\phn12.3$ & $\phn\phn0.73_{{-}0.03}^{{+}0.03}$ & $\phn\phn0.75_{{-}0.05}^{{+}0.10}$ & $\phn84$ & PM91,Ph03 \\
283.8${+}$02.2 & $\phn17.2$ & $\phn12.3$ & $\phn\phn1.57_{{-}0.02}^{{+}0.02}$ & $\phn\phn1.35_{{-}0.01}^{{+}0.01}$ & $113$ & Ph03 \\
285.4${-}$05.3 & $\phn15.5$ & $\phn11.6$ & $\phn\phn0.75_{{-}0.02}^{{+}0.03}$ & $\phn\phn0.68_{{-}0.01}^{{+}0.03}$ & $100$ & PM91,Ph03 \\
285.6${-}$02.7 & $\phn13.0$ & $\phn10.2$ & $\phn\phn2.11_{{-}0.02}^{{+}0.02}$ & $\phn\phn2.16_{{-}0.05}^{{+}0.06}$ & $\phn34$ & PM89,Ph03 \\
285.7${-}$14.9 & $\phn14.2$ & $\phn13.3$ & $\phn\phn0.14_{{-}0.03}^{{+}0.03}$ & $\phn\phn0.14_{{-}0.03}^{{+}0.02}$ & $\phn55$ & Me88 \\
291.6${-}$04.8 & $\phn15.4$ & $\phn10.5$ & $\phn\phn1.80_{{-}0.02}^{{+}0.02}$ & $\phn\phn1.79_{{-}0.03}^{{+}0.13}$ & $116$ & PM91,Ph03 \\
292.8${+}$01.1 & \nodata & $\phn12.5$ & $\phn\phn2.03_{{-}0.03}^{{+}0.02}$ & $\phn\phn2.16_{{-}0.07}^{{+}0.19}$ & $\phn95$ & PM91 \\
294.9${-}$04.3 & \nodata & $\phn11.5$ & $\phn\phn1.80_{{-}0.02}^{{+}0.02}$ & $\phn\phn1.80_{{-}0.05}^{{+}0.08}$ & $\phn41$ & PM89 \\
296.3${-}$03.0 & $\phn16.1$ & $\phn11.6$ & $\phn\phn2.32_{{-}0.05}^{{+}0.05}$ & $\phn\phn2.38_{{-}0.06}^{{+}0.35}$ & $103$ & PM91,Ph03 \\
304.5${-}$04.8 & $\phn16.4$ & $\phn10.8$ & $\phn\phn1.34_{{-}0.04}^{{+}0.04}$ & $\phn\phn1.34_{{-}0.02}^{{+}0.05}$ & $102$ & PM91,Ph03 \\
305.1${+}$01.4 & $\phn15.6$ & $\phn7.93$ & $\phn\phn\phn3.5_{{-}0.1}^{{+}0.1}$ & $\phn\phn3.50_{{-}0.05}^{{+}0.02}$ & $\phn59$ & PM91,KJ91,Ph03 \\
307.2${-}$09.0 & $\phn15.3$ & $\phn11.6$ & $\phn\phn1.16_{{-}0.02}^{{+}0.02}$ & $\phn\phn1.10_{{-}0.03}^{{+}0.05}$ & $\phn50$ & PM89,KJ91,Ph03 \\
307.5${-}$04.9 & \nodata & $\phn10.8$ & $\phn\phn\phn1.6_{{-}0.1}^{{+}0.2}$ & $\phn\phn1.64_{{-}0.03}^{{+}0.07}$ & $\phn38$ & PM89,Ph03 \\
312.6${-}$01.8 & $\phn15.1$ & $\phn11.9$ & $\phn\phn3.17_{{-}0.03}^{{+}0.03}$ & $\phn\phn3.35_{{-}0.08}^{{+}0.47}$ & $\phn42$ & PM89,KJ91,Ph03 \\
315.1${-}$13.0 & $\phn11.0$ & $\phn9.36$ & $\phn\phn\phn0.5_{{-}0.1}^{{+}0.1}$ & $\phn\phn\phn\phn\phn9_{{-}5}^{{+}1}$ & $\phn27$ & PM89,Ka78,Ph03 \\
320.1${-}$09.6 & $\phn10.9$ & $\phn10.1$ & $\phn\phn0.72_{{-}0.07}^{{+}0.08}$ & $\phn\phn0.63_{{-}0.05}^{{+}3.31}$ & $\phn27$ & Me88 \\
320.9${+}$02.0 & ${>}17.9$ & $\phn10.2$ & $\phn\phn4.96_{{-}0.09}^{{+}0.09}$ & $\phn\phn\phn\phn\phn7_{{-}2}^{{+}2}$ & $\phn44$ & PM89 \\
322.5${-}$05.2 & $\phn15.3$ & $\phn12.7$ & $\phn\phn0.80_{{-}0.03}^{{+}0.03}$ & $\phn\phn0.89_{{-}0.04}^{{+}0.19}$ & $127$ & Ph03 \\
323.9${+}$02.4 & $\phn16.8$ & $\phn11.8$ & $\phn\phn3.05_{{-}0.03}^{{+}0.03}$ & $\phn\phn3.01_{{-}0.22}^{{+}0.05}$ & $\phn42$ & PM89,Ph03 \\
324.8${-}$01.1 & \nodata & $\phn10.6$ & $\phn\phn6.34_{{-}0.03}^{{+}0.02}$ & $\phn\phn\phn6.5_{{-}0.2}^{{+}0.3}$ & \nodata & \nodata \\
325.8${-}$12.8 & $\phn13.4$ & $\phn11.1$ & $\phn\phn0.50_{{-}0.03}^{{+}0.03}$ & $\phn\phn\phn0.9_{{-}0.1}^{{+}0.5}$ & $\phn36$ & Me88 \\
326.0${-}$06.5 & $\phn13.1$ & $\phn11.8$ & $\phn\phn1.23_{{-}0.07}^{{+}0.07}$ & $\phn\phn1.16_{{-}0.05}^{{+}0.04}$ & $\phn25$ & Me88 \\
327.1${-}$01.8 & $\phn17.2$ & $\phn11.5$ & $\phn\phn3.83_{{-}0.03}^{{+}0.04}$ & $\phn\phn\phn4.2_{{-}0.3}^{{+}0.2}$ & $\phn35$ & PM89,Ph03 \\
327.8${-}$01.6 & \nodata & $\phn11.0$ & $\phn\phn5.10_{{-}0.03}^{{+}0.03}$ & $\phn\phn\phn5.7_{{-}0.5}^{{+}0.7}$ & \nodata & \nodata \\
331.1${-}$05.7 & $\phn12.7$ & $\phn10.3$ & $\phn\phn\phn1.4_{{-}0.1}^{{+}0.1}$ & $\phn\phn1.36_{{-}0.03}^{{+}3.35}$ & $\phn31$ & PM89,Ph03 \\
331.3${+}$16.8 & $\phn15.5$ & $\phn12.5$ & $\phn\phn0.10_{{-}0.03}^{{+}0.03}$ & $\phn\phn0.78_{{-}0.46}^{{+}0.01}$ & $\phn86$ & PM89,Ph03 \\
336.3${-}$05.6 & $\phn16.6$ & $\phn12.2$ & $\phn\phn1.32_{{-}0.06}^{{+}0.07}$ & $\phn\phn1.36_{{-}0.03}^{{+}0.02}$ & $101$ & PM91,Ph03 \\
342.1${+}$27.5 & \nodata & $\phn13.2$ & $\phn\phn0.25_{{-}0.02}^{{+}0.02}$ & $\phn\phn0.27_{{-}0.04}^{{+}0.47}$ & $119$ & PM89,Ph03 \\
349.8${+}$04.4 & $\phn17.0$ & $\phn12.1$ & $\phn\phn\phn1.9_{{-}0.1}^{{+}0.2}$ & $\phn\phn\phn2.1_{{-}0.1}^{{+}0.1}$ & $\phn76$ & PM89 \\
350.9${+}$04.4 & $\phn13.3$ & $\phn10.3$ & $\phn\phn1.83_{{-}0.08}^{{+}0.07}$ & $\phn\phn1.83_{{-}0.02}^{{+}2.23}$ & $\phn33$ & Me88 \\
356.1${+}$02.7 & \nodata & $\phn11.9$ & $\phn\phn\phn5.8_{{-}0.5}^{{+}1.0}$ & $\phn\phn5.54_{{-}0.01}^{{+}0.16}$ & $\phn73$ & PM89 \\
357.6${+}$02.6 & \nodata & $\phn12.2$ & $\phn\phn\phn4.7_{{-}0.6}^{{+}1.8}$ & $\phn\phn4.71_{{-}0.45}^{{+}0.06}$ & $107$ & PM91 \\
\enddata

\tablerefs{$^{(a)}$\citet{acker_strasbourg-eso_1992}; $^{(b)}$ \citet{skrutskie_two_2006}; HF83: \citet{harrington_planetary_1983}, Ka76: \citet{kaler_exciting_1976}, Ka78: \citet{kaler_forbidden_1978}, KJ91: \citet{kaler_central_1991}, Lu01: \citet{lumsden_infrared_2001}, Me88: \citet{mendez_high_1988}, Ph03: \citet{phillips_relation_2003}, PM89: \citet{preite-martinez_energy-balance_1989}, PM91: \citet{preite-martinez_energy-balance_1991}.}

\end{deluxetable}

\setcounter{table}{4}
\begin{deluxetable}{cccccccc}
\rotate
\tabletypesize{\small}
\tablewidth{0pt}

\tablecaption{Extinction Corrected Line Intensity \label{tab:intensity}}
\tablehead{
  \colhead{PN\,G} &
  \colhead{Bracket-$\alpha$} &
  \colhead{\ion{He}{2}{\footnotesize$(3.09{\rm\mu m})$}} &
  \colhead{\ion{He}{1}{\footnotesize$(4.30{\rm\mu m})$}} &
  \colhead{{PAH}{\footnotesize$(3.3{\rm\mu m})$}} &
  \colhead{{C{\sbond}H{$_{al}$}}{\footnotesize$(3.4$-$3.5{\rm\mu m})$}} &
  \colhead{[\ion{Mg}{4}]{\footnotesize$(4.49{\rm\mu m})$}} &
  \colhead{[\ion{Ar}{6}]{\footnotesize$(4.53{\rm\mu m})$}} \\
  \colhead{} &
  \colhead{$10^{-15}\,{\rm W\,m^{-2}}$} &
  \colhead{$10^{-15}\,{\rm W\,m^{-2}}$} &
  \colhead{$10^{-15}\,{\rm W\,m^{-2}}$} &
  \colhead{$10^{-15}\,{\rm W\,m^{-2}}$} &
  \colhead{$10^{-15}\,{\rm W\,m^{-2}}$} &
  \colhead{$10^{-15}\,{\rm W\,m^{-2}}$} &
  \colhead{$10^{-15}\,{\rm W\,m^{-2}}$}
}
\startdata
000.3${+}$12.2 & $\phn\phn\phn2.74_{{-}0.04}^{{+}0.05}$ & \nodata & $\phn\phn\phn0.14_{{-}0.03}^{{+}0.03}$ & ${<}   0.11$ & \nodata & \nodata & \nodata \\
002.0${-}$13.4 & $\phn\phn\phn2.05_{{-}0.03}^{{+}0.04}$ & \nodata & $\phn\phn\phn0.31_{{-}0.02}^{{+}0.03}$ & $\phn\phn\phn0.25_{{-}0.05}^{{+}0.05}$ & ${<}   0.10$ & \nodata & \nodata \\
003.1${+}$02.9 & $\phn\phn\phn3.16_{{-}0.04}^{{+}0.04}$ & ${<}   0.53$ & $\phn\phn\phn0.43_{{-}0.04}^{{+}0.03}$ & $\phn\phn\phn0.94_{{-}0.16}^{{+}0.25}$ & $\phn\phn\phn0.49_{{-}0.08}^{{+}0.09}$ & $\phn\phn\phn0.35_{{-}0.04}^{{+}0.05}$ & \nodata \\
011.0${+}$05.8 & $\phn\phn\phn1.09_{{-}0.04}^{{+}0.03}$ & $\phn\phn\phn0.13_{{-}0.03}^{{+}0.03}$ & ${<}   0.13$ & \nodata & \nodata & $\phn\phn\phn1.11_{{-}0.08}^{{+}0.05}$ & $\phn\phn\phn0.20_{{-}0.05}^{{+}0.04}$ \\
027.6${-}$09.6 & $\phn\phn\phn1.12_{{-}0.02}^{{+}0.04}$ & \nodata & $\phn\phn\phn0.10_{{-}0.01}^{{+}0.01}$ & \nodata & \nodata & \nodata & \nodata \\
037.8${-}$06.3 & $\phn\phn\phn4.61_{{-}0.10}^{{+}0.24}$ & \nodata & $\phn\phn\phn1.80_{{-}0.09}^{{+}0.13}$ & $\phn\phn\phn6.64_{{-}0.21}^{{+}0.40}$ & $\phn\phn\phn2.60_{{-}0.14}^{{+}0.21}$ & \nodata & \nodata \\
038.2${+}$12.0 & $\phn\phn\phn2.18_{{-}0.06}^{{+}0.04}$ & \nodata & ${<}   0.13$ & $\phn\phn\phn0.45_{{-}0.05}^{{+}0.04}$ & $\phn\phn\phn0.12_{{-}0.02}^{{+}0.03}$ & \nodata & \nodata \\
043.1${+}$03.8 & $\phn\phn\phn0.45_{{-}0.01}^{{+}0.01}$ & \nodata & \nodata & ${<}   0.06$ & ${<}   0.01$ & \nodata & \nodata \\
046.4${-}$04.1 & $\phn\phn\phn2.43_{{-}0.01}^{{+}0.02}$ & $\phn\phn\phn0.11_{{-}0.02}^{{+}0.02}$ & $\phn\phn\phn0.32_{{-}0.01}^{{+}0.02}$ & $\phn\phn\phn0.40_{{-}0.05}^{{+}0.04}$ & ${<}   0.16$ & \nodata & \nodata \\
051.4${+}$09.6 & $\phn\phn\phn3.25_{{-}0.02}^{{+}0.06}$ & \nodata & $\phn\phn\phn0.52_{{-}0.02}^{{+}0.02}$ & $\phn\phn\phn0.97_{{-}0.07}^{{+}0.05}$ & $\phn\phn\phn0.23_{{-}0.05}^{{+}0.04}$ & \nodata & \nodata \\
052.2${-}$04.0 & $\phn\phn\phn0.87_{{-}0.01}^{{+}0.01}$ & \nodata & $\phn\phn\phn0.14_{{-}0.01}^{{+}0.01}$ & $\phn\phn\phn0.05_{{-}0.01}^{{+}0.01}$ & ${<}   0.03$ & \nodata & \nodata \\
058.3${-}$10.9 & $\phn\phn\phn4.62_{{-}0.07}^{{+}0.10}$ & \nodata & $\phn\phn\phn0.99_{{-}0.04}^{{+}0.07}$ & ${<}   0.55$ & ${<}   0.26$ & \nodata & \nodata \\
060.1${-}$07.7 & $\phn\phn\phn1.41_{{-}0.01}^{{+}0.02}$ & $\phn\phn\phn0.19_{{-}0.01}^{{+}0.01}$ & $\phn\phn\phn0.05_{{-}0.01}^{{+}0.01}$ & $\phn\phn\phn2.21_{{-}0.03}^{{+}0.03}$ & $\phn\phn\phn1.34_{{-}0.03}^{{+}0.03}$ & $\phn\phn\phn5.04_{{-}0.06}^{{+}0.06}$ & $\phn\phn\phn1.90_{{-}0.03}^{{+}0.04}$ \\
060.5${+}$01.8 & $\phn\phn\phn1.12_{{-}0.01}^{{+}0.03}$ & \nodata & $\phn\phn\phn0.16_{{-}0.01}^{{+}0.01}$ & $\phn\phn\phn0.23_{{-}0.04}^{{+}0.02}$ & $\phn\phn\phn0.07_{{-}0.02}^{{+}0.01}$ & \nodata & \nodata \\
064.7${+}$05.0 & $\phn\phn34.69_{{-}1.38}^{{+}1.49}$ & \nodata & $\phn\phn\phn5.17_{{-}1.38}^{{+}2.18}$ & $\phn145.36_{{-}4.23}^{{+}4.77}$ & $\phn\phn50.92_{{-}1.79}^{{+}2.87}$ & \nodata & \nodata \\
071.6${-}$02.3 & $\phn\phn\phn3.50_{{-}0.09}^{{+}0.22}$ & \nodata & $\phn\phn\phn0.75_{{-}0.09}^{{+}0.13}$ & $\phn\phn\phn7.13_{{-}0.58}^{{+}0.37}$ & $\phn\phn\phn2.07_{{-}0.44}^{{+}0.29}$ & \nodata & \nodata \\
074.5${+}$02.1 & $\phn\phn\phn1.82_{{-}0.06}^{{+}0.08}$ & $\phn\phn\phn0.41_{{-}0.03}^{{+}0.06}$ & $\phn\phn\phn0.12_{{-}0.04}^{{+}0.20}$ & $\phn\phn\phn2.86_{{-}0.22}^{{+}0.25}$ & $\phn\phn\phn1.68_{{-}0.10}^{{+}0.37}$ & $\phn\phn\phn4.43_{{-}0.25}^{{+}0.39}$ & $\phn\phn\phn6.18_{{-}0.34}^{{+}0.30}$ \\
082.1${+}$07.0 & $\phn\phn\phn3.85_{{-}0.04}^{{+}0.40}$ & $\phn\phn\phn0.40_{{-}0.02}^{{+}0.08}$ & $\phn\phn\phn0.34_{{-}0.02}^{{+}0.06}$ & \nodata & \nodata & $\phn\phn\phn2.25_{{-}0.06}^{{+}0.09}$ & $\phn\phn\phn0.10_{{-}0.02}^{{+}0.09}$ \\
082.5${+}$11.3 & $\phn\phn\phn0.52_{{-}0.02}^{{+}0.02}$ & \nodata & $\phn\phn\phn0.10_{{-}0.02}^{{+}0.02}$ & \nodata & \nodata & \nodata & \nodata \\
086.5${-}$08.8 & $\phn\phn\phn0.92_{{-}0.01}^{{+}0.01}$ & $\phn\phn\phn0.49_{{-}0.01}^{{+}0.02}$ & \nodata & ${<}   0.08$ & ${<}   0.07$ & $\phn\phn\phn1.40_{{-}0.03}^{{+}0.03}$ & $\phn\phn\phn2.75_{{-}0.06}^{{+}0.05}$ \\
089.3${-}$02.2 & $\phn\phn\phn0.79_{{-}0.07}^{{+}0.24}$ & \nodata & \nodata & $\phn\phn\phn1.96_{{-}0.17}^{{+}0.62}$ & ${<}   0.48$ & \nodata & \nodata \\
089.8${-}$05.1 & $\phn\phn\phn2.92_{{-}0.03}^{{+}0.15}$ & $\phn\phn\phn0.23_{{-}0.04}^{{+}0.09}$ & $\phn\phn\phn0.92_{{-}0.06}^{{+}0.11}$ & $\phn\phn14.75_{{-}0.19}^{{+}0.41}$ & $\phn\phn\phn4.86_{{-}0.09}^{{+}0.22}$ & $\phn\phn\phn2.91_{{-}0.05}^{{+}0.24}$ & $\phn\phn\phn0.43_{{-}0.06}^{{+}0.14}$ \\
095.2${+}$00.7 & $\phn\phn\phn2.52_{{-}0.06}^{{+}0.07}$ & \nodata & $\phn\phn\phn0.53_{{-}0.02}^{{+}0.06}$ & $\phn\phn\phn0.56_{{-}0.06}^{{+}0.08}$ & ${<}   0.18$ & \nodata & \nodata \\
100.6${-}$05.4 & $\phn\phn\phn0.93_{{-}0.01}^{{+}0.01}$ & $\phn\phn\phn0.05_{{-}0.01}^{{+}0.01}$ & $\phn\phn\phn0.05_{{-}0.01}^{{+}0.01}$ & \nodata & \nodata & $\phn\phn\phn0.06_{{-}0.01}^{{+}0.01}$ & \nodata \\
111.8${-}$02.8 & $\phn\phn\phn6.18_{{-}0.47}^{{+}0.01}$ & \nodata & ${<}   2.62$ & ${<}   3.55$ & $\phn\phn\phn1.65_{{-}0.29}^{{+}0.38}$ & \nodata & \nodata \\
118.0${-}$08.6 & $\phn\phn\phn0.35_{{-}0.02}^{{+}0.02}$ & \nodata & ${<}   0.05$ & \nodata & \nodata & \nodata & \nodata \\
123.6${+}$34.5 & $\phn\phn\phn2.02_{{-}0.05}^{{+}0.05}$ & \nodata & ${<}   0.15$ & ${<}   0.18$ & ${<}   0.05$ & \nodata & \nodata \\
146.7${+}$07.6 & $\phn\phn\phn0.56_{{-}0.01}^{{+}0.04}$ & \nodata & $\phn\phn\phn0.10_{{-}0.02}^{{+}0.03}$ & $\phn\phn\phn4.23_{{-}0.10}^{{+}0.14}$ & $\phn\phn\phn0.79_{{-}0.04}^{{+}0.08}$ & \nodata & \nodata \\
159.0${-}$15.1 & $\phn\phn\phn0.48_{{-}0.01}^{{+}0.01}$ & $\phn\phn\phn0.11_{{-}0.01}^{{+}0.01}$ & ${<}   0.06$ & \nodata & \nodata & $\phn\phn\phn0.86_{{-}0.03}^{{+}0.03}$ & \nodata \\
166.1${+}$10.4 & $\phn\phn\phn4.34_{{-}0.06}^{{+}0.05}$ & \nodata & $\phn\phn\phn0.35_{{-}0.06}^{{+}0.06}$ & $\phn\phn\phn0.23_{{-}0.06}^{{+}0.25}$ & ${<}   0.16$ & \nodata & \nodata \\
190.3${-}$17.7 & $\phn\phn\phn0.46_{{-}0.01}^{{+}0.01}$ & \nodata & $\phn\phn\phn0.04_{{-}0.01}^{{+}0.01}$ & ${<}   0.04$ & \nodata & \nodata & \nodata \\
194.2${+}$02.5 & $\phn\phn\phn1.72_{{-}0.08}^{{+}0.05}$ & $\phn\phn\phn0.41_{{-}0.04}^{{+}0.06}$ & ${<}   0.20$ & $\phn\phn\phn3.70_{{-}0.17}^{{+}0.25}$ & $\phn\phn\phn1.16_{{-}0.14}^{{+}0.17}$ & $\phn\phn\phn2.44_{{-}0.09}^{{+}0.11}$ & $\phn\phn\phn0.34_{{-}0.10}^{{+}0.11}$ \\
211.2${-}$03.5 & $\phn\phn\phn1.82_{{-}0.04}^{{+}0.05}$ & \nodata & $\phn\phn\phn0.26_{{-}0.05}^{{+}0.05}$ & $\phn\phn\phn0.47_{{-}0.07}^{{+}0.23}$ & ${<}   0.24$ & \nodata & \nodata \\
221.3${-}$12.3 & $\phn\phn\phn2.63_{{-}0.03}^{{+}0.03}$ & $\phn\phn\phn0.70_{{-}0.04}^{{+}0.04}$ & $\phn\phn\phn0.18_{{-}0.03}^{{+}0.04}$ & ${<}   0.77$ & ${<}   0.30$ & $\phn\phn\phn6.56_{{-}0.09}^{{+}0.08}$ & $\phn\phn\phn2.17_{{-}0.07}^{{+}0.07}$ \\
226.7${+}$05.6 & $\phn\phn\phn0.47_{{-}0.01}^{{+}0.01}$ & $\phn\phn\phn0.05_{{-}0.01}^{{+}0.01}$ & $\phn\phn\phn0.04_{{-}0.01}^{{+}0.01}$ & $\phn\phn\phn0.84_{{-}0.03}^{{+}0.05}$ & $\phn\phn\phn0.52_{{-}0.02}^{{+}0.02}$ & $\phn\phn\phn0.40_{{-}0.01}^{{+}0.01}$ & $\phn\phn\phn0.18_{{-}0.01}^{{+}0.01}$ \\
232.8${-}$04.7 & $\phn\phn\phn2.88_{{-}0.14}^{{+}0.19}$ & \nodata & ${<}   0.59$ & $\phn\phn\phn8.87_{{-}0.61}^{{+}0.69}$ & $\phn\phn\phn2.49_{{-}0.23}^{{+}0.47}$ & \nodata & \nodata \\
235.3${-}$03.9 & $\phn\phn\phn1.12_{{-}0.05}^{{+}0.05}$ & \nodata & $\phn\phn\phn0.16_{{-}0.04}^{{+}0.06}$ & $\phn\phn\phn0.57_{{-}0.08}^{{+}0.08}$ & ${<}   0.21$ & \nodata & \nodata \\
258.1${-}$00.3 & $\phn\phn\phn3.70_{{-}0.05}^{{+}0.06}$ & \nodata & $\phn\phn\phn0.56_{{-}0.04}^{{+}0.05}$ & $\phn\phn\phn0.35_{{-}0.07}^{{+}0.15}$ & ${<}   0.09$ & \nodata & \nodata \\
264.4${-}$12.7 & $\phn\phn\phn0.70_{{-}0.01}^{{+}0.02}$ & \nodata & $\phn\phn\phn0.07_{{-}0.01}^{{+}0.01}$ & ${<}   0.06$ & ${<}   0.04$ & \nodata & \nodata \\
268.4${+}$02.4 & $\phn\phn\phn1.52_{{-}0.03}^{{+}0.04}$ & $\phn\phn\phn0.29_{{-}0.02}^{{+}0.05}$ & $\phn\phn\phn0.27_{{-}0.04}^{{+}0.06}$ & $\phn\phn11.23_{{-}0.31}^{{+}0.38}$ & $\phn\phn\phn5.52_{{-}0.13}^{{+}0.19}$ & $\phn\phn\phn5.34_{{-}0.14}^{{+}0.14}$ & $\phn\phn\phn3.69_{{-}0.09}^{{+}0.15}$ \\
278.6${-}$06.7 & $\phn\phn\phn0.59_{{-}0.01}^{{+}0.01}$ & $\phn\phn\phn0.03_{{-}0.01}^{{+}0.01}$ & $\phn\phn\phn0.09_{{-}0.01}^{{+}0.01}$ & $\phn\phn\phn0.10_{{-}0.01}^{{+}0.02}$ & ${<}   0.04$ & $\phn\phn\phn0.10_{{-}0.01}^{{+}0.01}$ & \nodata \\
283.8${+}$02.2 & $\phn\phn\phn0.73_{{-}0.01}^{{+}0.01}$ & $\phn\phn\phn0.34_{{-}0.01}^{{+}0.01}$ & ${<}   0.01$ & \nodata & \nodata & $\phn\phn\phn5.46_{{-}0.01}^{{+}0.06}$ & $\phn\phn\phn0.96_{{-}0.01}^{{+}0.01}$ \\
285.4${-}$05.3 & $\phn\phn\phn2.68_{{-}0.01}^{{+}0.02}$ & $\phn\phn\phn0.42_{{-}0.02}^{{+}0.02}$ & $\phn\phn\phn0.18_{{-}0.02}^{{+}0.02}$ & $\phn\phn\phn0.21_{{-}0.06}^{{+}0.06}$ & ${<}   0.06$ & $\phn\phn\phn2.84_{{-}0.01}^{{+}0.02}$ & $\phn\phn\phn0.29_{{-}0.01}^{{+}0.02}$ \\
285.6${-}$02.7 & $\phn\phn\phn4.92_{{-}0.04}^{{+}0.05}$ & \nodata & $\phn\phn\phn0.24_{{-}0.02}^{{+}0.02}$ & $\phn\phn\phn1.41_{{-}0.04}^{{+}0.06}$ & $\phn\phn\phn0.21_{{-}0.04}^{{+}0.07}$ & \nodata & \nodata \\
285.7${-}$14.9 & $\phn\phn\phn1.28_{{-}0.02}^{{+}0.03}$ & $\phn\phn\phn0.22_{{-}0.03}^{{+}0.03}$ & ${<}   0.12$ & \nodata & \nodata & $\phn\phn\phn0.65_{{-}0.05}^{{+}0.05}$ & ${<}   0.08$ \\
291.6${-}$04.8 & $\phn\phn\phn2.91_{{-}0.03}^{{+}0.08}$ & $\phn\phn\phn0.55_{{-}0.03}^{{+}0.09}$ & $\phn\phn\phn0.38_{{-}0.01}^{{+}0.06}$ & $\phn\phn\phn7.18_{{-}0.08}^{{+}0.24}$ & $\phn\phn\phn3.61_{{-}0.04}^{{+}0.18}$ & $\phn\phn10.18_{{-}0.05}^{{+}0.40}$ & $\phn\phn\phn7.72_{{-}0.04}^{{+}0.38}$ \\
292.8${+}$01.1 & $\phn\phn\phn0.78_{{-}0.02}^{{+}0.02}$ & $\phn\phn\phn0.08_{{-}0.01}^{{+}0.01}$ & $\phn\phn\phn0.07_{{-}0.01}^{{+}0.01}$ & $\phn\phn\phn0.43_{{-}0.03}^{{+}0.04}$ & $\phn\phn\phn0.23_{{-}0.03}^{{+}0.04}$ & $\phn\phn\phn0.06_{{-}0.01}^{{+}0.02}$ & \nodata \\
294.9${-}$04.3 & $\phn\phn\phn0.92_{{-}0.02}^{{+}0.02}$ & \nodata & $\phn\phn\phn0.14_{{-}0.03}^{{+}0.03}$ & ${<}   0.24$ & ${<}   0.02$ & \nodata & \nodata \\
296.3${-}$03.0 & $\phn\phn\phn1.16_{{-}0.03}^{{+}0.03}$ & $\phn\phn\phn0.12_{{-}0.01}^{{+}0.02}$ & $\phn\phn\phn0.14_{{-}0.02}^{{+}0.02}$ & $\phn\phn\phn0.49_{{-}0.03}^{{+}0.04}$ & $\phn\phn\phn0.23_{{-}0.02}^{{+}0.04}$ & $\phn\phn\phn0.84_{{-}0.04}^{{+}0.04}$ & $\phn\phn\phn0.38_{{-}0.03}^{{+}0.03}$ \\
304.5${-}$04.8 & $\phn\phn\phn3.47_{{-}0.02}^{{+}0.04}$ & $\phn\phn\phn0.31_{{-}0.04}^{{+}0.04}$ & $\phn\phn\phn0.48_{{-}0.03}^{{+}0.04}$ & $\phn\phn\phn1.32_{{-}0.27}^{{+}0.10}$ & $\phn\phn\phn0.49_{{-}0.09}^{{+}0.10}$ & $\phn\phn\phn4.49_{{-}0.03}^{{+}0.08}$ & $\phn\phn\phn0.85_{{-}0.03}^{{+}0.06}$ \\
305.1${+}$01.4 & $\phn\phn12.65_{{-}1.67}^{{+}1.42}$ & \nodata & \nodata & $\phn\phn\phn7.43_{{-}1.64}^{{+}2.59}$ & $\phn\phn10.78_{{-}1.40}^{{+}3.23}$ & \nodata & \nodata \\
307.2${-}$09.0 & $\phn\phn\phn0.98_{{-}0.01}^{{+}0.02}$ & \nodata & $\phn\phn\phn0.19_{{-}0.01}^{{+}0.01}$ & $\phn\phn\phn0.20_{{-}0.05}^{{+}0.05}$ & ${<}   0.06$ & \nodata & \nodata \\
307.5${-}$04.9 & $\phn\phn\phn2.67_{{-}0.03}^{{+}0.03}$ & \nodata & $\phn\phn\phn0.27_{{-}0.02}^{{+}0.04}$ & $\phn\phn\phn2.06_{{-}0.08}^{{+}0.18}$ & $\phn\phn\phn0.28_{{-}0.09}^{{+}0.12}$ & \nodata & \nodata \\
312.6${-}$01.8 & $\phn\phn\phn1.38_{{-}0.02}^{{+}0.03}$ & \nodata & $\phn\phn\phn0.11_{{-}0.02}^{{+}0.02}$ & $\phn\phn\phn0.11_{{-}0.03}^{{+}0.05}$ & ${<}   0.04$ & \nodata & \nodata \\
315.1${-}$13.0 & $\phn\phn11.90_{{-}0.91}^{{+}1.05}$ & \nodata & $\phn\phn\phn0.78_{{-}0.20}^{{+}0.27}$ & $\phn\phn\phn5.15_{{-}1.02}^{{+}1.08}$ & $\phn\phn\phn1.48_{{-}0.45}^{{+}0.50}$ & \nodata & \nodata \\
320.1${-}$09.6 & $\phn\phn\phn3.26_{{-}0.04}^{{+}0.40}$ & \nodata & ${<}   0.13$ & $\phn\phn\phn3.58_{{-}0.08}^{{+}0.33}$ & $\phn\phn\phn0.81_{{-}0.06}^{{+}0.19}$ & \nodata & \nodata \\
320.9${+}$02.0 & $\phn\phn\phn5.99_{{-}0.27}^{{+}0.46}$ & \nodata & $\phn\phn\phn1.20_{{-}0.09}^{{+}0.31}$ & $\phn\phn\phn3.23_{{-}0.19}^{{+}0.12}$ & $\phn\phn\phn1.31_{{-}0.09}^{{+}0.16}$ & \nodata & \nodata \\
322.5${-}$05.2 & $\phn\phn\phn1.12_{{-}0.02}^{{+}0.02}$ & $\phn\phn\phn0.78_{{-}0.03}^{{+}0.04}$ & ${<}   0.16$ & \nodata & \nodata & $\phn\phn13.18_{{-}0.11}^{{+}0.14}$ & \nodata \\
323.9${+}$02.4 & $\phn\phn\phn2.02_{{-}0.03}^{{+}0.04}$ & \nodata & $\phn\phn\phn0.25_{{-}0.02}^{{+}0.03}$ & $\phn\phn\phn2.20_{{-}0.09}^{{+}0.04}$ & $\phn\phn\phn0.95_{{-}0.03}^{{+}0.04}$ & \nodata & \nodata \\
324.8${-}$01.1 & $\phn\phn\phn4.11_{{-}0.08}^{{+}0.12}$ & \nodata & $\phn\phn\phn0.86_{{-}0.05}^{{+}0.07}$ & $\phn\phn\phn1.06_{{-}0.12}^{{+}0.14}$ & $\phn\phn\phn0.35_{{-}0.08}^{{+}0.11}$ & \nodata & \nodata \\
325.8${-}$12.8 & $\phn\phn\phn1.50_{{-}0.03}^{{+}0.06}$ & \nodata & $\phn\phn\phn0.17_{{-}0.01}^{{+}0.03}$ & \nodata & \nodata & \nodata & \nodata \\
326.0${-}$06.5 & $\phn\phn\phn0.36_{{-}0.01}^{{+}0.01}$ & \nodata & ${<}   0.01$ & $\phn\phn\phn0.07_{{-}0.01}^{{+}0.06}$ & ${<}   0.04$ & \nodata & \nodata \\
327.1${-}$01.8 & $\phn\phn\phn1.69_{{-}0.07}^{{+}0.07}$ & \nodata & $\phn\phn\phn0.17_{{-}0.01}^{{+}0.04}$ & $\phn\phn\phn1.20_{{-}0.16}^{{+}0.10}$ & $\phn\phn\phn0.60_{{-}0.02}^{{+}0.03}$ & \nodata & \nodata \\
327.8${-}$01.6 & $\phn\phn\phn1.97_{{-}0.03}^{{+}0.07}$ & $\phn\phn\phn0.36_{{-}0.03}^{{+}0.06}$ & $\phn\phn\phn0.28_{{-}0.03}^{{+}0.05}$ & $\phn\phn\phn2.24_{{-}0.35}^{{+}0.21}$ & $\phn\phn\phn1.33_{{-}0.11}^{{+}0.13}$ & $\phn\phn\phn3.12_{{-}0.14}^{{+}0.08}$ & $\phn\phn\phn3.68_{{-}0.28}^{{+}0.04}$ \\
331.1${-}$05.7 & $\phn\phn\phn1.13_{{-}0.03}^{{+}0.22}$ & \nodata & $\phn\phn\phn0.29_{{-}0.02}^{{+}0.08}$ & \nodata & \nodata & \nodata & \nodata \\
331.3${+}$16.8 & $\phn\phn\phn0.74_{{-}0.02}^{{+}0.02}$ & $\phn\phn\phn0.21_{{-}0.02}^{{+}0.03}$ & ${<}   0.04$ & $\phn\phn\phn0.16_{{-}0.03}^{{+}0.04}$ & $\phn\phn\phn0.05_{{-}0.01}^{{+}0.03}$ & $\phn\phn\phn1.20_{{-}0.04}^{{+}0.02}$ & \nodata \\
336.3${-}$05.6 & $\phn\phn\phn0.21_{{-}0.01}^{{+}0.01}$ & $\phn\phn\phn0.06_{{-}0.01}^{{+}0.01}$ & ${<}   0.03$ & $\phn\phn\phn0.43_{{-}0.03}^{{+}0.08}$ & $\phn\phn\phn0.23_{{-}0.02}^{{+}0.04}$ & $\phn\phn\phn0.39_{{-}0.02}^{{+}0.02}$ & $\phn\phn\phn0.41_{{-}0.02}^{{+}0.02}$ \\
342.1${+}$27.5 & $\phn\phn\phn0.47_{{-}0.01}^{{+}0.01}$ & $\phn\phn\phn0.25_{{-}0.01}^{{+}0.02}$ & ${<}   0.03$ & \nodata & \nodata & $\phn\phn\phn2.91_{{-}0.05}^{{+}0.05}$ & $\phn\phn\phn0.44_{{-}0.02}^{{+}0.03}$ \\
349.8${+}$04.4 & $\phn\phn\phn0.87_{{-}0.02}^{{+}0.02}$ & \nodata & $\phn\phn\phn0.13_{{-}0.01}^{{+}0.01}$ & $\phn\phn\phn0.41_{{-}0.02}^{{+}0.03}$ & $\phn\phn\phn0.19_{{-}0.03}^{{+}0.04}$ & \nodata & \nodata \\
350.9${+}$04.4 & $\phn\phn\phn2.03_{{-}0.04}^{{+}0.21}$ & \nodata & $\phn\phn\phn0.14_{{-}0.03}^{{+}0.08}$ & \nodata & \nodata & \nodata & \nodata \\
356.1${+}$02.7 & $\phn\phn\phn0.74_{{-}0.04}^{{+}0.01}$ & \nodata & $\phn\phn\phn0.22_{{-}0.02}^{{+}0.01}$ & $\phn\phn\phn0.99_{{-}0.02}^{{+}0.03}$ & $\phn\phn\phn0.48_{{-}0.01}^{{+}0.08}$ & \nodata & \nodata \\
357.6${+}$02.6 & $\phn\phn\phn0.73_{{-}0.04}^{{+}0.05}$ & \nodata & $\phn\phn\phn0.16_{{-}0.04}^{{+}0.04}$ & $\phn\phn\phn0.54_{{-}0.07}^{{+}0.08}$ & $\phn\phn\phn0.15_{{-}0.01}^{{+}0.05}$ & \nodata & \nodata \\
\enddata
\end{deluxetable}

\setcounter{table}{5}
\begin{deluxetable}{cccccccc}
\rotate
\tabletypesize{\small}
\tablewidth{0pt}

\tablecaption{Extinction Corrected Line Equivalent Width \label{tab:equivw}}
\tablehead{
  \colhead{PN\,G} &
  \colhead{Bracket-$\alpha$} &
  \colhead{\ion{He}{2}{\footnotesize$(3.09{\rm\mu m})$}} &
  \colhead{\ion{He}{1}{\footnotesize$(4.30{\rm\mu m})$}} &
  \colhead{{PAH}{\footnotesize$(3.3{\rm\mu m})$}} &
  \colhead{{C{\sbond}H{$_{al}$}}{\footnotesize$(3.4$-$3.5{\rm\mu m})$}} &
  \colhead{[\ion{Mg}{4}]{\footnotesize$(4.49{\rm\mu m})$}} &
  \colhead{[\ion{Ar}{6}]{\footnotesize$(4.53{\rm\mu m})$}} \\
  \colhead{} &
  \colhead{{\AA}} &
  \colhead{{\AA}} &
  \colhead{{\AA}} &
  \colhead{{\AA}} &
  \colhead{{\AA}} &
  \colhead{{\AA}} &
  \colhead{{\AA}}
}
\startdata
000.3${+}$12.2 & $\phn2825.6_{{-}38.2}^{{+}47.3}$ & \nodata & $\phn\phn154.5_{{-}32.7}^{{+}35.5}$ & ${<}  130.7$ & \nodata & \nodata & \nodata \\
002.0${-}$13.4 & $\phn2938.4_{{-}39.9}^{{+}64.0}$ & \nodata & $\phn\phn481.9_{{-}27.1}^{{+}42.1}$ & $\phn\phn435.3_{{-}84.6}^{{+}86.8}$ & ${<}  181.4$ & \nodata & \nodata \\
003.1${+}$02.9 & $\phn2105.7_{{-}25.5}^{{+}26.5}$ & ${<}  236.7$ & $\phn\phn343.7_{{-}30.2}^{{+}25.6}$ & $\phn\phn665.7_{{-}113.7}^{{+}176.0}$ & $\phn\phn365.1_{{-}59.5}^{{+}68.0}$ & $\phn\phn324.3_{{-}42.0}^{{+}51.4}$ & \nodata \\
011.0${+}$05.8 & $\phn2005.1_{{-}68.8}^{{+}61.4}$ & $\phn\phn144.8_{{-}34.2}^{{+}32.1}$ & ${<}  266.9$ & \nodata & \nodata & $\phn2605.4_{{-}179.7}^{{+}127.9}$ & $\phn\phn486.2_{{-}111.4}^{{+}101.8}$ \\
027.6${-}$09.6 & $\phn3229.6_{{-}44.8}^{{+}128.8}$ & \nodata & $\phn\phn328.8_{{-}27.6}^{{+}35.8}$ & \nodata & \nodata & \nodata & \nodata \\
037.8${-}$06.3 & $\phn\phn482.1_{{-}10.0}^{{+}25.4}$ & \nodata & $\phn\phn182.2_{{-}\phn8.6}^{{+}13.6}$ & $\phn1342.1_{{-}42.1}^{{+}80.2}$ & $\phn\phn531.1_{{-}29.1}^{{+}42.6}$ & \nodata & \nodata \\
038.2${+}$12.0 & $\phn2863.3_{{-}84.4}^{{+}54.5}$ & \nodata & ${<}  192.5$ & $\phn\phn709.0_{{-}75.8}^{{+}55.0}$ & $\phn\phn204.3_{{-}38.7}^{{+}47.2}$ & \nodata & \nodata \\
043.1${+}$03.8 & $\phn1316.2_{{-}37.5}^{{+}42.9}$ & \nodata & \nodata & ${<}  230.0$ & \nodata & \nodata & \nodata \\
046.4${-}$04.1 & $\phn3171.2_{{-}\phn5.0}^{{+}29.3}$ & $\phn\phn\phn99.6_{{-}18.0}^{{+}17.8}$ & $\phn\phn475.1_{{-}\phn9.3}^{{+}29.6}$ & $\phn\phn595.7_{{-}75.1}^{{+}65.3}$ & ${<}  250.8$ & \nodata & \nodata \\
051.4${+}$09.6 & $\phn1882.4_{{-}13.5}^{{+}36.4}$ & \nodata & $\phn\phn315.8_{{-}12.1}^{{+}15.0}$ & $\phn\phn740.7_{{-}56.1}^{{+}41.7}$ & $\phn\phn180.1_{{-}42.6}^{{+}32.7}$ & \nodata & \nodata \\
052.2${-}$04.0 & $\phn2778.2_{{-}41.1}^{{+}38.8}$ & \nodata & $\phn\phn478.9_{{-}40.3}^{{+}33.0}$ & $\phn\phn192.2_{{-}49.9}^{{+}47.2}$ & ${<}  138.0$ & \nodata & \nodata \\
058.3${-}$10.9 & $\phn1580.9_{{-}23.7}^{{+}33.7}$ & \nodata & $\phn\phn354.5_{{-}14.5}^{{+}26.0}$ & ${<}  252.0$ & ${<}  124.7$ & \nodata & \nodata \\
060.1${-}$07.7 & $\phn1489.3_{{-}13.4}^{{+}18.8}$ & $\phn\phn179.9_{{-}\phn9.8}^{{+}\phn8.4}$ & $\phn\phn\phn50.0_{{-}10.6}^{{+}\phn9.9}$ & $\phn3259.8_{{-}47.6}^{{+}47.8}$ & $\phn2085.7_{{-}43.7}^{{+}46.6}$ & $\phn5697.5_{{-}65.0}^{{+}71.7}$ & $\phn2172.0_{{-}39.7}^{{+}47.5}$ \\
060.5${+}$01.8 & $\phn2334.5_{{-}\phn7.7}^{{+}54.5}$ & \nodata & $\phn\phn379.5_{{-}13.2}^{{+}\phn7.7}$ & $\phn\phn543.1_{{-}85.9}^{{+}48.1}$ & $\phn\phn175.0_{{-}47.2}^{{+}\phn0.4}$ & \nodata & \nodata \\
064.7${+}$05.0 & $\phn\phn479.5_{{-}19.0}^{{+}20.6}$ & \nodata & $\phn\phn\phn72.0_{{-}19.3}^{{+}30.4}$ & $\phn5717.7_{{-}166.3}^{{+}187.7}$ & $\phn1790.1_{{-}62.9}^{{+}100.9}$ & \nodata & \nodata \\
071.6${-}$02.3 & $\phn\phn623.6_{{-}15.8}^{{+}39.5}$ & \nodata & $\phn\phn139.0_{{-}17.5}^{{+}24.8}$ & $\phn1853.9_{{-}152.0}^{{+}95.9}$ & $\phn\phn559.1_{{-}120.2}^{{+}77.4}$ & \nodata & \nodata \\
074.5${+}$02.1 & $\phn1507.3_{{-}47.9}^{{+}63.7}$ & $\phn\phn317.6_{{-}25.1}^{{+}48.7}$ & $\phn\phn104.7_{{-}31.3}^{{+}170.7}$ & $\phn3458.0_{{-}263.3}^{{+}298.1}$ & $\phn2146.1_{{-}131.8}^{{+}477.7}$ & $\phn3936.3_{{-}217.8}^{{+}348.3}$ & $\phn5582.0_{{-}310.8}^{{+}274.4}$ \\
082.1${+}$07.0 & $\phn2773.7_{{-}27.6}^{{+}291.2}$ & $\phn\phn203.6_{{-}\phn9.5}^{{+}38.5}$ & $\phn\phn279.1_{{-}17.6}^{{+}47.0}$ & \nodata & \nodata & $\phn2007.7_{{-}52.5}^{{+}82.0}$ & $\phn\phn\phn93.2_{{-}19.5}^{{+}80.4}$ \\
082.5${+}$11.3 & $\phn1850.5_{{-}56.8}^{{+}69.2}$ & \nodata & $\phn\phn418.8_{{-}66.7}^{{+}91.3}$ & \nodata & \nodata & \nodata & \nodata \\
086.5${-}$08.8 & $\phn1225.3_{{-}16.8}^{{+}18.4}$ & $\phn\phn503.8_{{-}14.4}^{{+}16.0}$ & \nodata & ${<}  140.6$ & ${<}  123.9$ & $\phn2075.1_{{-}47.1}^{{+}45.4}$ & $\phn4149.5_{{-}88.3}^{{+}73.8}$ \\
089.3${-}$02.2 & $\phn\phn\phn58.2_{{-}\phn4.8}^{{+}18.0}$ & \nodata & \nodata & $\phn\phn187.6_{{-}15.8}^{{+}59.6}$ & ${<}   48.8$ & \nodata & \nodata \\
089.8${-}$05.1 & $\phn\phn284.4_{{-}\phn2.9}^{{+}14.9}$ & $\phn\phn\phn31.3_{{-}\phn5.1}^{{+}11.7}$ & $\phn\phn\phn88.8_{{-}\phn5.4}^{{+}10.2}$ & $\phn2869.5_{{-}36.7}^{{+}79.0}$ & $\phn\phn953.0_{{-}17.2}^{{+}44.0}$ & $\phn\phn276.9_{{-}\phn4.9}^{{+}23.1}$ & $\phn\phn\phn40.8_{{-}\phn5.7}^{{+}13.6}$ \\
095.2${+}$00.7 & $\phn1430.9_{{-}32.0}^{{+}38.1}$ & \nodata & $\phn\phn316.8_{{-}14.4}^{{+}35.7}$ & $\phn\phn415.1_{{-}43.4}^{{+}55.9}$ & ${<}  138.7$ & \nodata & \nodata \\
100.6${-}$05.4 & $\phn2648.6_{{-}42.1}^{{+}41.4}$ & $\phn\phn\phn86.8_{{-}17.6}^{{+}23.1}$ & $\phn\phn174.4_{{-}36.0}^{{+}32.5}$ & \nodata & \nodata & $\phn\phn207.7_{{-}41.2}^{{+}40.0}$ & \nodata \\
111.8${-}$02.8 & $\phn\phn422.8_{{-}31.8}^{{+}\phn0.7}$ & \nodata & ${<}  177.5$ & ${<}  376.8$ & $\phn\phn179.2_{{-}32.0}^{{+}41.0}$ & \nodata & \nodata \\
118.0${-}$08.6 & $\phn\phn876.0_{{-}59.9}^{{+}57.3}$ & \nodata & ${<}  124.1$ & \nodata & \nodata & \nodata & \nodata \\
123.6${+}$34.5 & $\phn3217.8_{{-}71.6}^{{+}77.1}$ & \nodata & ${<}  271.0$ & ${<}  321.3$ & \nodata & \nodata & \nodata \\
146.7${+}$07.6 & $\phn\phn317.0_{{-}\phn5.0}^{{+}20.9}$ & \nodata & $\phn\phn\phn57.3_{{-}\phn9.4}^{{+}19.8}$ & $\phn3562.1_{{-}84.5}^{{+}121.4}$ & $\phn\phn676.7_{{-}38.6}^{{+}69.3}$ & \nodata & \nodata \\
159.0${-}$15.1 & $\phn1660.3_{{-}49.9}^{{+}43.4}$ & $\phn\phn270.2_{{-}36.4}^{{+}37.1}$ & ${<}  234.5$ & \nodata & \nodata & $\phn3545.4_{{-}136.3}^{{+}109.5}$ & \nodata \\
166.1${+}$10.4 & $\phn1723.6_{{-}23.4}^{{+}20.6}$ & \nodata & $\phn\phn154.5_{{-}24.6}^{{+}25.5}$ & $\phn\phn112.8_{{-}31.5}^{{+}122.4}$ & ${<}   84.6$ & \nodata & \nodata \\
190.3${-}$17.7 & $\phn2110.9_{{-}27.2}^{{+}32.7}$ & \nodata & $\phn\phn215.4_{{-}44.8}^{{+}41.2}$ & ${<}  202.7$ & \nodata & \nodata & \nodata \\
194.2${+}$02.5 & $\phn\phn372.3_{{-}17.7}^{{+}11.4}$ & $\phn\phn\phn85.0_{{-}\phn9.3}^{{+}12.7}$ & ${<}   44.5$ & $\phn1157.9_{{-}53.9}^{{+}79.8}$ & $\phn\phn379.3_{{-}46.7}^{{+}56.9}$ & $\phn\phn585.6_{{-}21.7}^{{+}26.1}$ & $\phn\phn\phn82.2_{{-}23.5}^{{+}27.1}$ \\
211.2${-}$03.5 & $\phn\phn476.4_{{-}10.3}^{{+}12.3}$ & \nodata & $\phn\phn\phn71.3_{{-}12.3}^{{+}12.4}$ & $\phn\phn173.7_{{-}25.6}^{{+}84.5}$ & ${<}   95.3$ & \nodata & \nodata \\
221.3${-}$12.3 & $\phn1873.6_{{-}18.8}^{{+}23.2}$ & $\phn\phn371.5_{{-}21.0}^{{+}22.7}$ & $\phn\phn148.6_{{-}28.1}^{{+}30.7}$ & ${<}  655.3$ & ${<}  273.5$ & $\phn5881.3_{{-}78.8}^{{+}74.2}$ & $\phn1983.8_{{-}62.0}^{{+}63.3}$ \\
226.7${+}$05.6 & $\phn1578.7_{{-}19.0}^{{+}22.1}$ & $\phn\phn142.7_{{-}13.3}^{{+}14.8}$ & $\phn\phn147.0_{{-}20.4}^{{+}32.7}$ & $\phn3534.1_{{-}134.5}^{{+}206.2}$ & $\phn2311.3_{{-}98.9}^{{+}100.0}$ & $\phn1501.2_{{-}37.7}^{{+}45.9}$ & $\phn\phn659.9_{{-}26.9}^{{+}42.4}$ \\
232.8${-}$04.7 & $\phn\phn191.3_{{-}\phn9.0}^{{+}12.9}$ & \nodata & ${<}   40.3$ & $\phn\phn865.2_{{-}59.6}^{{+}67.6}$ & $\phn\phn250.9_{{-}22.8}^{{+}47.1}$ & \nodata & \nodata \\
235.3${-}$03.9 & $\phn\phn339.2_{{-}15.4}^{{+}14.4}$ & \nodata & $\phn\phn\phn50.9_{{-}11.2}^{{+}17.7}$ & $\phn\phn243.1_{{-}35.8}^{{+}32.0}$ & ${<}   92.8$ & \nodata & \nodata \\
258.1${-}$00.3 & $\phn1577.4_{{-}19.8}^{{+}23.8}$ & \nodata & $\phn\phn248.9_{{-}16.0}^{{+}21.1}$ & $\phn\phn195.1_{{-}42.1}^{{+}81.7}$ & \nodata & \nodata & \nodata \\
264.4${-}$12.7 & $\phn2768.5_{{-}47.2}^{{+}62.1}$ & \nodata & $\phn\phn273.6_{{-}47.5}^{{+}54.6}$ & ${<}  289.7$ & ${<}  223.4$ & \nodata & \nodata \\
268.4${+}$02.4 & $\phn\phn354.8_{{-}\phn7.8}^{{+}\phn8.2}$ & $\phn\phn\phn86.7_{{-}\phn6.4}^{{+}13.9}$ & $\phn\phn\phn64.2_{{-}\phn9.2}^{{+}15.1}$ & $\phn4927.5_{{-}135.8}^{{+}166.2}$ & $\phn2485.7_{{-}57.7}^{{+}86.2}$ & $\phn1292.7_{{-}32.9}^{{+}34.2}$ & $\phn\phn894.1_{{-}21.5}^{{+}37.5}$ \\
278.6${-}$06.7 & $\phn2063.3_{{-}36.9}^{{+}39.5}$ & $\phn\phn\phn80.3_{{-}24.2}^{{+}26.9}$ & $\phn\phn336.7_{{-}32.5}^{{+}31.8}$ & $\phn\phn423.5_{{-}64.1}^{{+}82.9}$ & ${<}  167.5$ & $\phn\phn433.0_{{-}38.6}^{{+}46.3}$ & \nodata \\
283.8${+}$02.2 & $\phn1568.7_{{-}\phn9.8}^{{+}21.4}$ & $\phn\phn581.3_{{-}12.8}^{{+}15.3}$ & \nodata & \nodata & \nodata & $14919.7_{{-}\phn0.8}^{{+}154.9}$ & $\phn2690.9_{{-}\phn2.4}^{{+}38.0}$ \\
285.4${-}$05.3 & $\phn3817.1_{{-}13.1}^{{+}25.0}$ & $\phn\phn411.9_{{-}17.5}^{{+}23.7}$ & $\phn\phn308.6_{{-}27.2}^{{+}28.5}$ & $\phn\phn325.8_{{-}85.8}^{{+}85.1}$ & \nodata & $\phn5558.5_{{-}27.3}^{{+}33.6}$ & $\phn\phn593.4_{{-}19.5}^{{+}46.2}$ \\
285.6${-}$02.7 & $\phn2316.2_{{-}20.5}^{{+}22.3}$ & \nodata & $\phn\phn120.8_{{-}10.8}^{{+}\phn9.4}$ & $\phn\phn808.8_{{-}24.1}^{{+}33.0}$ & $\phn\phn128.8_{{-}24.2}^{{+}41.1}$ & \nodata & \nodata \\
285.7${-}$14.9 & $\phn1226.6_{{-}20.7}^{{+}33.5}$ & $\phn\phn175.2_{{-}25.6}^{{+}22.9}$ & ${<}  121.2$ & \nodata & \nodata & $\phn\phn696.6_{{-}57.0}^{{+}52.1}$ & ${<}   85.6$ \\
291.6${-}$04.8 & $\phn1133.6_{{-}11.0}^{{+}32.8}$ & $\phn\phn212.1_{{-}10.2}^{{+}36.5}$ & $\phn\phn152.4_{{-}\phn1.1}^{{+}24.2}$ & $\phn4125.6_{{-}43.9}^{{+}135.3}$ & $\phn2129.4_{{-}22.0}^{{+}105.9}$ & $\phn4046.0_{{-}18.4}^{{+}160.5}$ & $\phn3068.7_{{-}15.6}^{{+}150.2}$ \\
292.8${+}$01.1 & $\phn2695.2_{{-}73.4}^{{+}73.5}$ & $\phn\phn219.4_{{-}25.2}^{{+}37.6}$ & $\phn\phn278.3_{{-}57.0}^{{+}54.5}$ & $\phn1736.0_{{-}129.6}^{{+}168.7}$ & $\phn1005.7_{{-}137.5}^{{+}164.8}$ & $\phn\phn232.8_{{-}60.8}^{{+}64.1}$ & \nodata \\
294.9${-}$04.3 & $\phn1092.7_{{-}22.4}^{{+}26.7}$ & \nodata & $\phn\phn179.9_{{-}33.8}^{{+}40.2}$ & ${<}  349.8$ & \nodata & \nodata & \nodata \\
296.3${-}$03.0 & $\phn1920.9_{{-}45.3}^{{+}53.9}$ & $\phn\phn148.3_{{-}15.0}^{{+}22.3}$ & $\phn\phn255.0_{{-}31.2}^{{+}34.9}$ & $\phn\phn958.8_{{-}68.0}^{{+}83.4}$ & $\phn\phn489.3_{{-}49.1}^{{+}88.5}$ & $\phn1575.0_{{-}80.5}^{{+}77.2}$ & $\phn\phn719.9_{{-}54.3}^{{+}48.8}$ \\
304.5${-}$04.8 & $\phn2554.8_{{-}14.7}^{{+}25.8}$ & $\phn\phn154.7_{{-}18.9}^{{+}20.4}$ & $\phn\phn409.0_{{-}23.1}^{{+}33.5}$ & $\phn1051.8_{{-}213.0}^{{+}79.3}$ & $\phn\phn415.2_{{-}76.7}^{{+}88.9}$ & $\phn4319.5_{{-}29.1}^{{+}72.4}$ & $\phn\phn840.2_{{-}28.2}^{{+}61.8}$ \\
305.1${+}$01.4 & $\phn\phn\phn60.5_{{-}\phn8.0}^{{+}\phn6.8}$ & \nodata & \nodata & $\phn\phn\phn76.2_{{-}16.8}^{{+}26.5}$ & $\phn\phn103.7_{{-}13.4}^{{+}31.1}$ & \nodata & \nodata \\
307.2${-}$09.0 & $\phn2398.0_{{-}20.1}^{{+}50.2}$ & \nodata & $\phn\phn518.9_{{-}16.9}^{{+}39.9}$ & $\phn\phn634.2_{{-}159.4}^{{+}161.6}$ & ${<}  212.6$ & \nodata & \nodata \\
307.5${-}$04.9 & $\phn2930.3_{{-}30.3}^{{+}35.5}$ & \nodata & $\phn\phn325.5_{{-}25.7}^{{+}48.3}$ & $\phn3231.1_{{-}123.0}^{{+}287.3}$ & $\phn\phn450.4_{{-}141.3}^{{+}196.3}$ & \nodata & \nodata \\
312.6${-}$01.8 & $\phn3717.1_{{-}56.8}^{{+}72.7}$ & \nodata & $\phn\phn334.5_{{-}57.3}^{{+}66.0}$ & $\phn\phn310.7_{{-}86.5}^{{+}142.1}$ & ${<}  130.4$ & \nodata & \nodata \\
315.1${-}$13.0 & $\phn2011.0_{{-}153.6}^{{+}176.7}$ & \nodata & $\phn\phn148.8_{{-}38.8}^{{+}50.3}$ & $\phn1027.3_{{-}202.8}^{{+}215.8}$ & $\phn\phn318.9_{{-}96.5}^{{+}107.4}$ & \nodata & \nodata \\
320.1${-}$09.6 & $\phn1647.3_{{-}22.0}^{{+}201.6}$ & \nodata & ${<}   69.8$ & $\phn2336.5_{{-}50.0}^{{+}212.7}$ & $\phn\phn566.8_{{-}43.7}^{{+}133.6}$ & \nodata & \nodata \\
320.9${+}$02.0 & $\phn2469.2_{{-}111.9}^{{+}191.4}$ & \nodata & $\phn\phn530.5_{{-}38.2}^{{+}136.2}$ & $\phn1650.4_{{-}94.8}^{{+}61.4}$ & $\phn\phn724.1_{{-}47.5}^{{+}87.9}$ & \nodata & \nodata \\
322.5${-}$05.2 & $\phn3553.0_{{-}58.9}^{{+}70.9}$ & $\phn2671.8_{{-}88.3}^{{+}124.1}$ & ${<}  540.1$ & \nodata & \nodata & $51038.3_{{-}417.1}^{{+}559.8}$ & \nodata \\
323.9${+}$02.4 & $\phn2326.6_{{-}29.6}^{{+}46.8}$ & \nodata & $\phn\phn298.4_{{-}27.3}^{{+}40.4}$ & $\phn3466.2_{{-}134.4}^{{+}63.6}$ & $\phn1567.2_{{-}56.4}^{{+}63.0}$ & \nodata & \nodata \\
324.8${-}$01.1 & $\phn3142.8_{{-}63.1}^{{+}88.2}$ & \nodata & $\phn\phn697.5_{{-}43.7}^{{+}52.8}$ & $\phn1571.2_{{-}176.5}^{{+}206.2}$ & $\phn\phn488.3_{{-}107.4}^{{+}158.1}$ & \nodata & \nodata \\
325.8${-}$12.8 & $\phn1794.9_{{-}31.8}^{{+}72.0}$ & \nodata & $\phn\phn228.1_{{-}15.1}^{{+}37.1}$ & \nodata & \nodata & \nodata & \nodata \\
326.0${-}$06.5 & $\phn1304.5_{{-}38.8}^{{+}44.7}$ & \nodata & ${<}   51.7$ & $\phn\phn255.9_{{-}46.4}^{{+}205.4}$ & ${<}  154.9$ & \nodata & \nodata \\
327.1${-}$01.8 & $\phn2261.4_{{-}87.8}^{{+}97.3}$ & \nodata & $\phn\phn264.9_{{-}21.1}^{{+}68.2}$ & $\phn1435.7_{{-}187.8}^{{+}121.8}$ & $\phn\phn806.4_{{-}29.4}^{{+}40.5}$ & \nodata & \nodata \\
327.8${-}$01.6 & $\phn1645.0_{{-}27.2}^{{+}54.4}$ & $\phn\phn203.9_{{-}18.3}^{{+}35.8}$ & $\phn\phn253.8_{{-}31.9}^{{+}41.9}$ & $\phn2054.3_{{-}319.2}^{{+}194.1}$ & $\phn1326.9_{{-}108.0}^{{+}129.8}$ & $\phn3042.7_{{-}139.6}^{{+}73.8}$ & $\phn3621.1_{{-}273.8}^{{+}36.6}$ \\
331.1${-}$05.7 & $\phn\phn812.3_{{-}22.4}^{{+}156.7}$ & \nodata & $\phn\phn222.7_{{-}16.4}^{{+}60.4}$ & \nodata & \nodata & \nodata & \nodata \\
331.3${+}$16.8 & $\phn1783.6_{{-}45.2}^{{+}51.3}$ & $\phn\phn419.0_{{-}33.0}^{{+}51.2}$ & ${<}  103.9$ & $\phn\phn489.5_{{-}86.0}^{{+}140.0}$ & $\phn\phn159.6_{{-}40.3}^{{+}103.3}$ & $\phn3389.7_{{-}108.1}^{{+}54.5}$ & \nodata \\
336.3${-}$05.6 & $\phn\phn890.3_{{-}35.2}^{{+}54.9}$ & $\phn\phn193.2_{{-}20.0}^{{+}28.6}$ & ${<}  128.2$ & $\phn2160.7_{{-}156.5}^{{+}418.2}$ & $\phn1279.7_{{-}132.8}^{{+}197.5}$ & $\phn1925.4_{{-}102.0}^{{+}102.4}$ & $\phn2132.4_{{-}124.6}^{{+}91.4}$ \\
342.1${+}$27.5 & $\phn1960.9_{{-}60.0}^{{+}51.6}$ & $\phn\phn720.0_{{-}41.6}^{{+}54.5}$ & ${<}  145.2$ & \nodata & \nodata & $15844.8_{{-}282.7}^{{+}292.1}$ & $\phn2427.5_{{-}129.0}^{{+}145.3}$ \\
349.8${+}$04.4 & $\phn2499.7_{{-}64.9}^{{+}61.3}$ & \nodata & $\phn\phn415.9_{{-}45.4}^{{+}49.1}$ & $\phn1403.2_{{-}79.1}^{{+}111.5}$ & $\phn\phn721.8_{{-}99.5}^{{+}161.8}$ & \nodata & \nodata \\
350.9${+}$04.4 & $\phn\phn744.9_{{-}15.4}^{{+}76.0}$ & \nodata & $\phn\phn\phn56.6_{{-}12.0}^{{+}32.5}$ & \nodata & \nodata & \nodata & \nodata \\
356.1${+}$02.7 & $\phn\phn831.7_{{-}42.4}^{{+}\phn6.7}$ & \nodata & $\phn\phn243.8_{{-}23.0}^{{+}\phn0.1}$ & $\phn1854.2_{{-}35.3}^{{+}59.7}$ & $\phn\phn893.9_{{-}\phn7.2}^{{+}141.8}$ & \nodata & \nodata \\
357.6${+}$02.6 & $\phn1567.0_{{-}91.4}^{{+}97.8}$ & \nodata & $\phn\phn397.6_{{-}92.9}^{{+}89.1}$ & $\phn1177.0_{{-}149.0}^{{+}176.0}$ & $\phn\phn383.6_{{-}37.1}^{{+}129.1}$ & \nodata & \nodata \\
\enddata
\end{deluxetable}

\clearpage
\setcounter{figure}{6}
\begin{figure*}[p]
  \centering
  \plotone{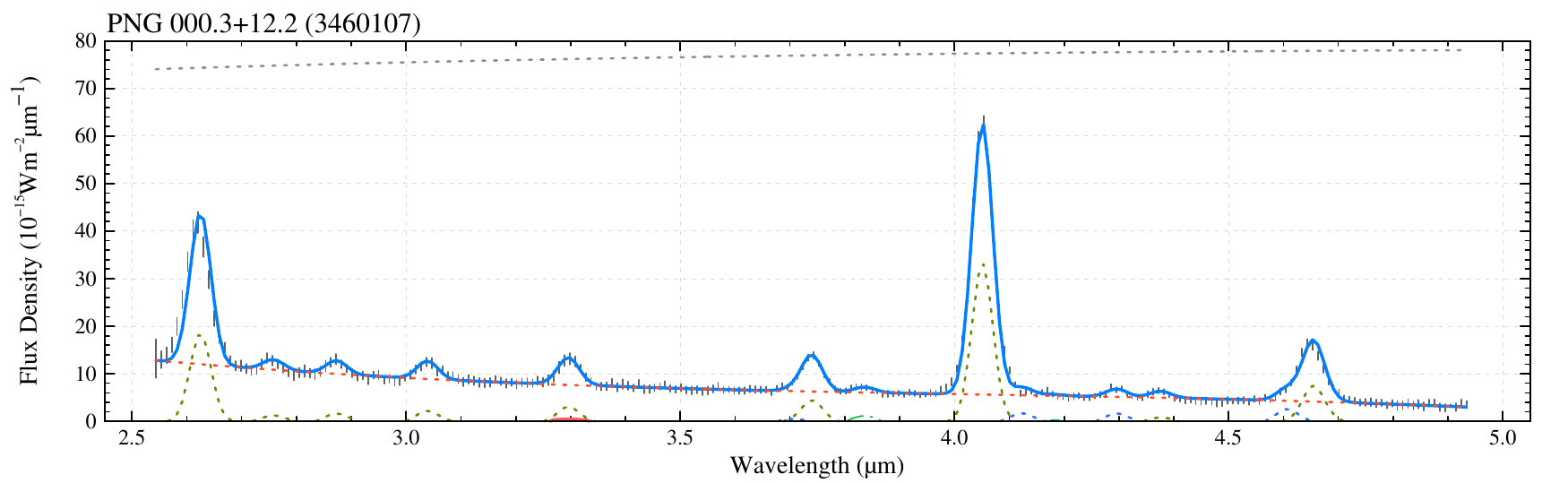}
  \plotone{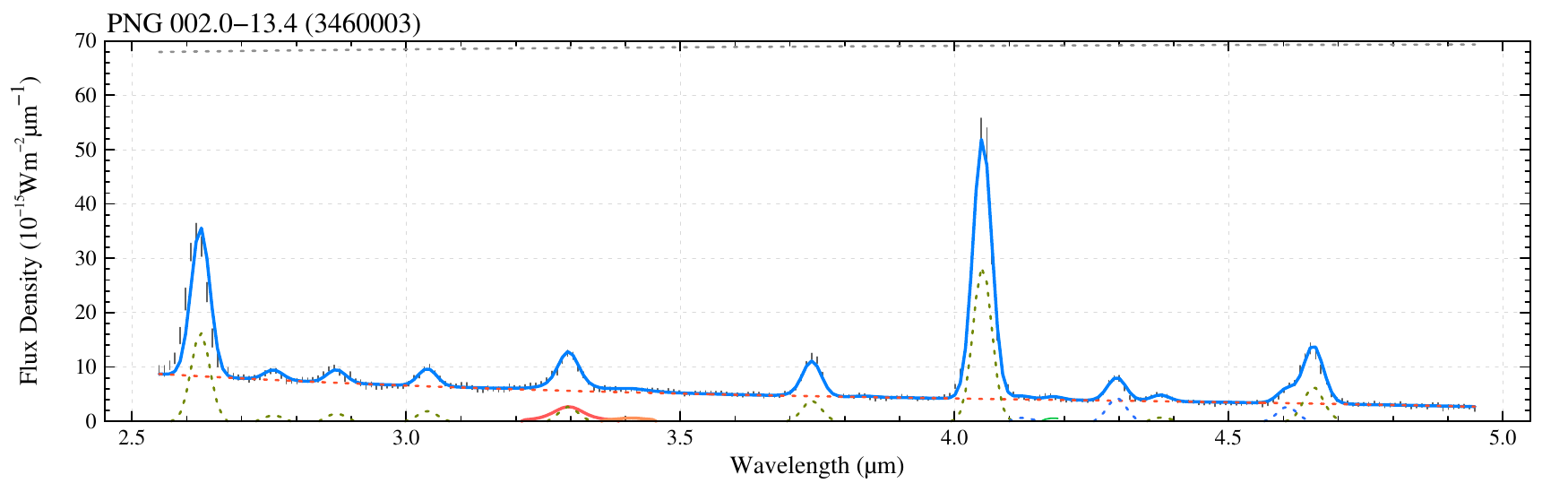}
  \plotone{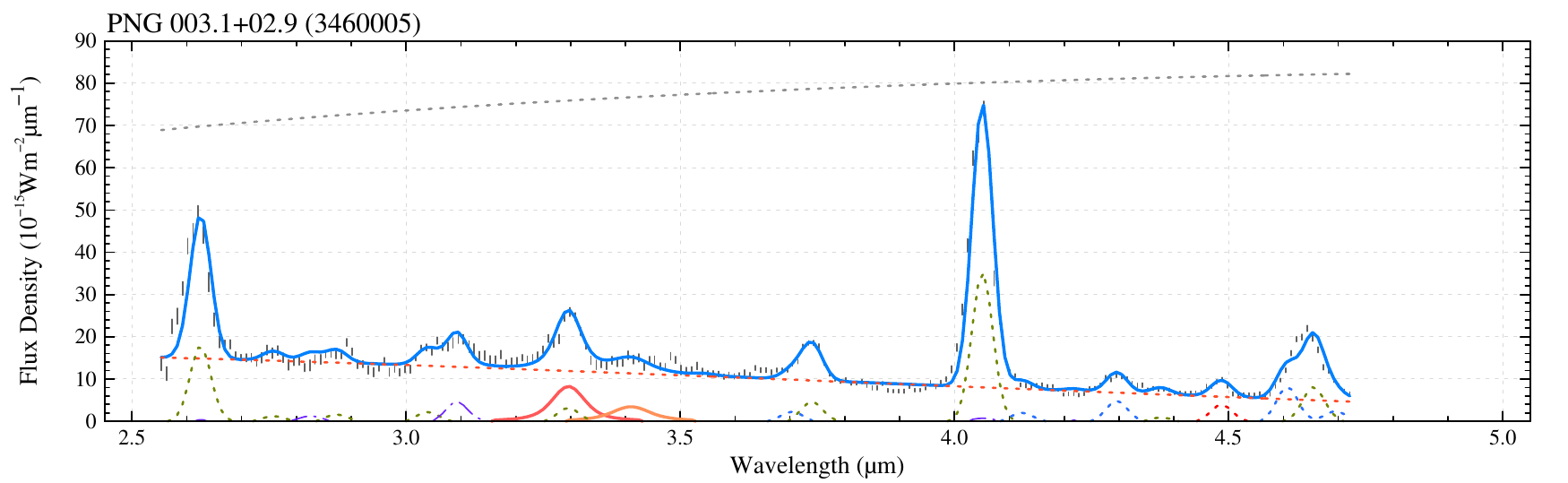}
  \includegraphics[width=1.0\linewidth]{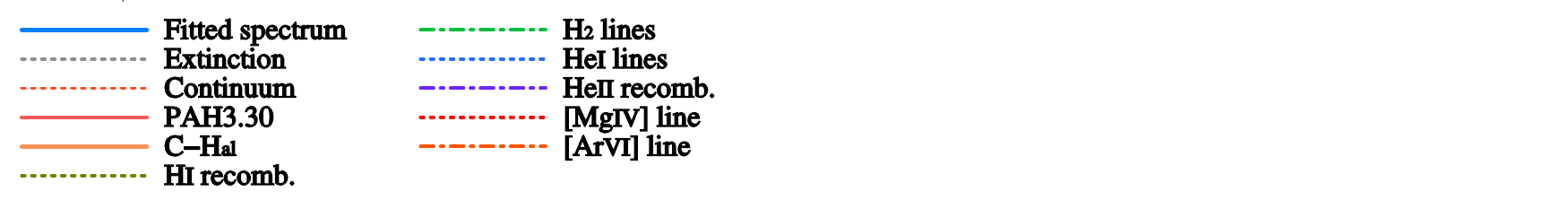}
  \caption{The \textit{AKARI}/IRC 2.5--5.0$\,\mu$m spectra. Explanations of the lines are shown in the bottom of the figures.}
  \label{fig:allspectrum}
\end{figure*}

\clearpage\addtocounter{figure}{-1}
\begin{figure*}[p]
  \centering
  \plotone{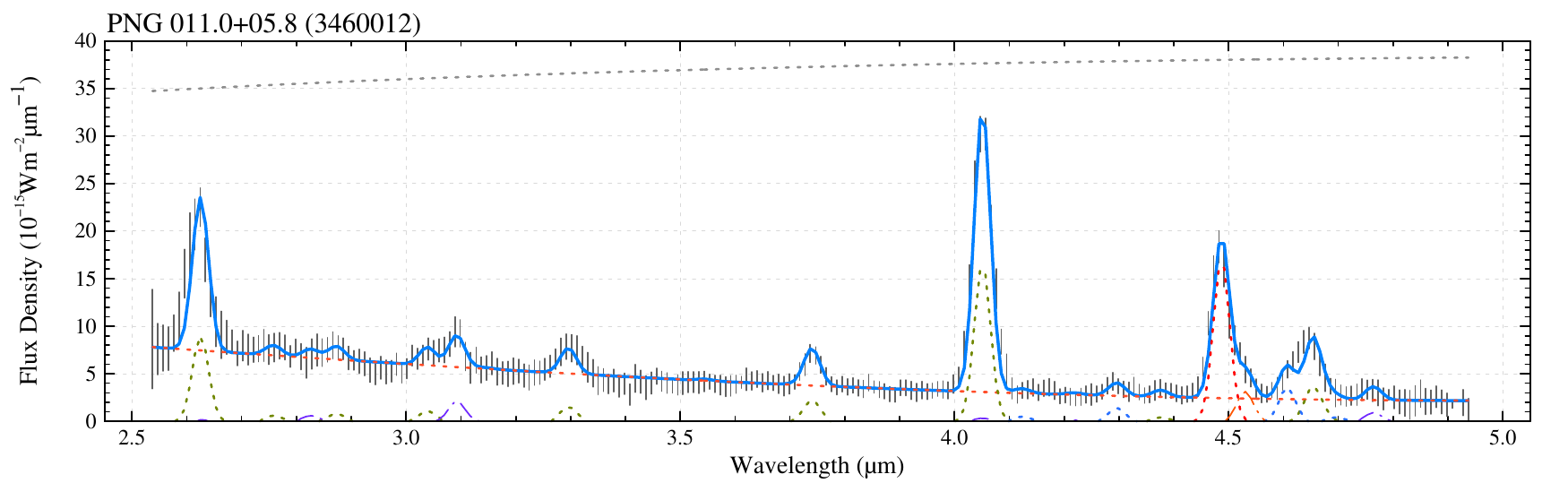}
  \plotone{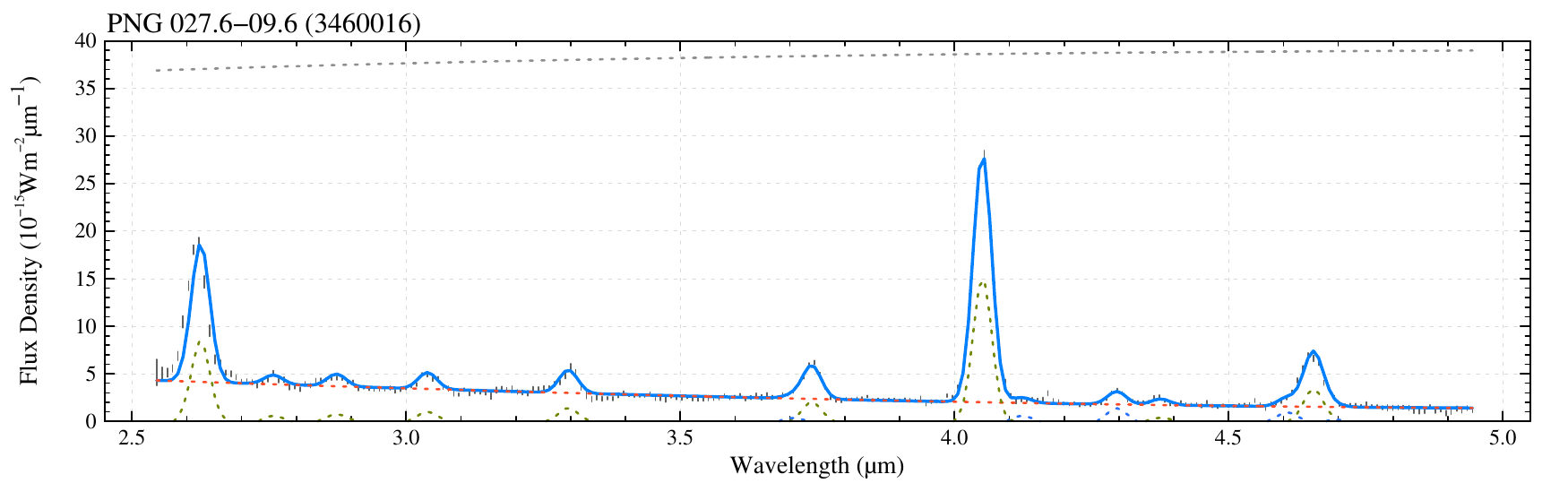}
  \plotone{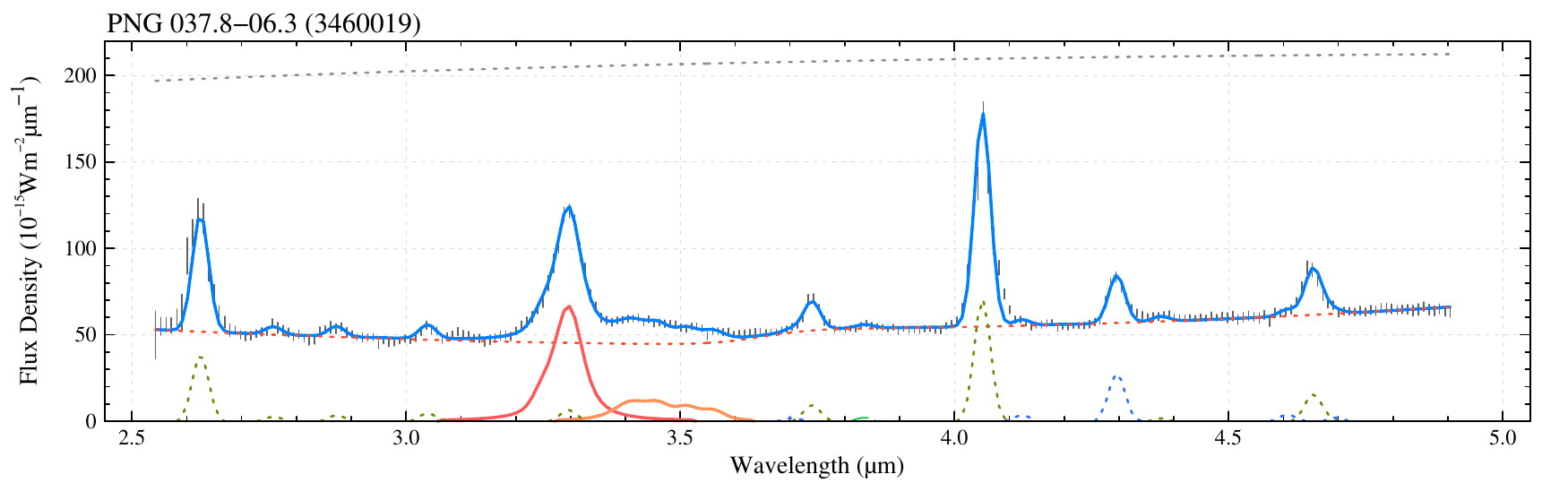}
  \caption{\textit{Cont.}---}
  \label{fig:allspectrum2}
\end{figure*}
\clearpage\addtocounter{figure}{-1}
\begin{figure*}[p]
  \centering
  \plotone{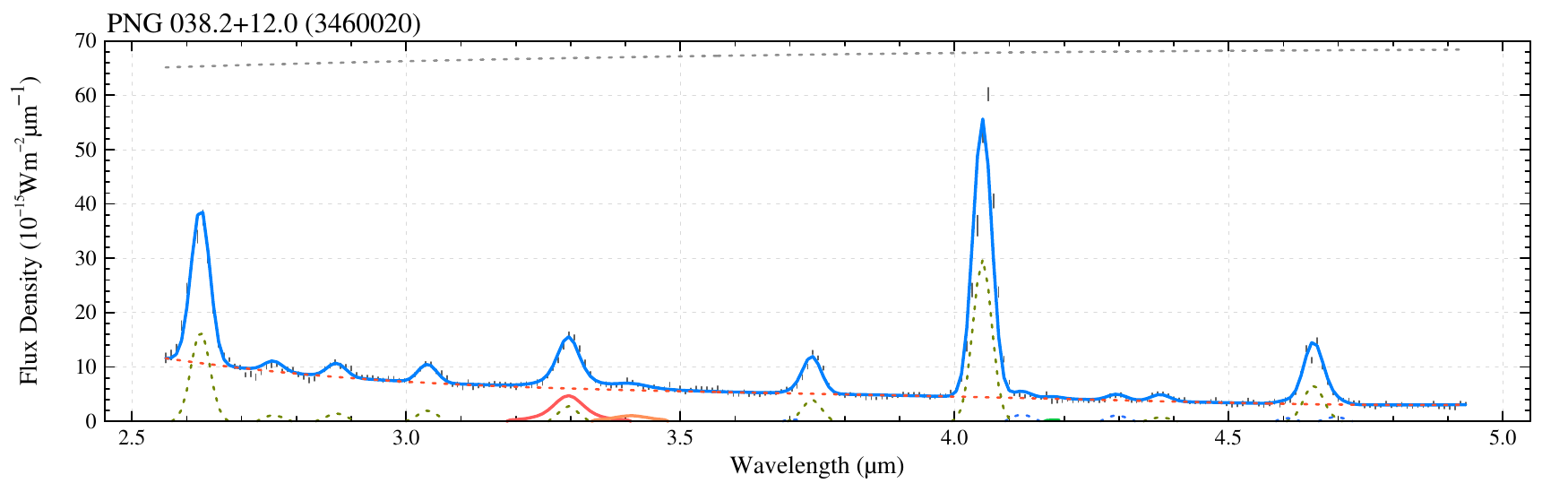}
  \plotone{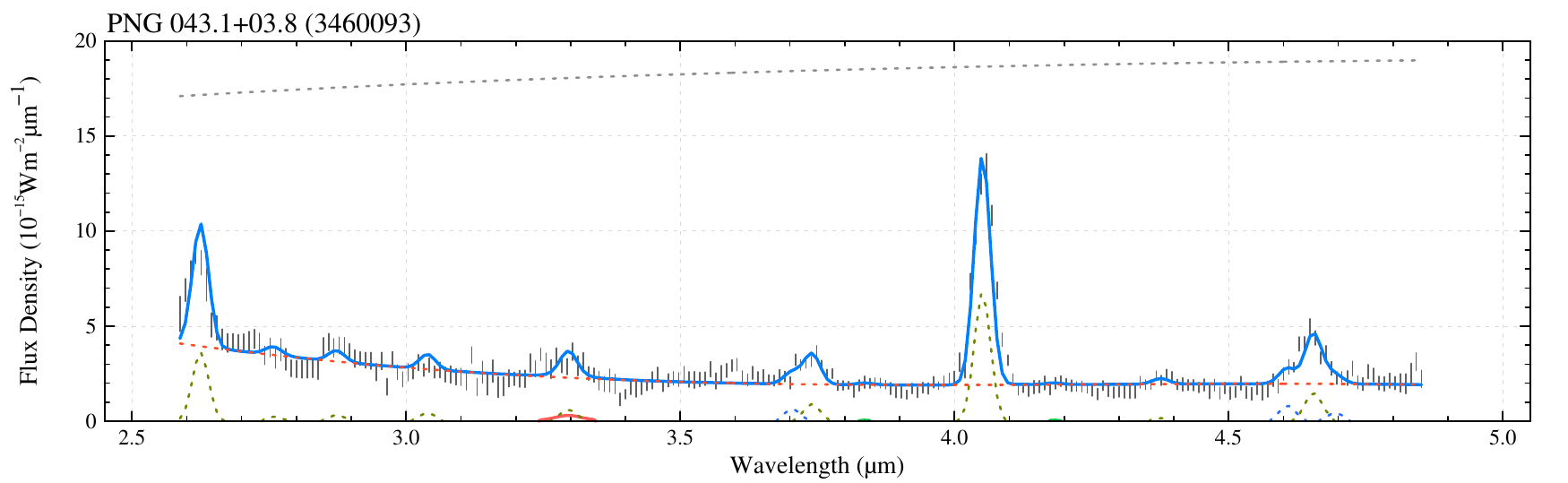}
  \plotone{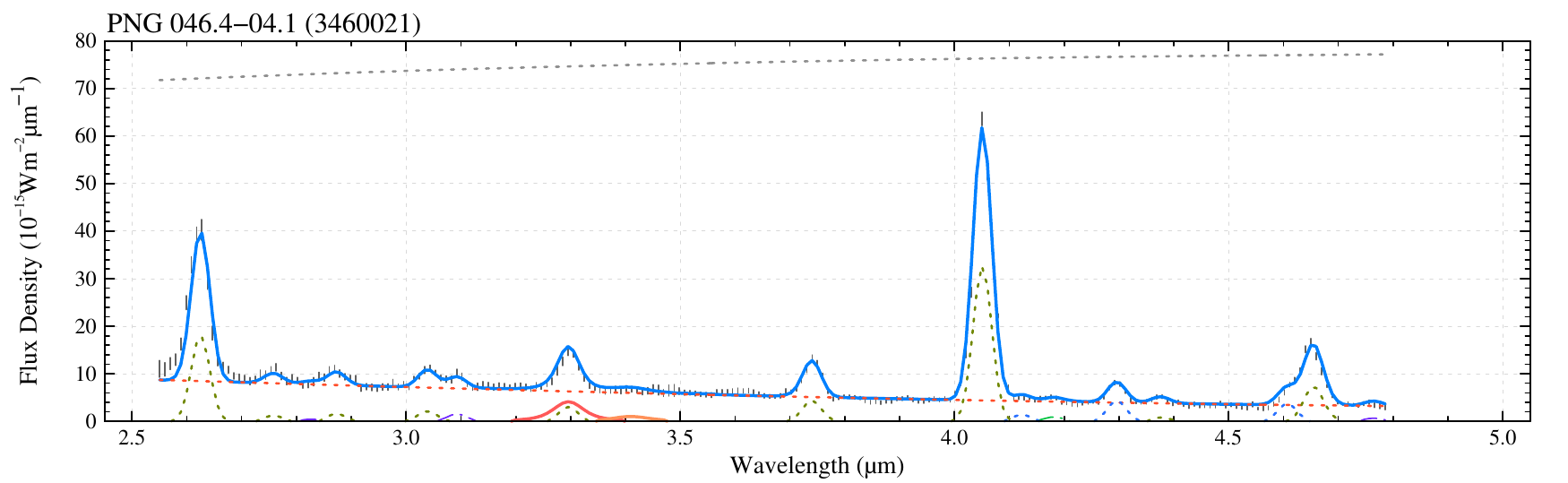}
  \caption{\textit{Cont.}---}
  \label{fig:allspectrum3}
\end{figure*}
\clearpage\addtocounter{figure}{-1}
\begin{figure*}[p]
  \centering
  \plotone{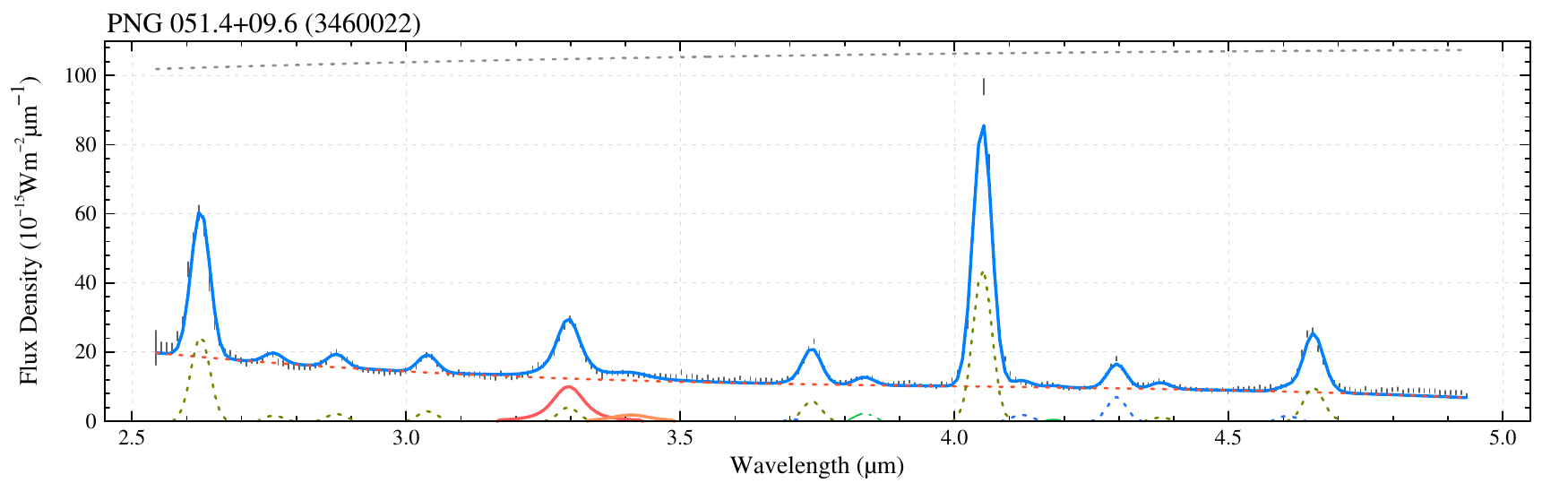}
  \plotone{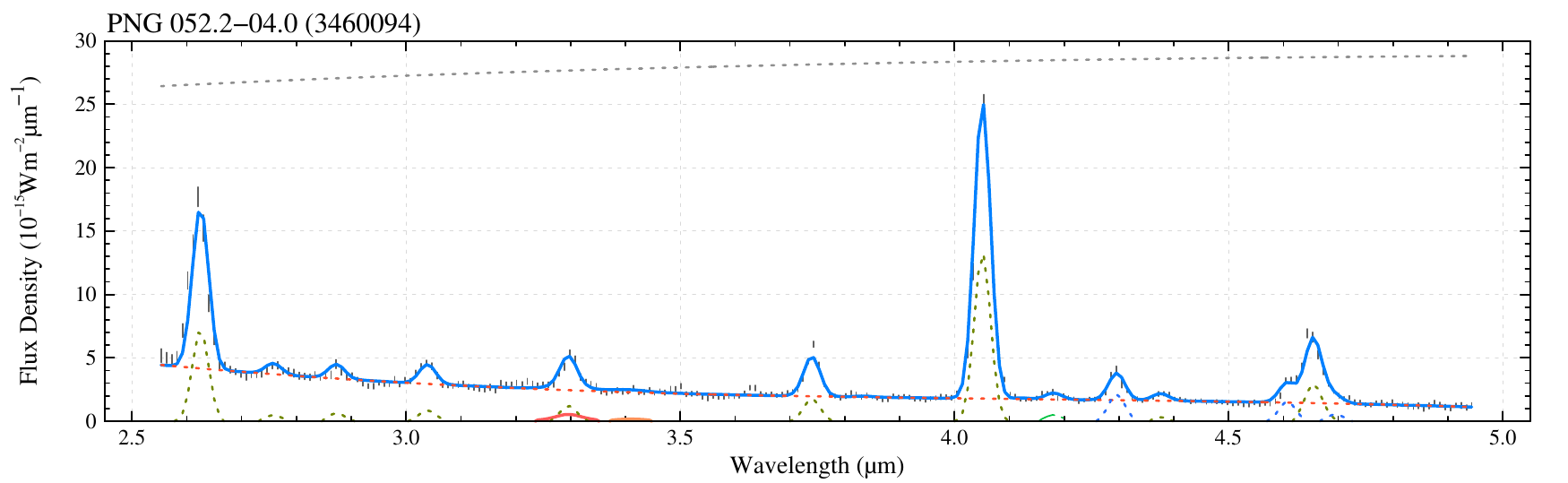}
  \plotone{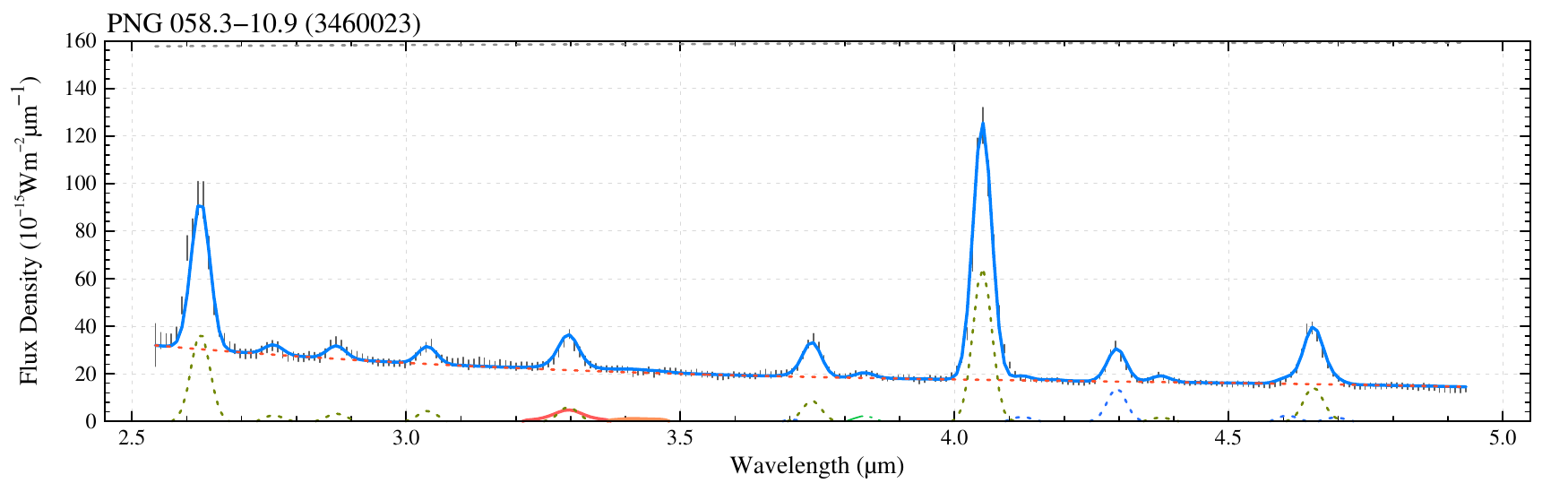}
  \caption{\textit{Cont.}---}
  \label{fig:allspectrum4}
\end{figure*}
\clearpage\addtocounter{figure}{-1}
\begin{figure*}[p]
  \centering
  \plotone{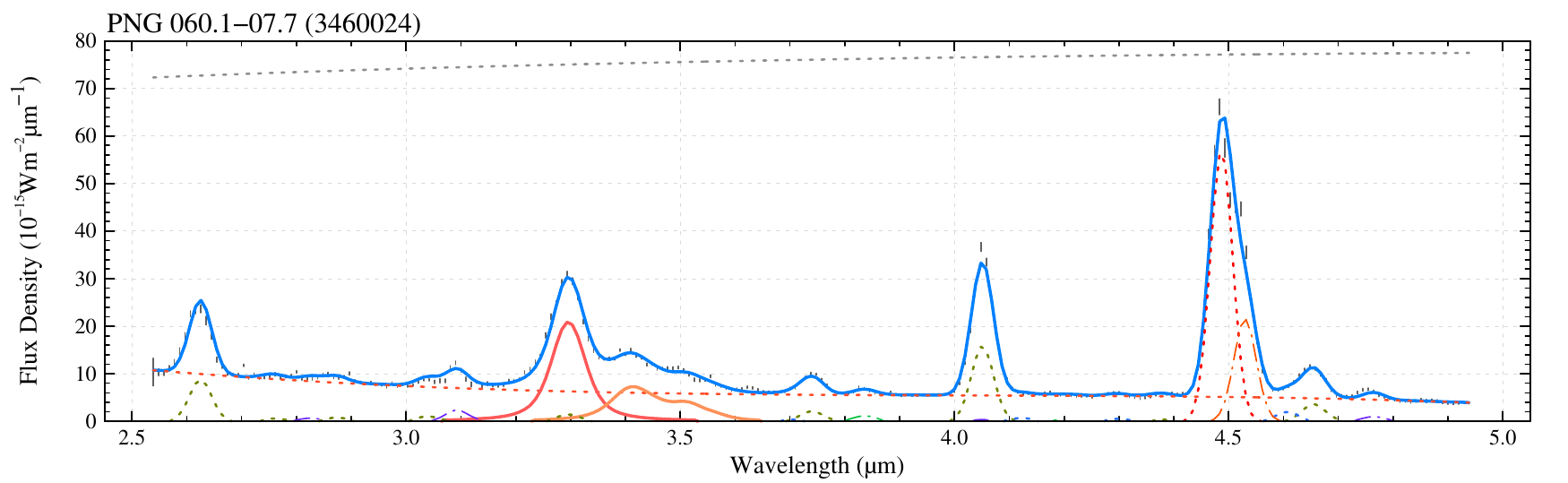}
  \plotone{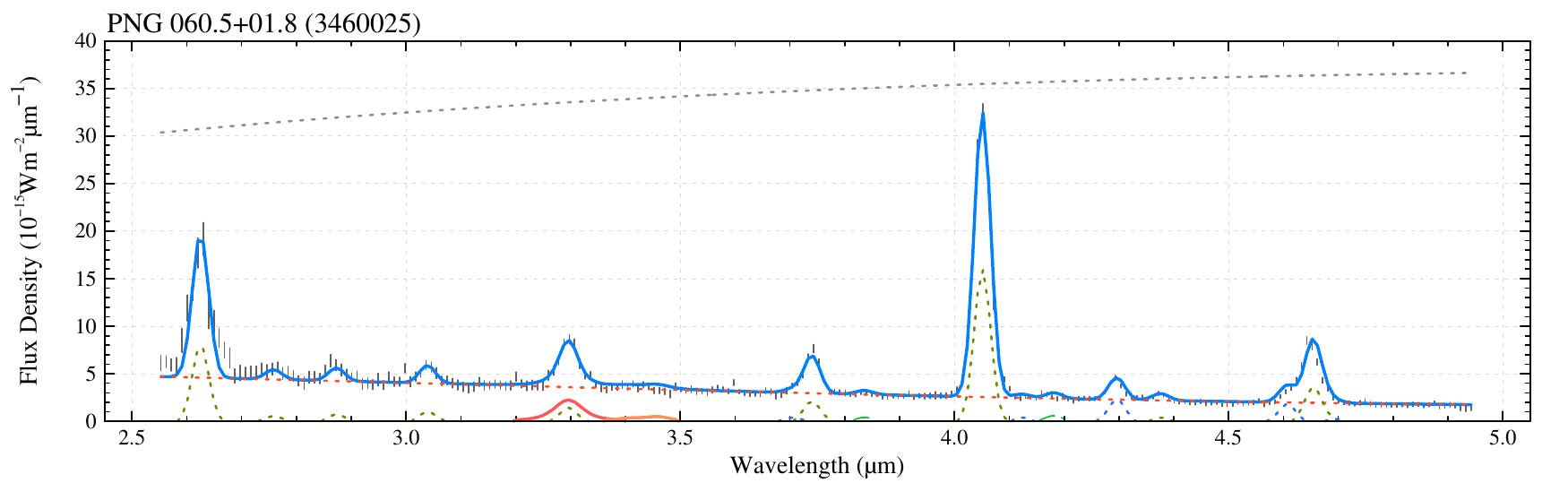}
  \plotone{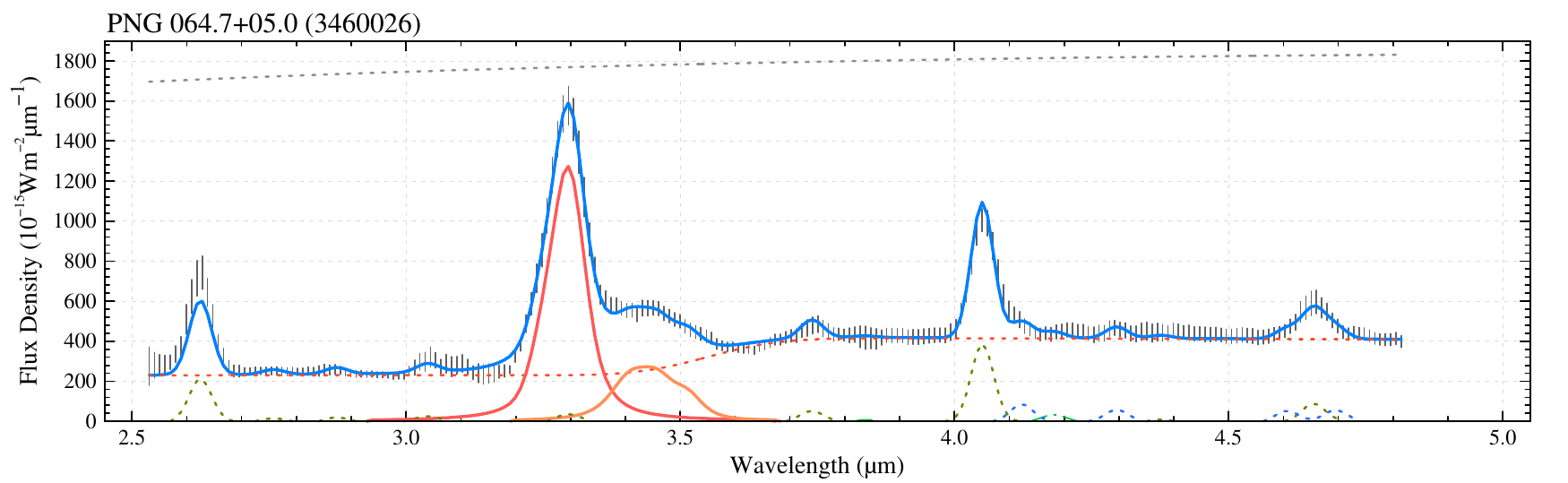}
  \caption{\textit{Cont.}---}
  \label{fig:allspectrum5}
\end{figure*}

\clearpage\addtocounter{figure}{-1}
\begin{figure*}[p]
  \centering
  \plotone{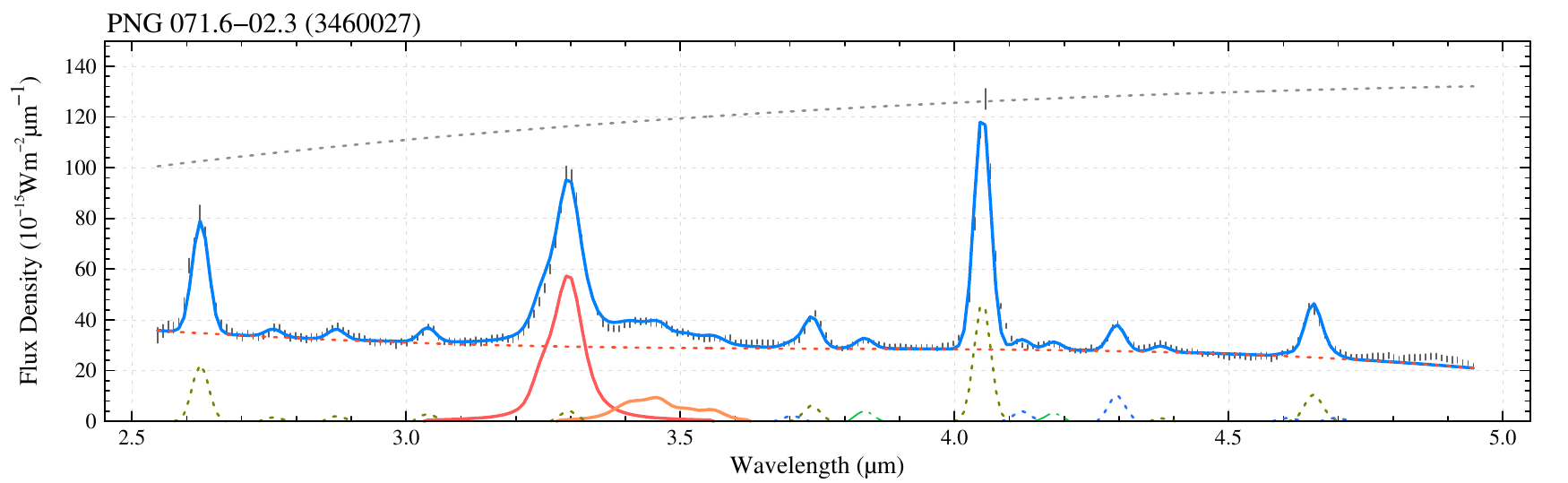}
  \plotone{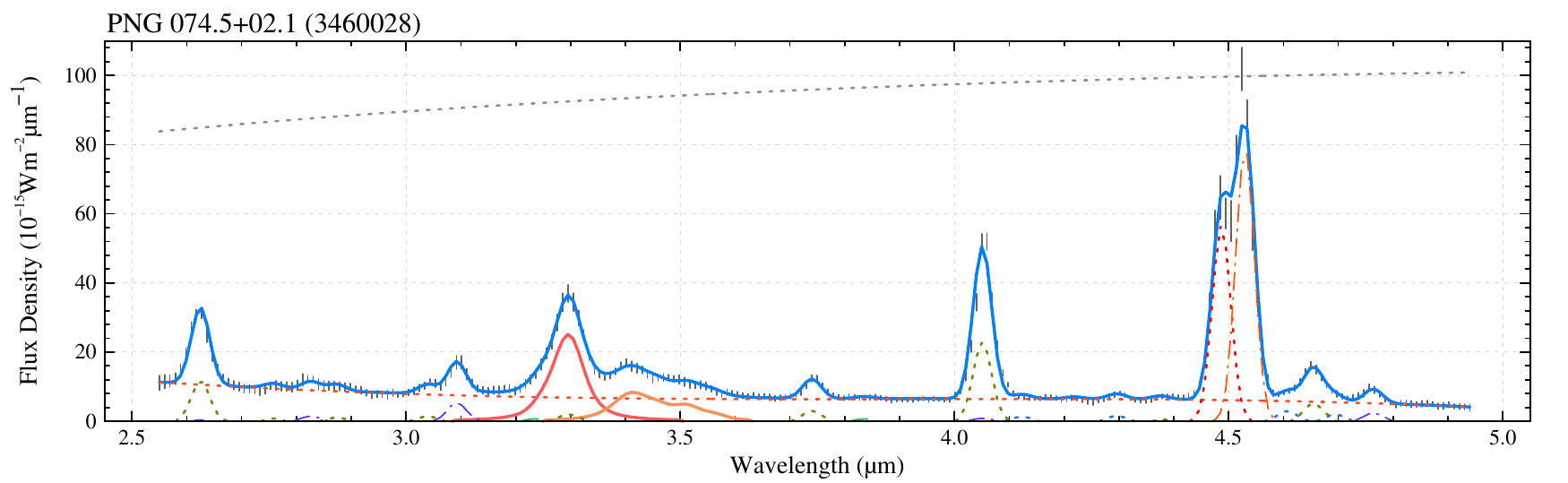}
  \plotone{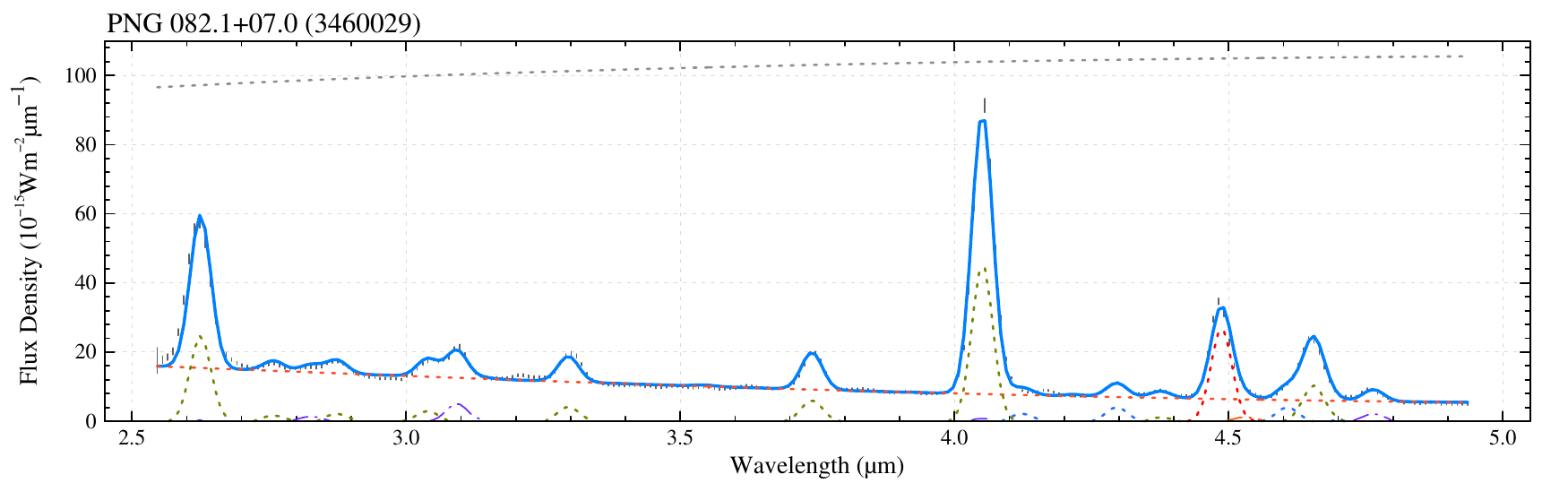}
  \caption{\textit{Cont.}---}
  \label{fig:allspectrum6}
\end{figure*}
\clearpage\addtocounter{figure}{-1}
\begin{figure*}[p]
  \centering
  \plotone{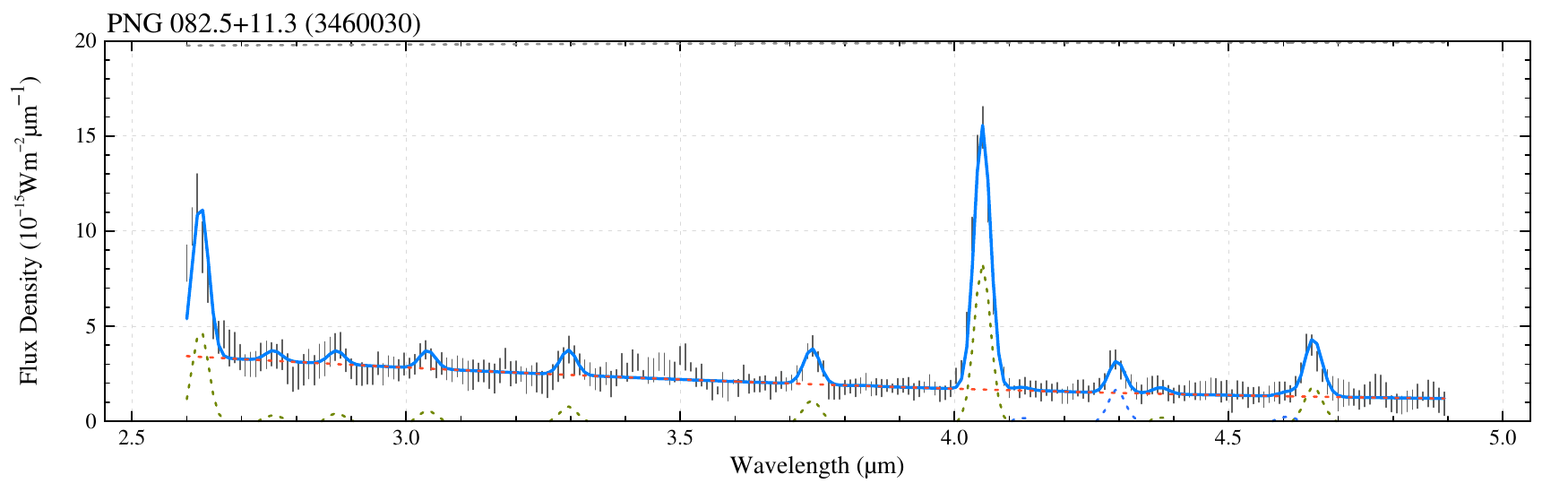}
  \plotone{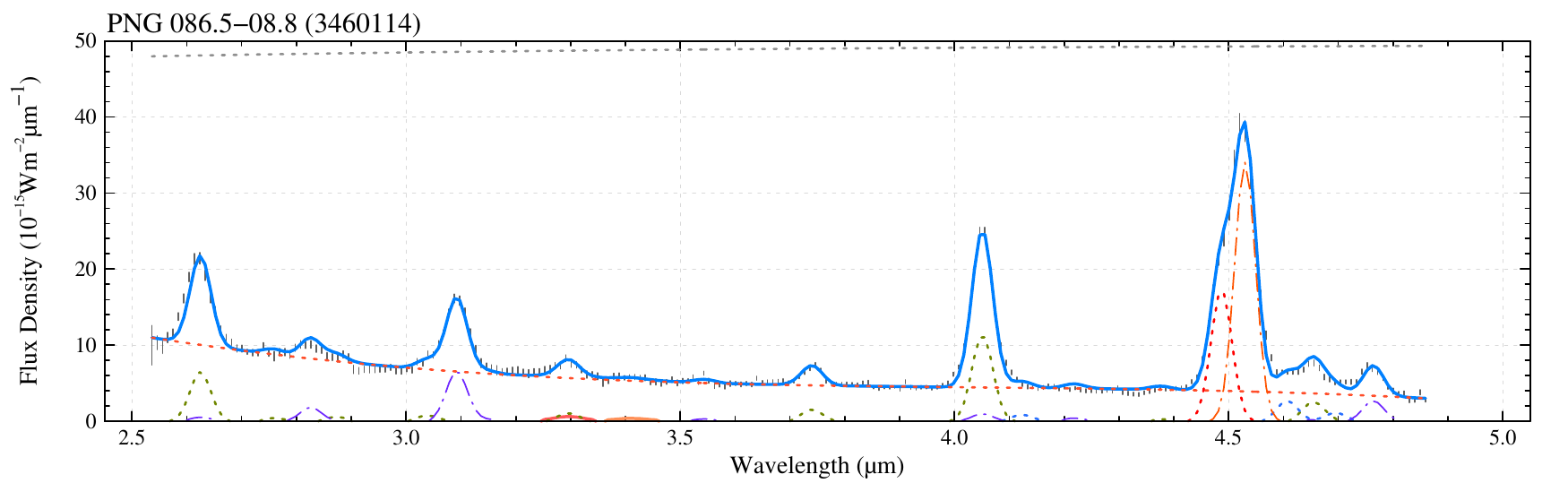}
  \plotone{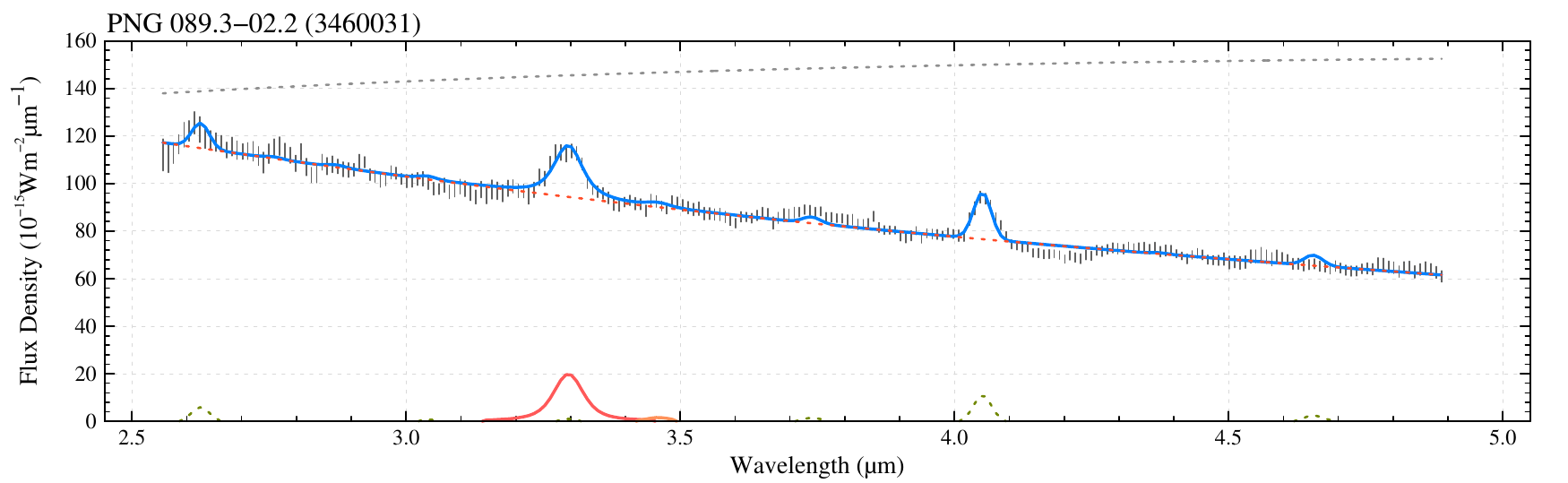}
  \caption{\textit{Cont.}---}
  \label{fig:allspectrum7}
\end{figure*}
\clearpage\addtocounter{figure}{-1}
\begin{figure*}[p]
  \centering
  \plotone{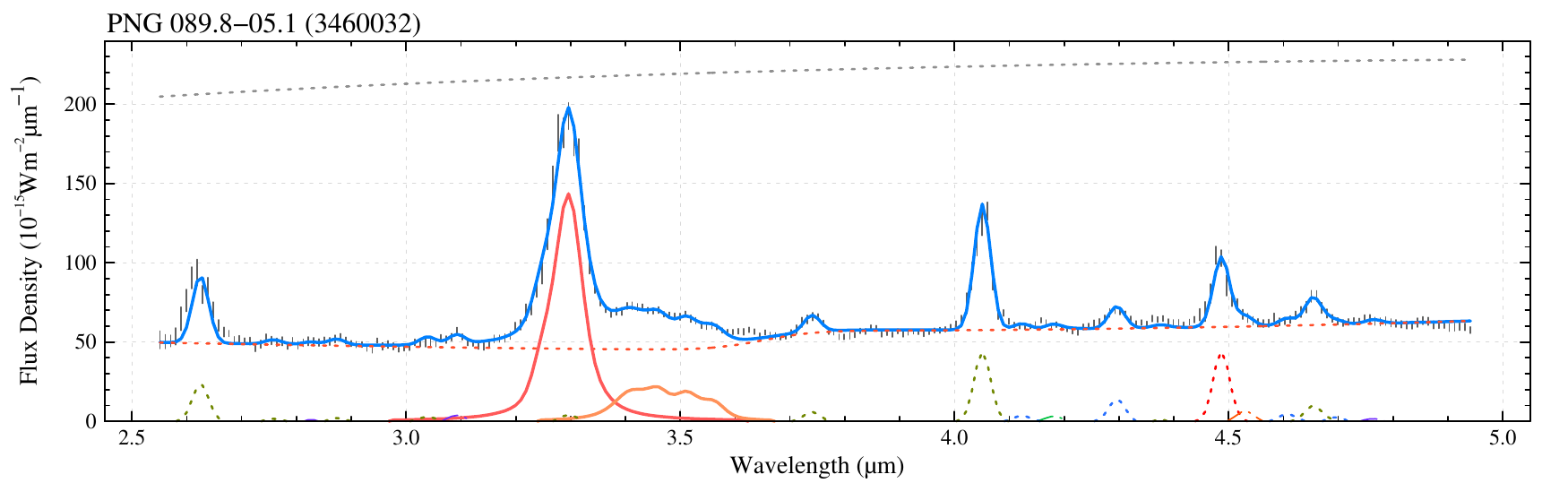}
  \plotone{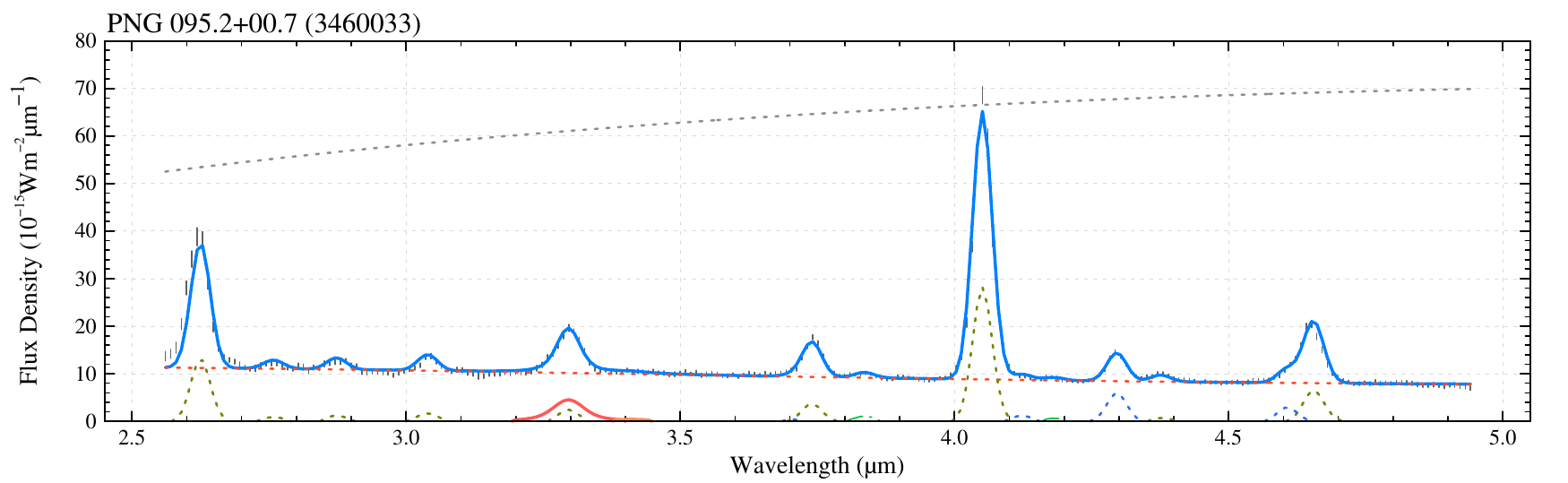}
  \plotone{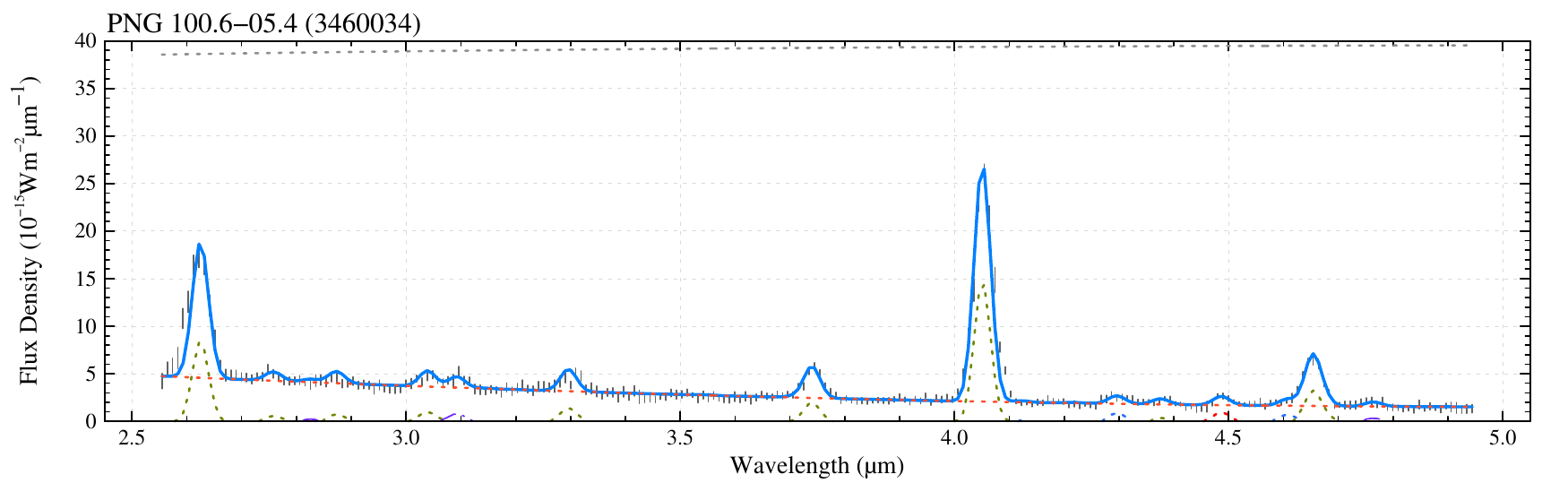}
  \caption{\textit{Cont.}---}
  \label{fig:allspectrum8}
\end{figure*}
\clearpage\addtocounter{figure}{-1}
\begin{figure*}[p]
  \centering
  \plotone{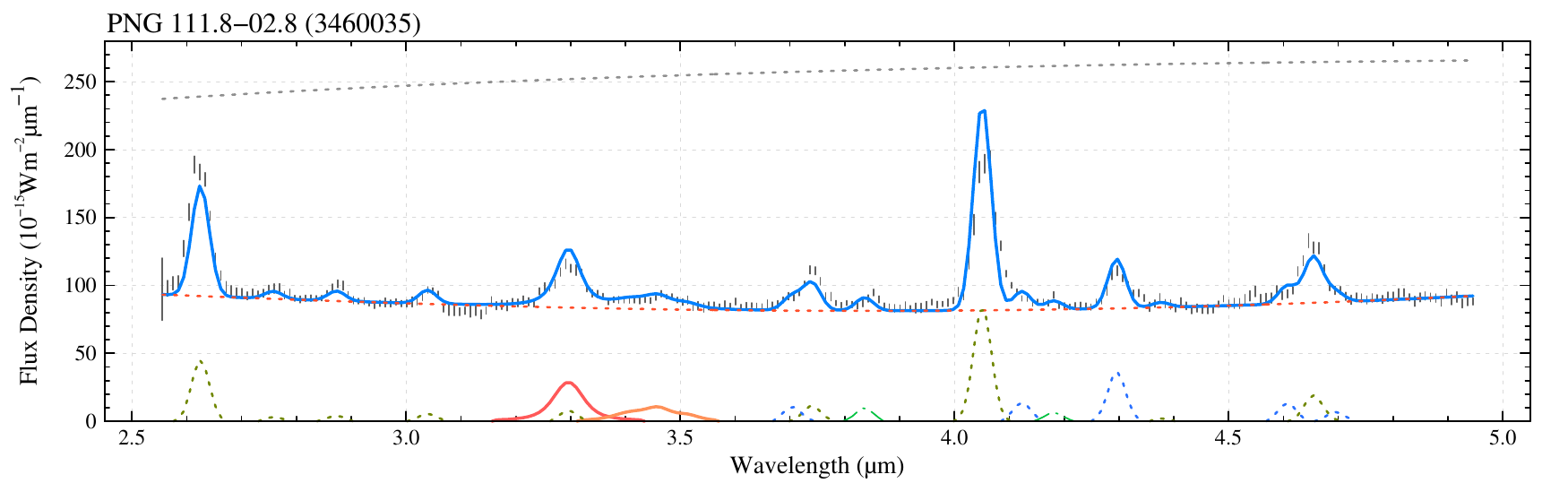}
  \plotone{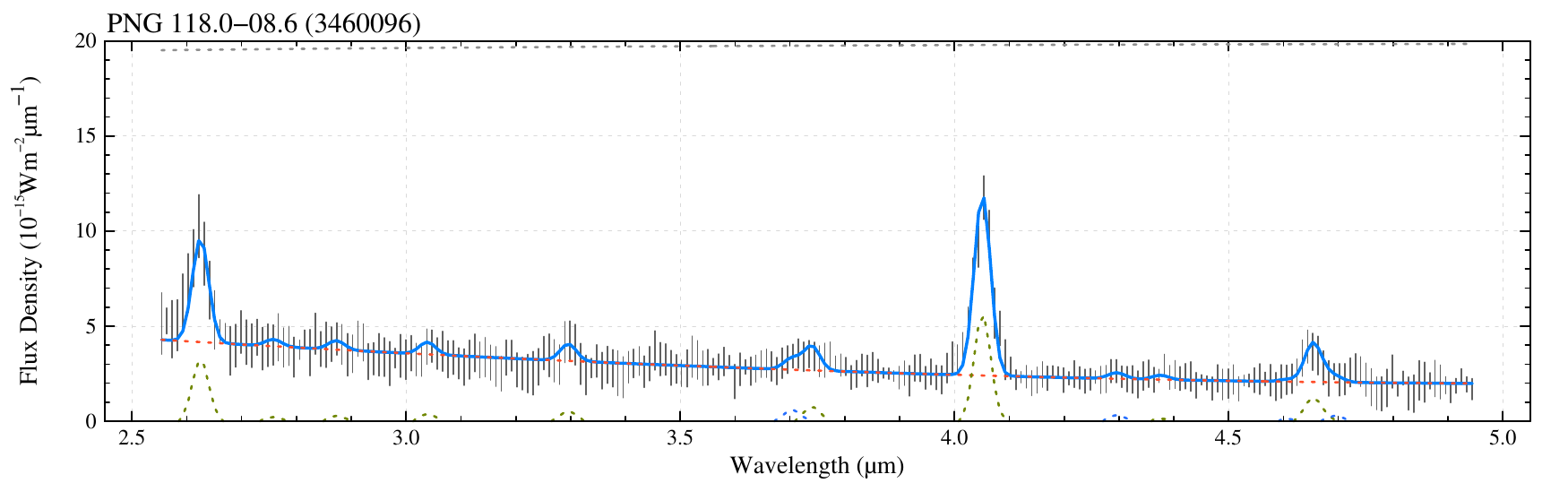}
  \plotone{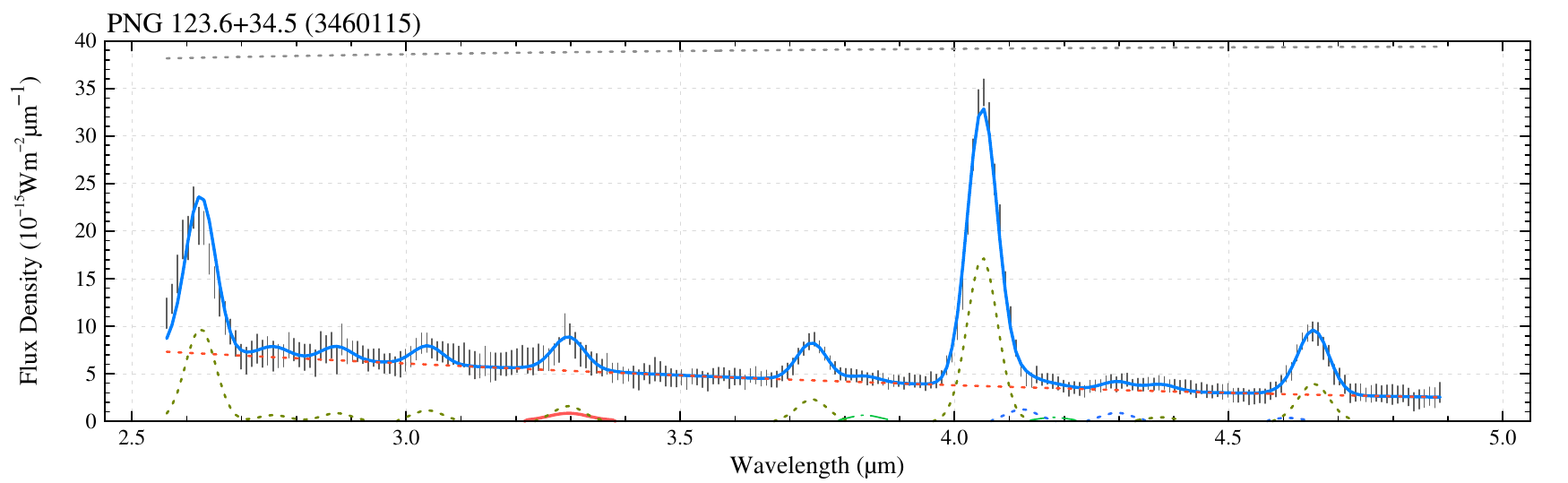}
  \caption{\textit{Cont.}---}
  \label{fig:allspectrum9}
\end{figure*}
\clearpage\addtocounter{figure}{-1}
\begin{figure*}[p]
  \centering
  \plotone{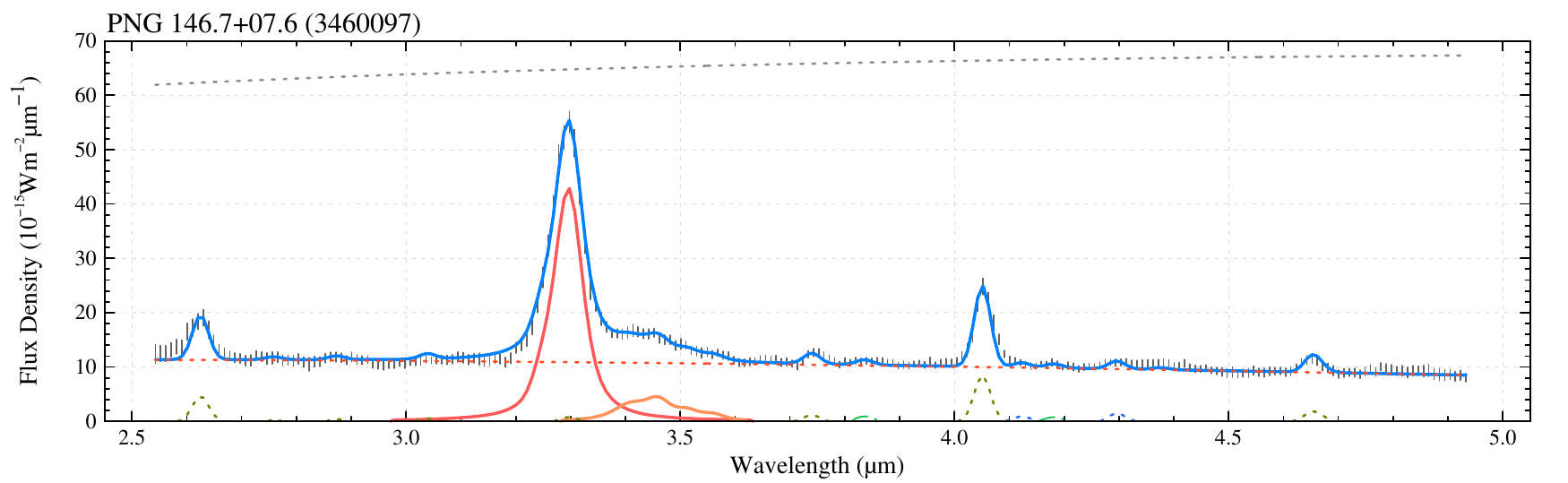}
  \plotone{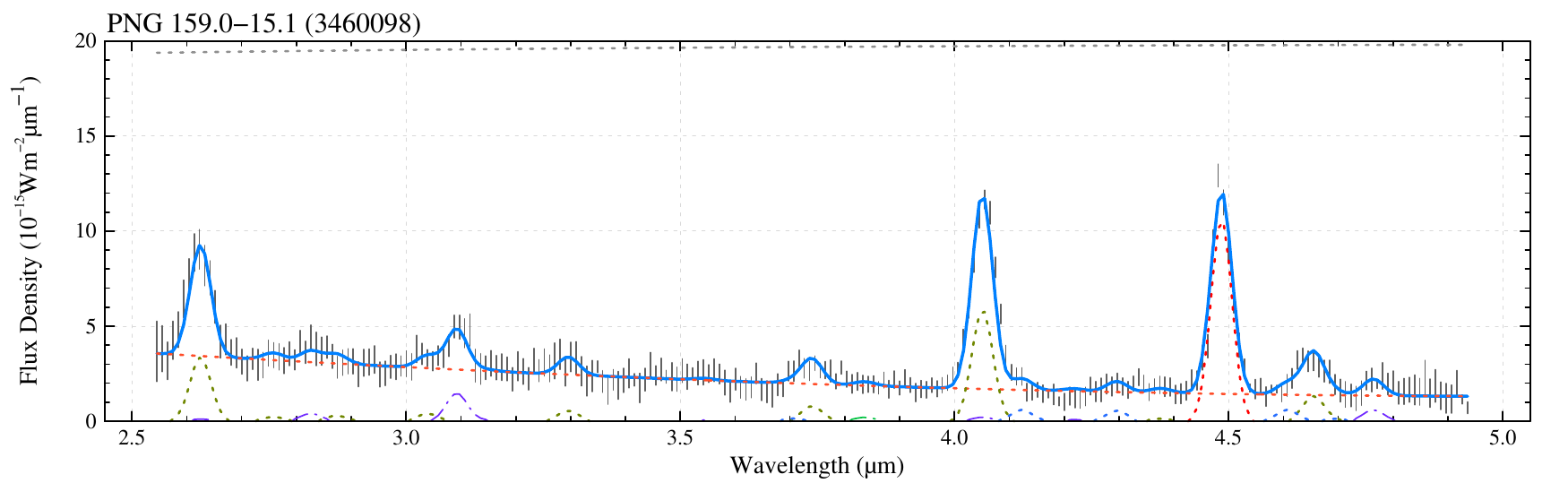}
  \plotone{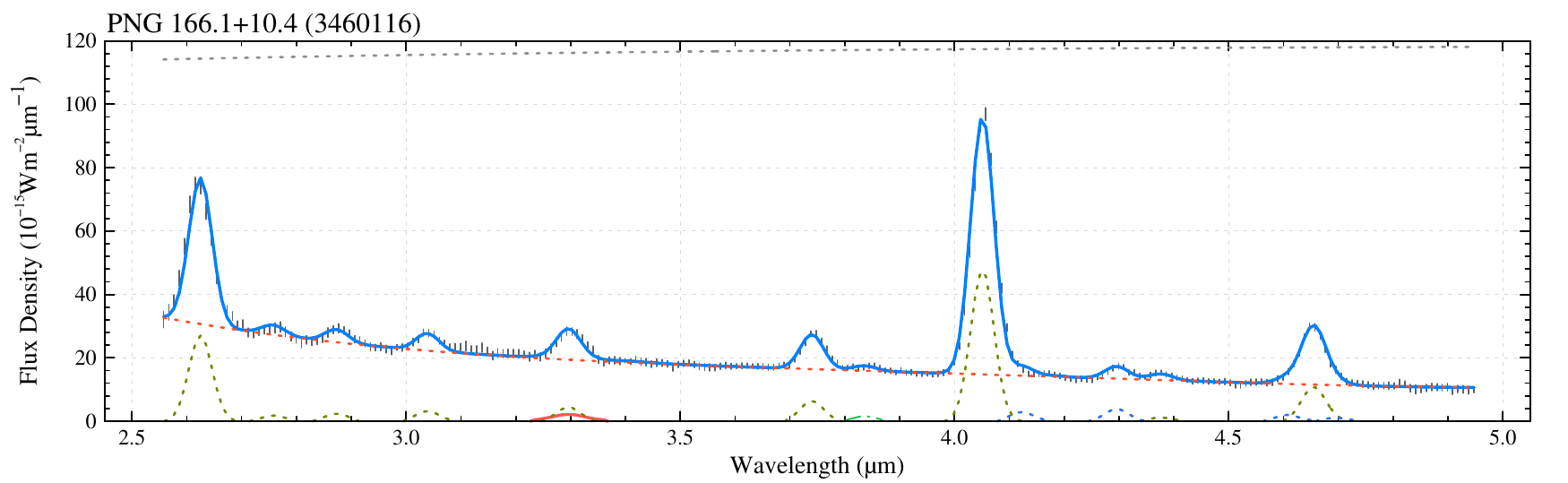}
  \caption{\textit{Cont.}---}
  \label{fig:allspectrum10}
\end{figure*}

\clearpage\addtocounter{figure}{-1}
\begin{figure*}[p]
  \centering
  \plotone{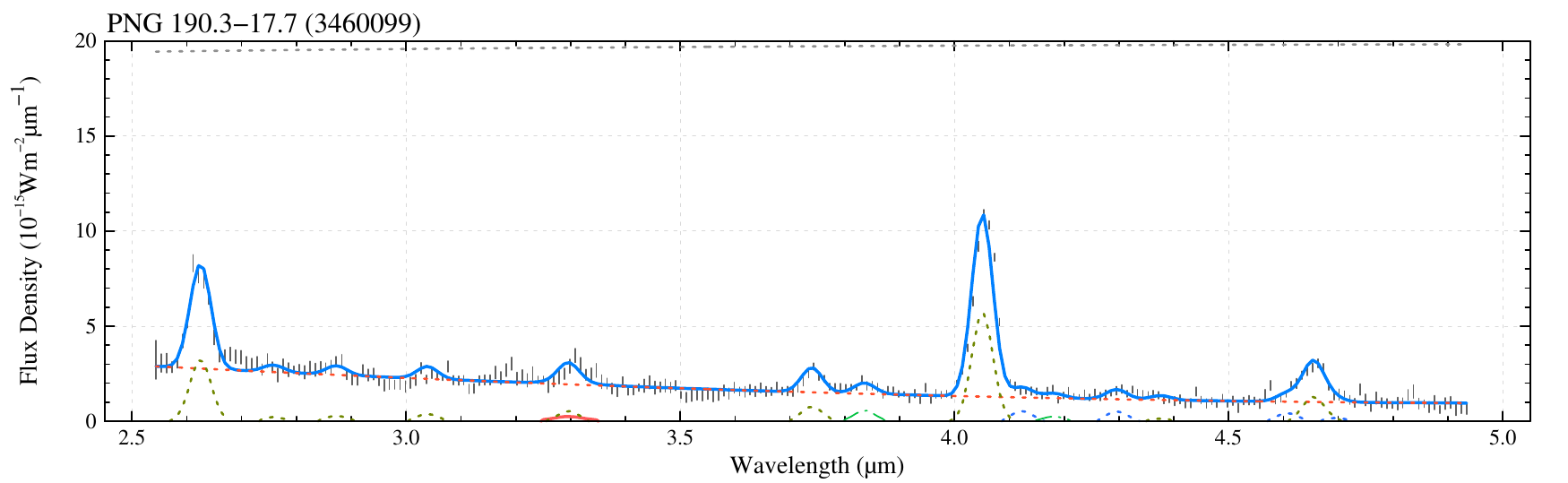}
  \plotone{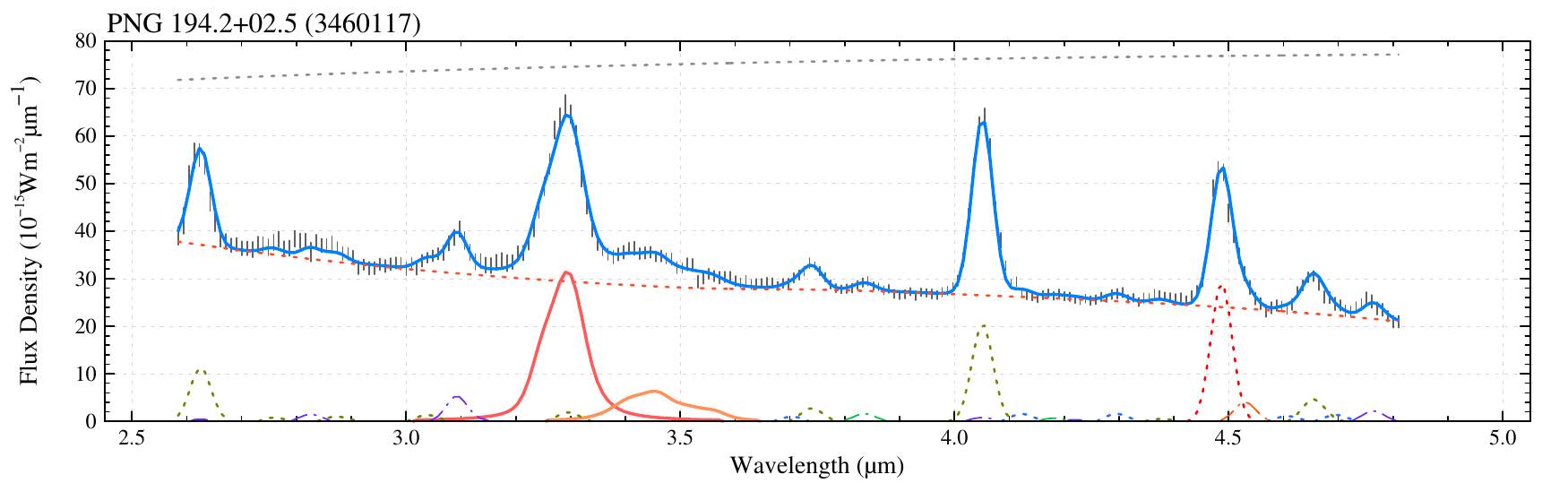}
  \plotone{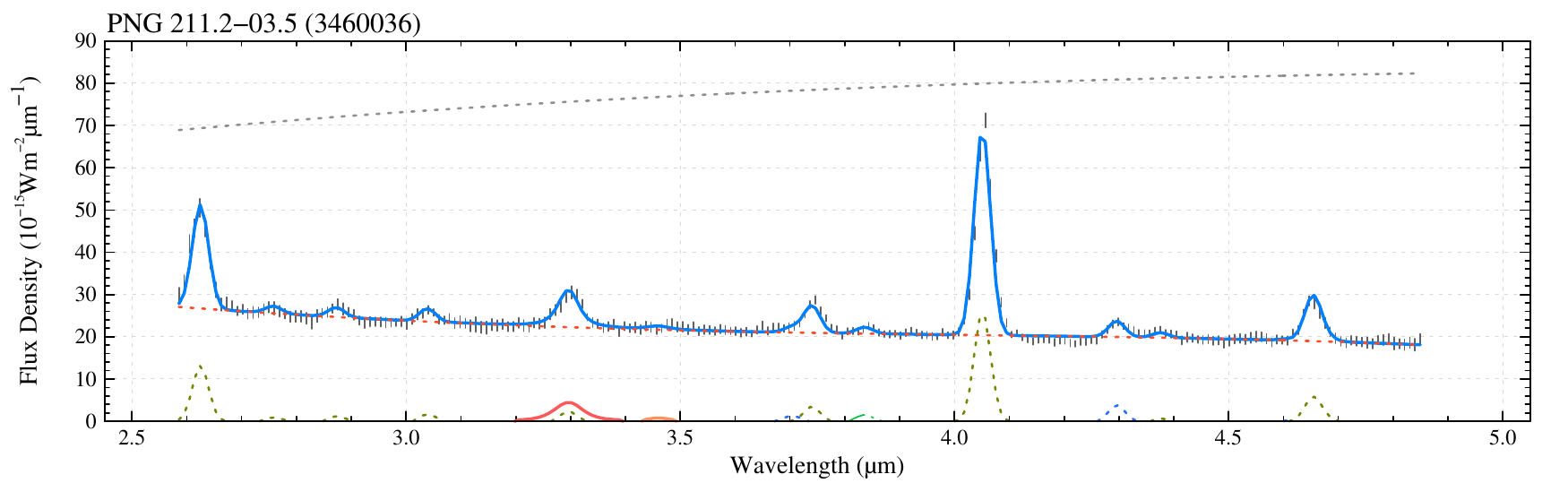}
  \caption{\textit{Cont.}---}
  \label{fig:allspectrum11}
\end{figure*}
\clearpage\addtocounter{figure}{-1}
\begin{figure*}[p]
  \centering
  \plotone{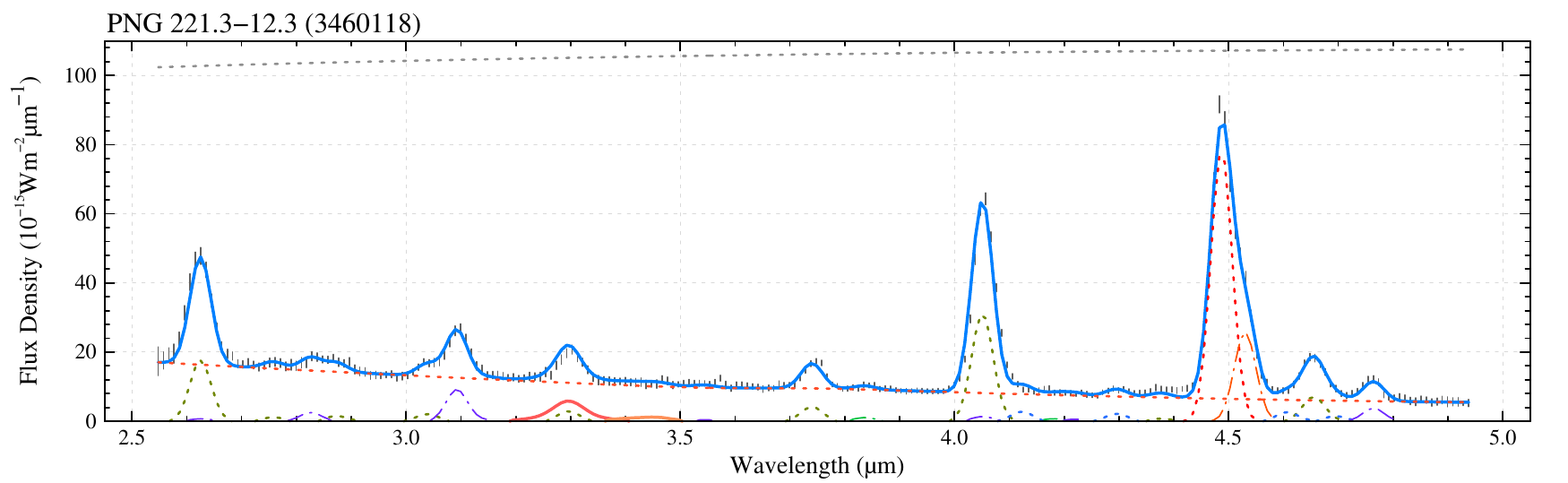}
  \plotone{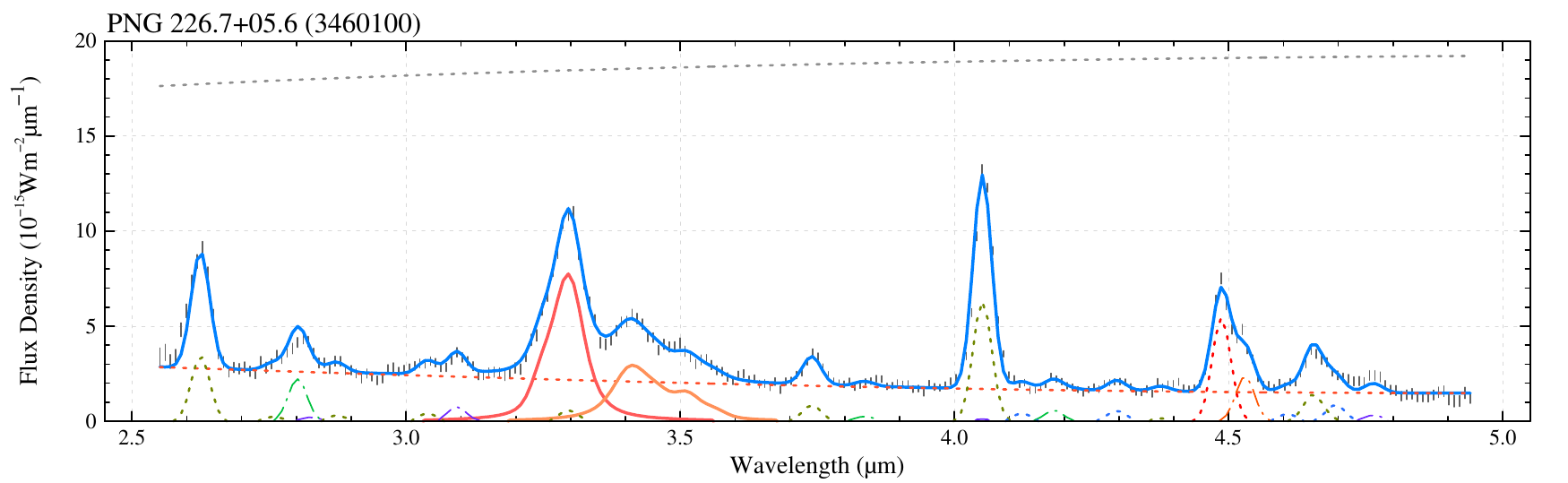}
  \plotone{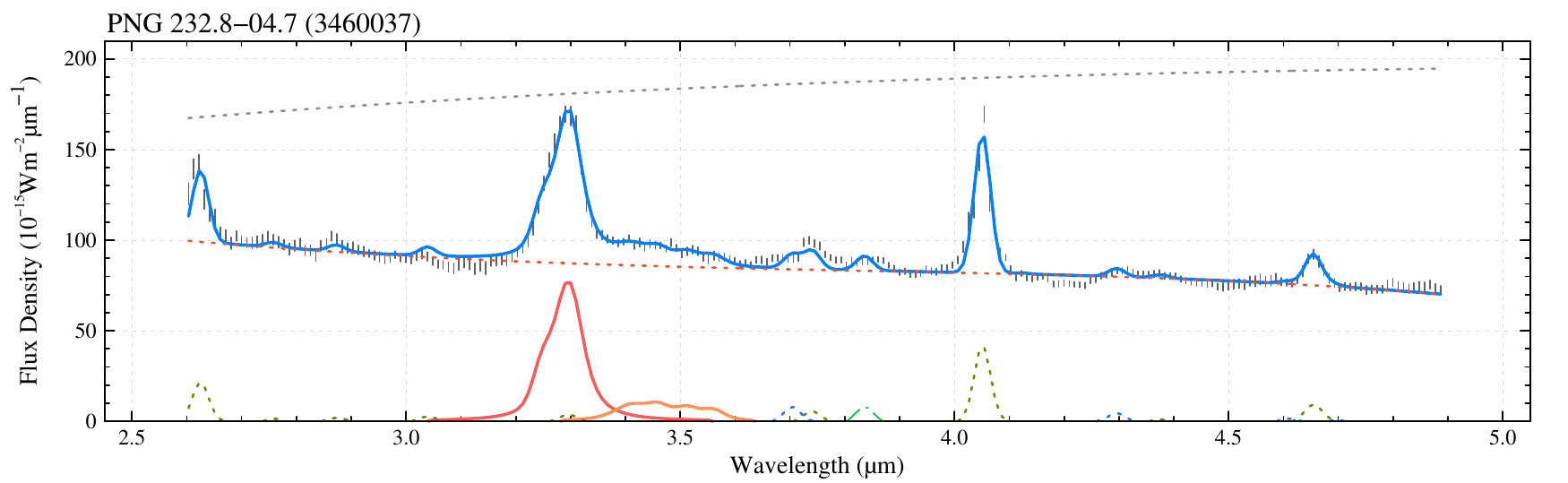}
  \caption{\textit{Cont.}---}
  \label{fig:allspectrum12}
\end{figure*}
\clearpage\addtocounter{figure}{-1}
\begin{figure*}[p]
  \centering
  \plotone{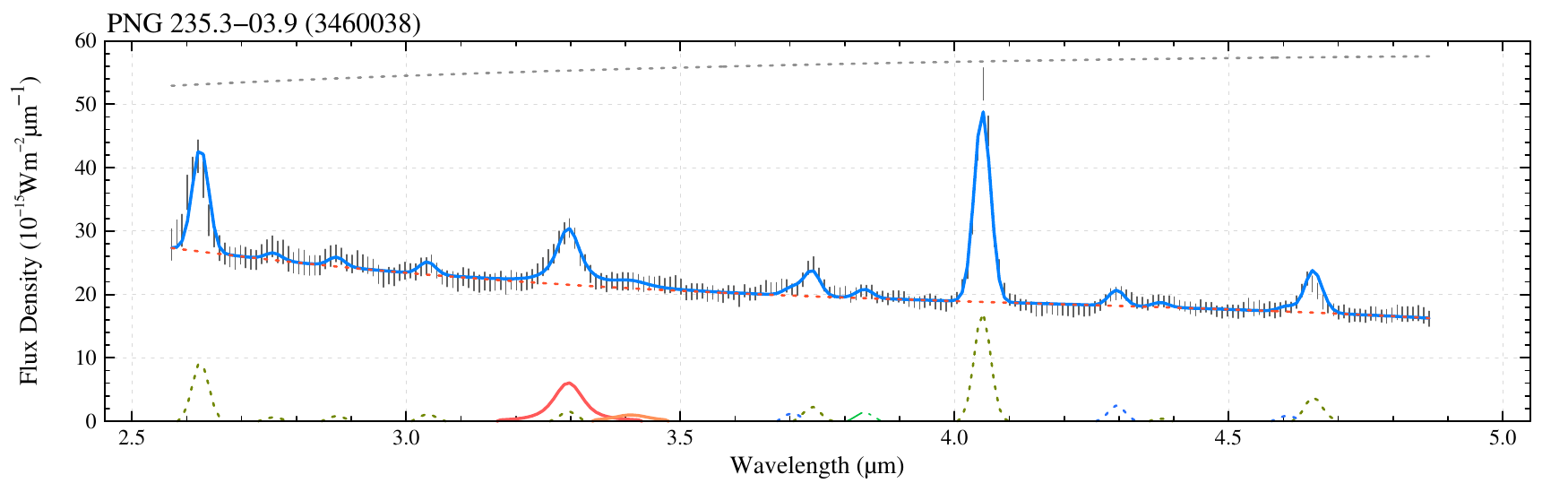}
  \plotone{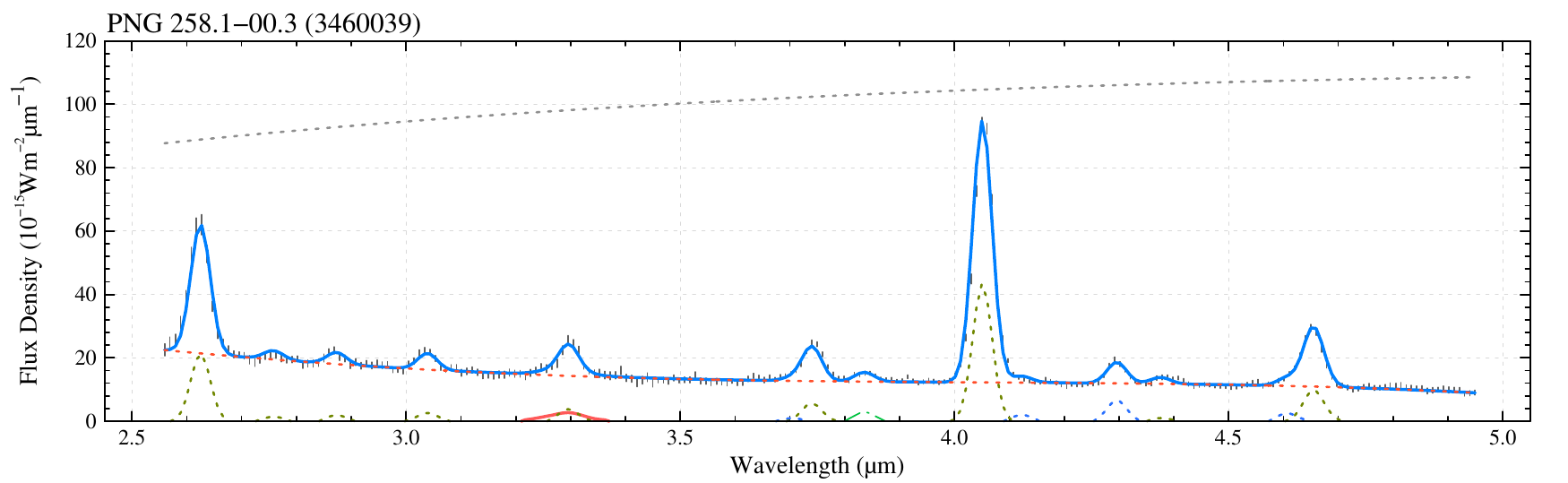}
  \plotone{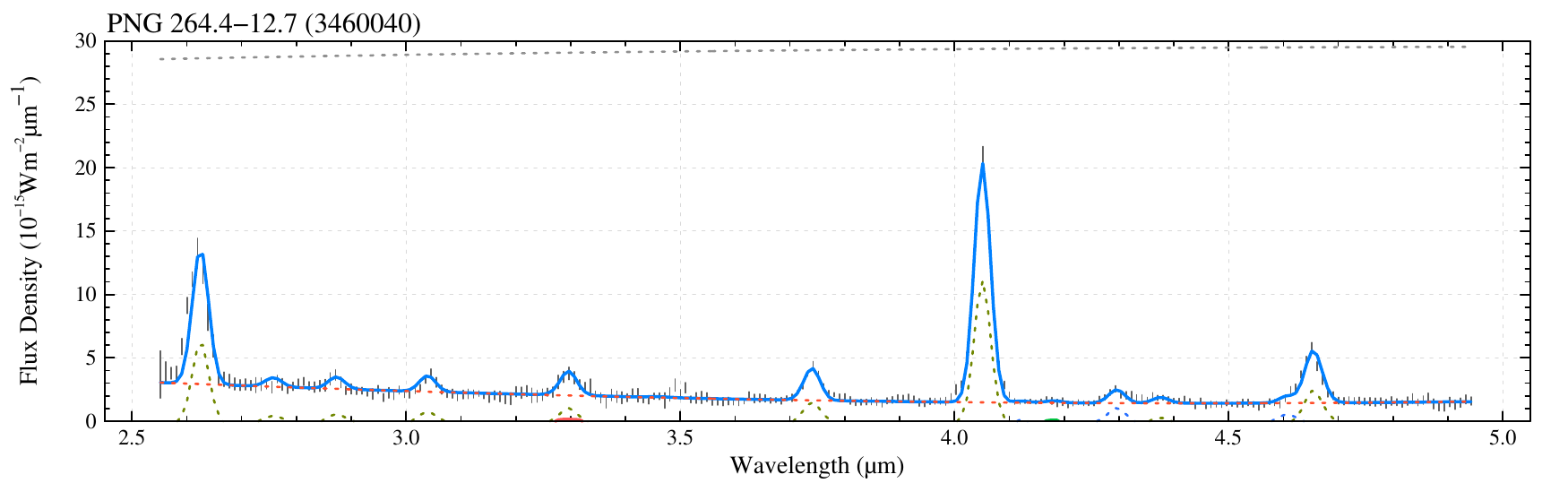}
  \caption{\textit{Cont.}---}
  \label{fig:allspectrum13}
\end{figure*}
\clearpage\addtocounter{figure}{-1}
\begin{figure*}[p]
  \centering
  \plotone{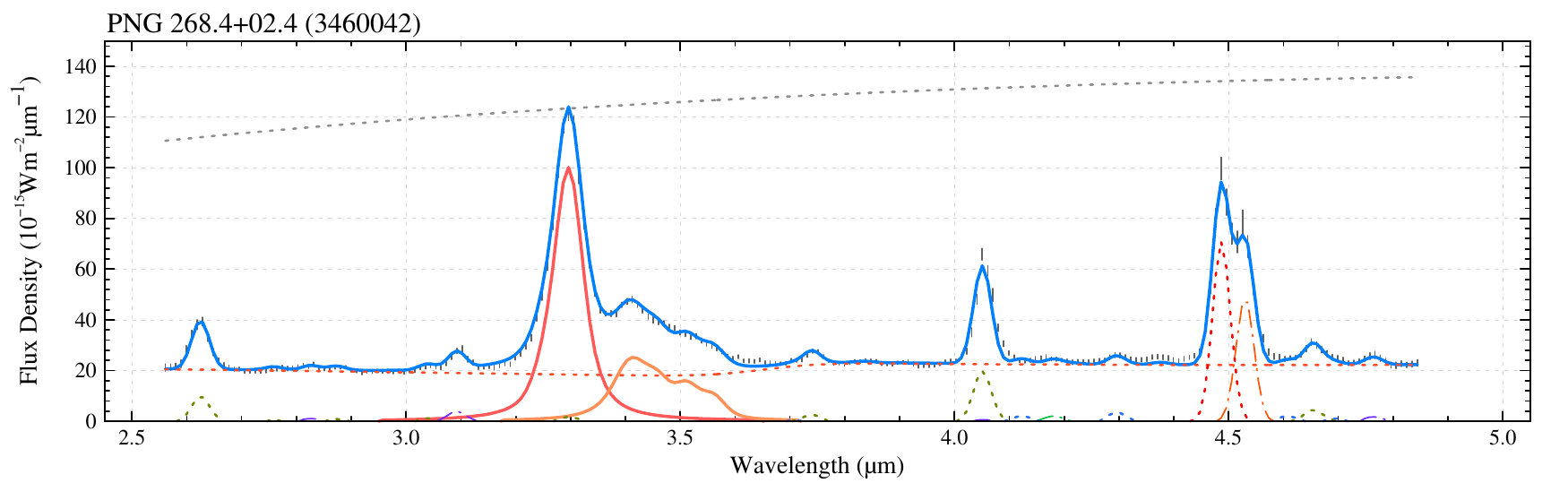}
  \plotone{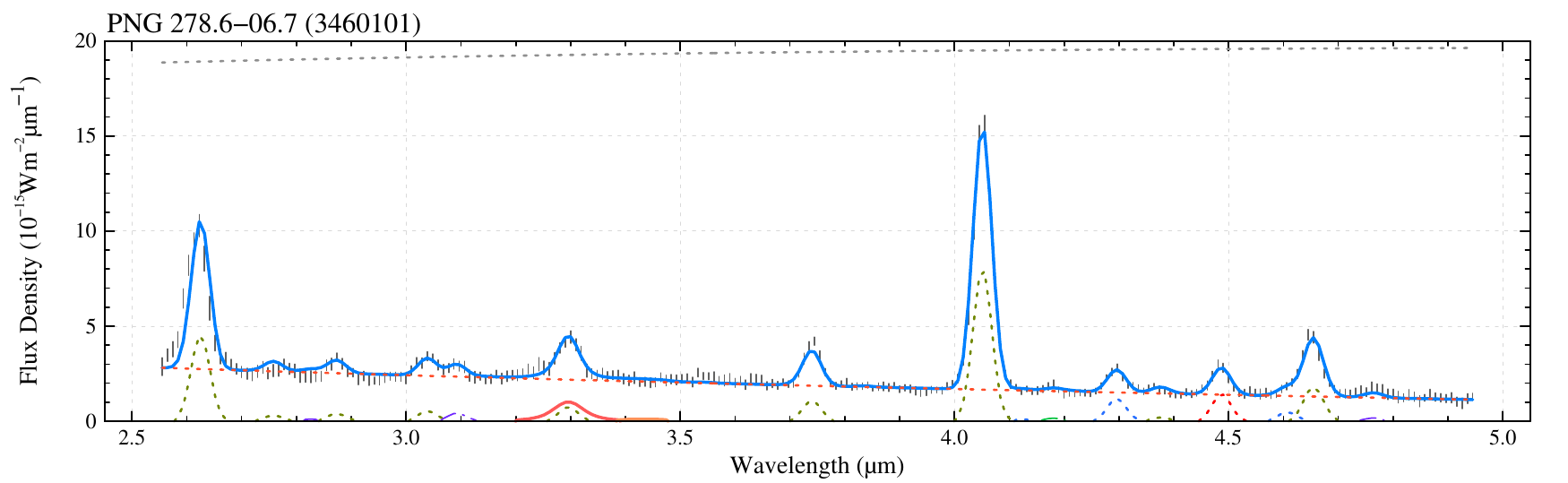}
  \plotone{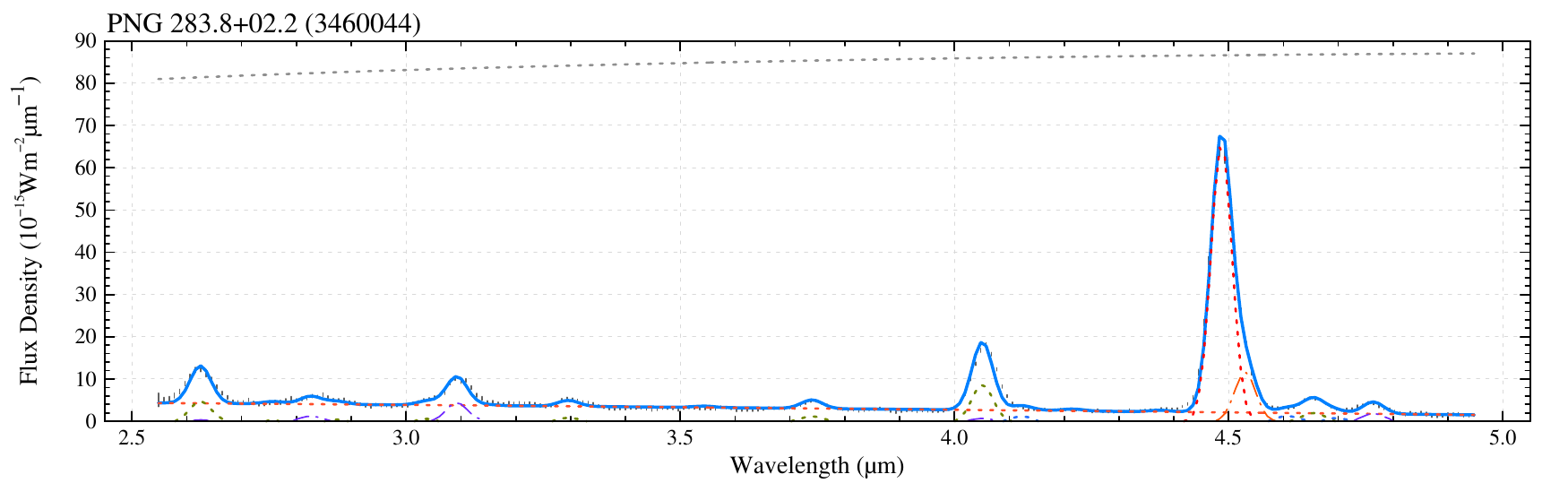}
  \caption{\textit{Cont.}---}
  \label{fig:allspectrum14}
\end{figure*}
\clearpage\addtocounter{figure}{-1}
\begin{figure*}[p]
  \centering
  \plotone{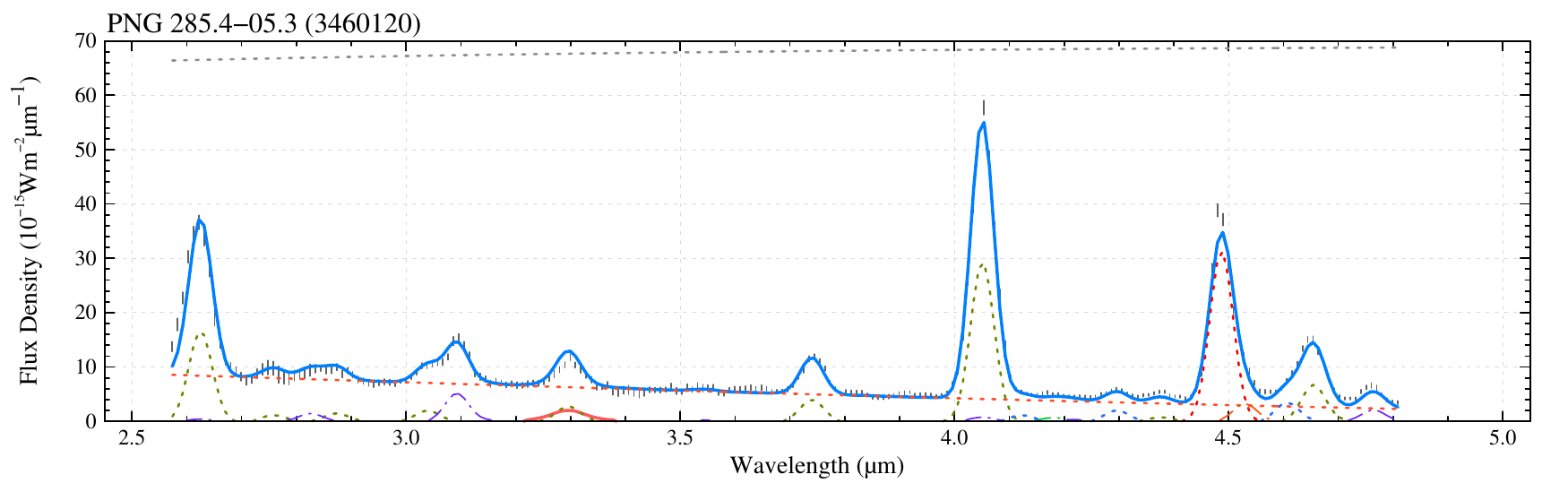}
  \plotone{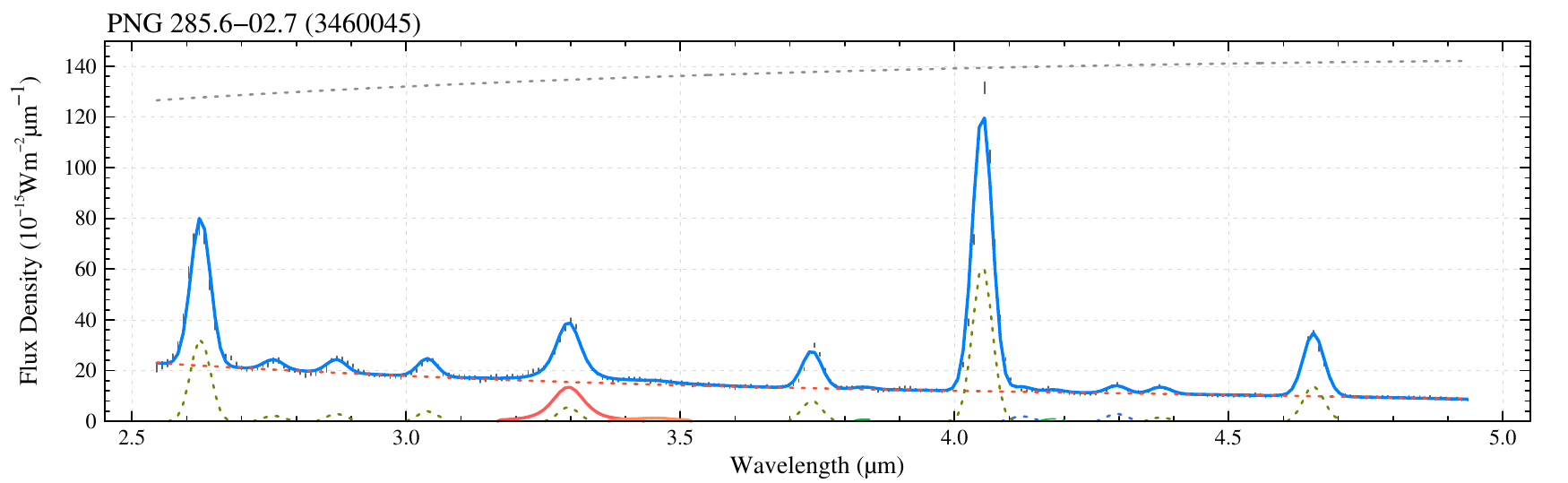}
  \plotone{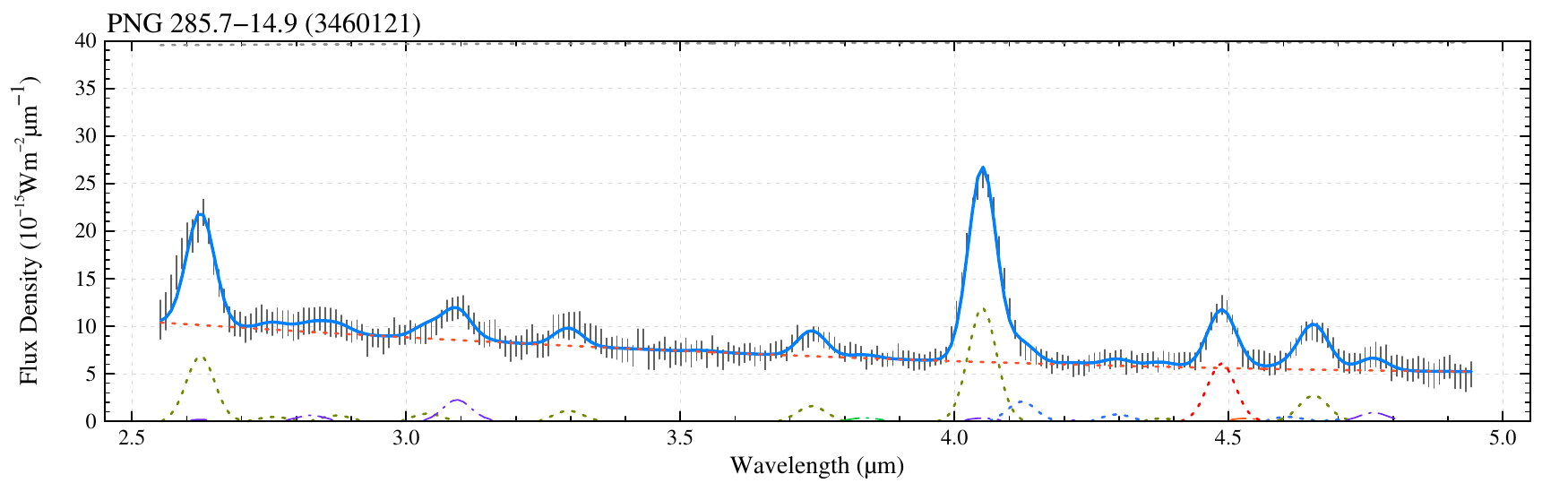}
  \caption{\textit{Cont.}---}
  \label{fig:allspectrum15}
\end{figure*}
\clearpage\addtocounter{figure}{-1}
\begin{figure*}[p]
  \centering
  \plotone{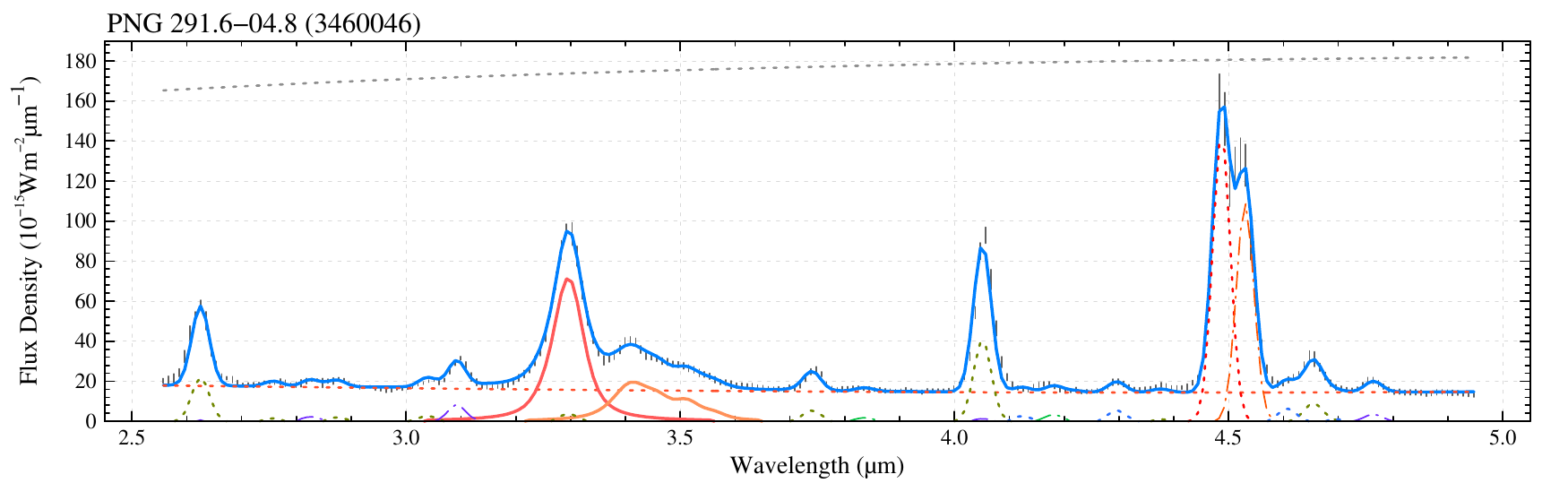}
  \plotone{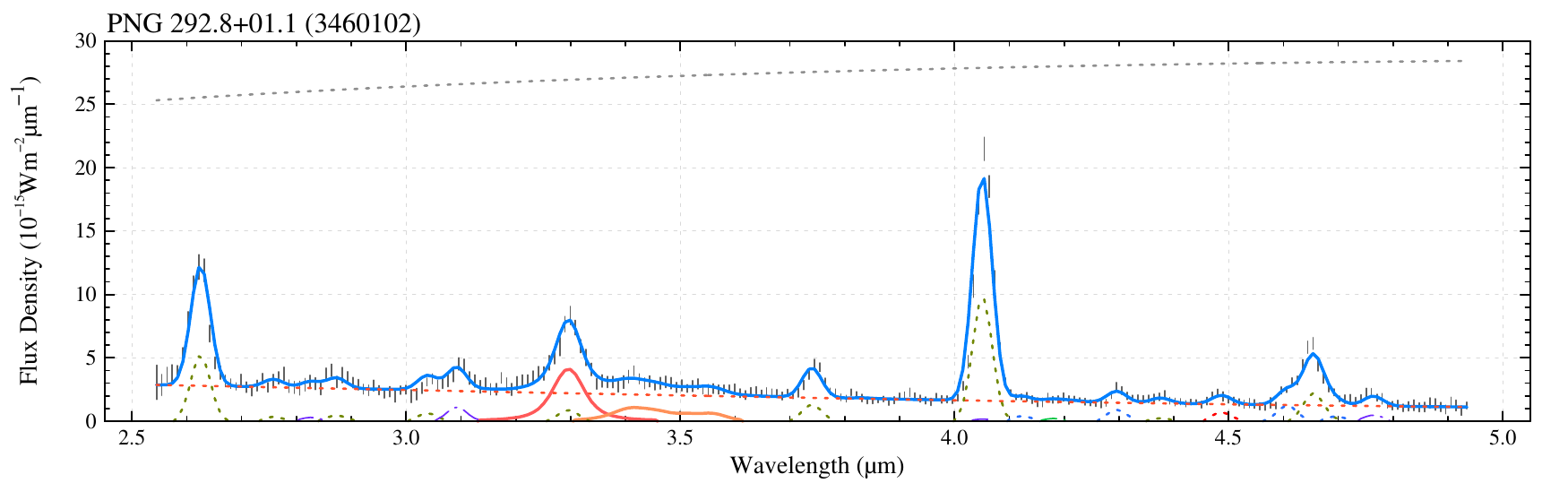}
  \plotone{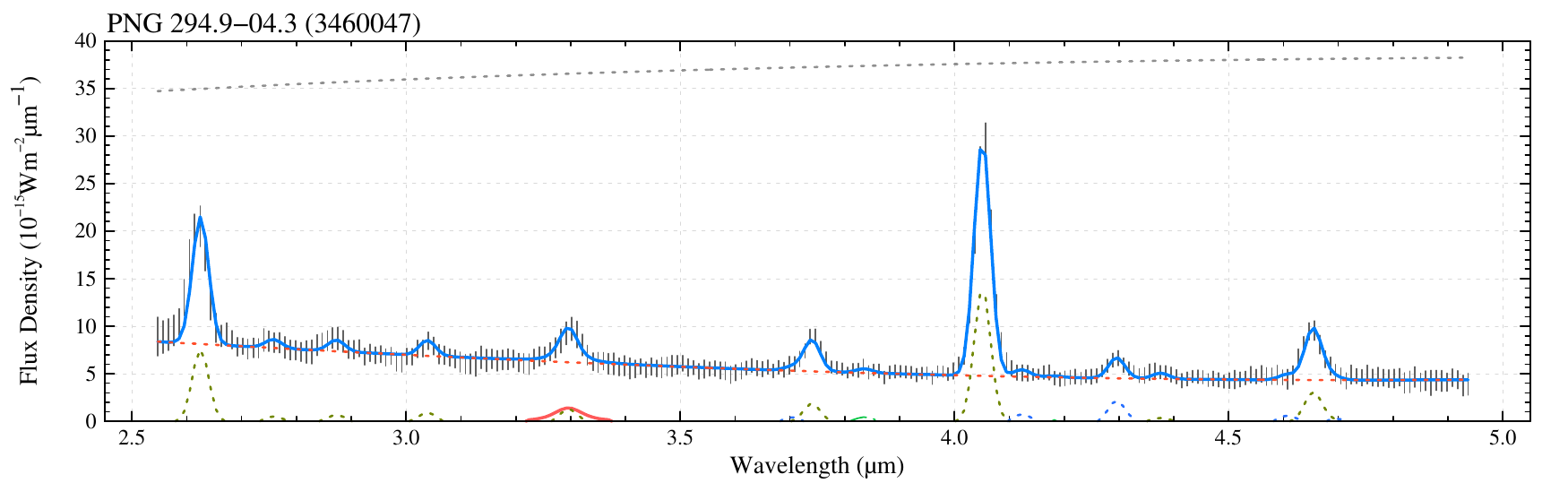}
  \caption{\textit{Cont.}---}
  \label{fig:allspectrum16}
\end{figure*}
\clearpage\addtocounter{figure}{-1}
\begin{figure*}[p]
  \centering
  \plotone{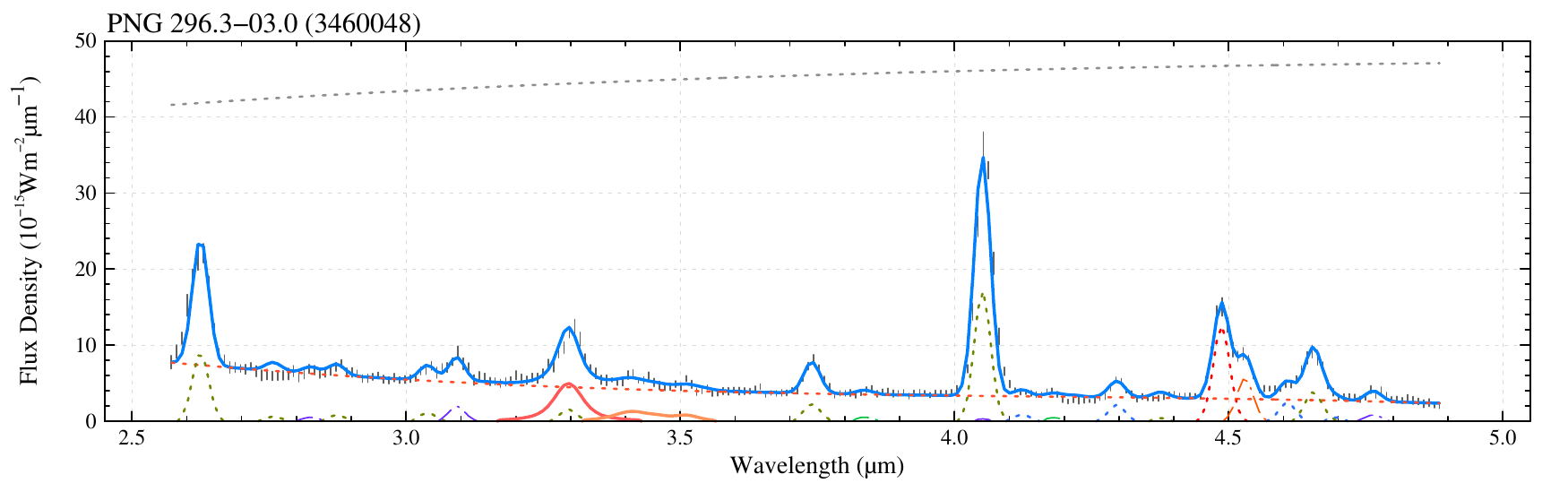}
  \plotone{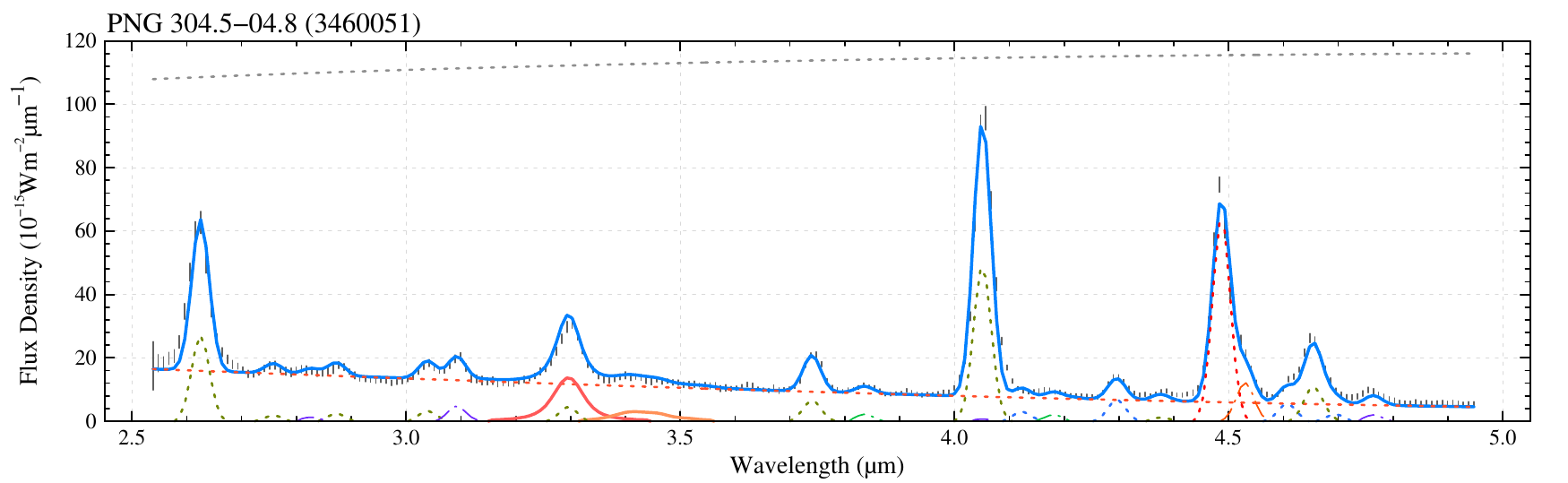}
  \plotone{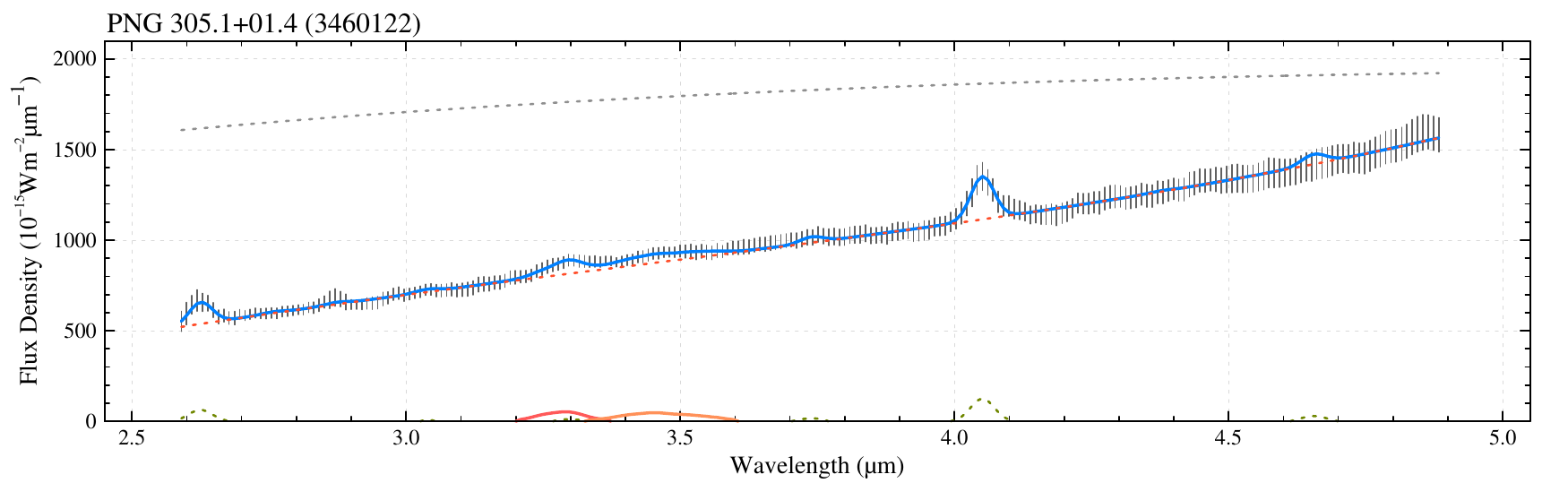}
  \caption{\textit{Cont.}---}
  \label{fig:allspectrum17}
\end{figure*}
\clearpage\addtocounter{figure}{-1}
\begin{figure*}[p]
  \centering
  \plotone{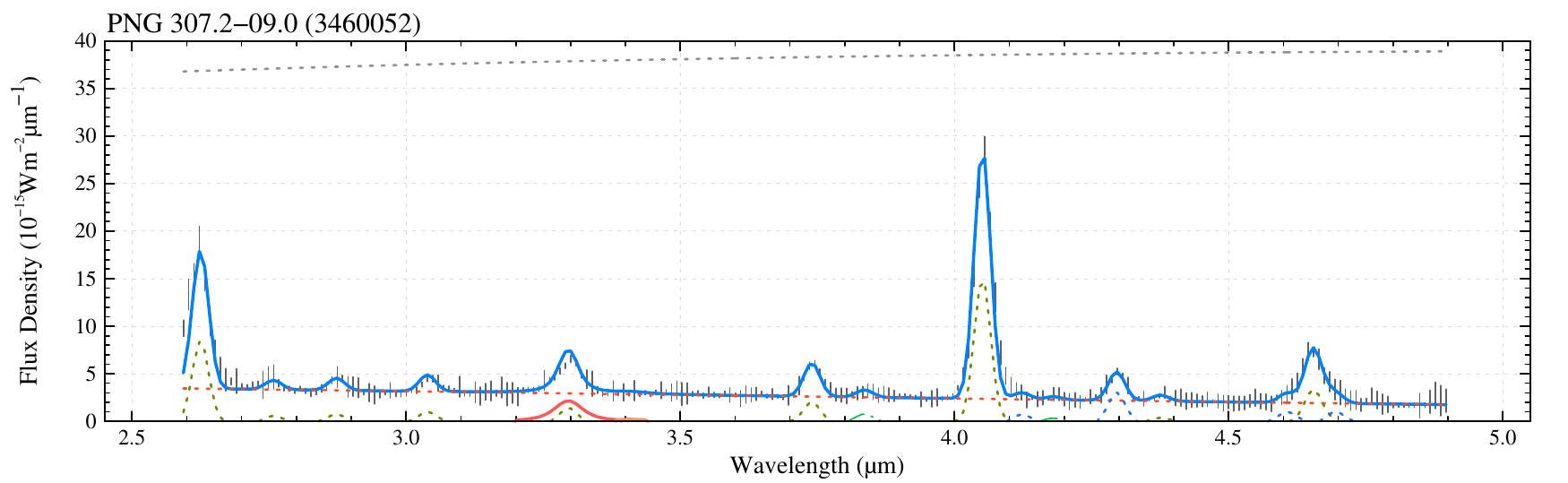}
  \plotone{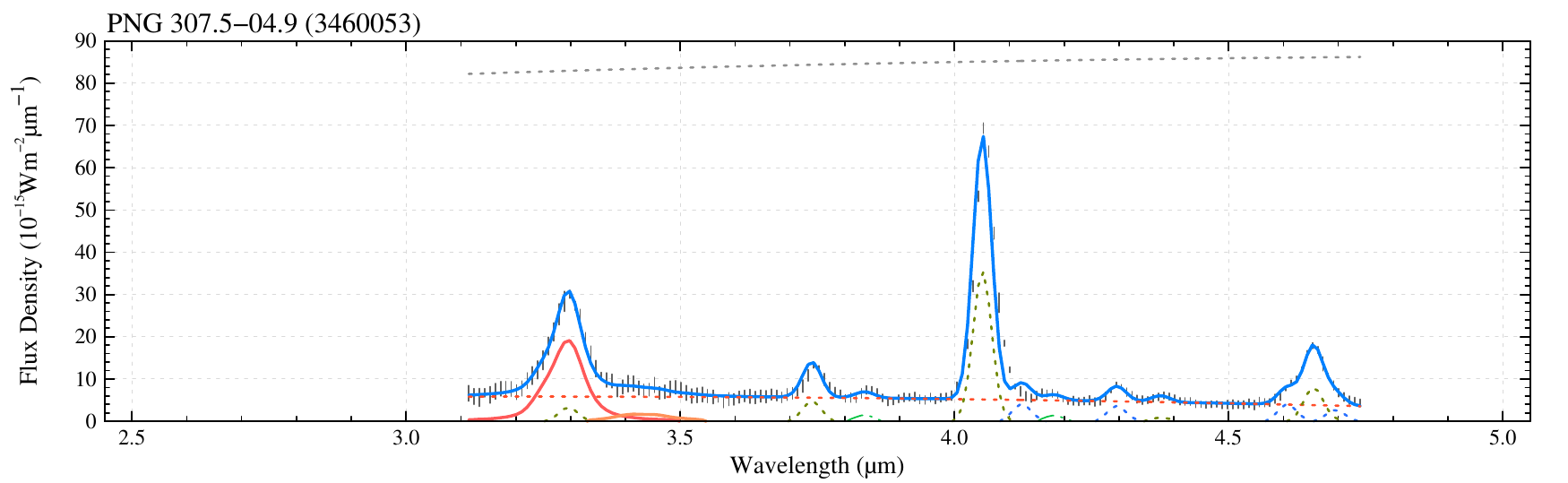}
  \plotone{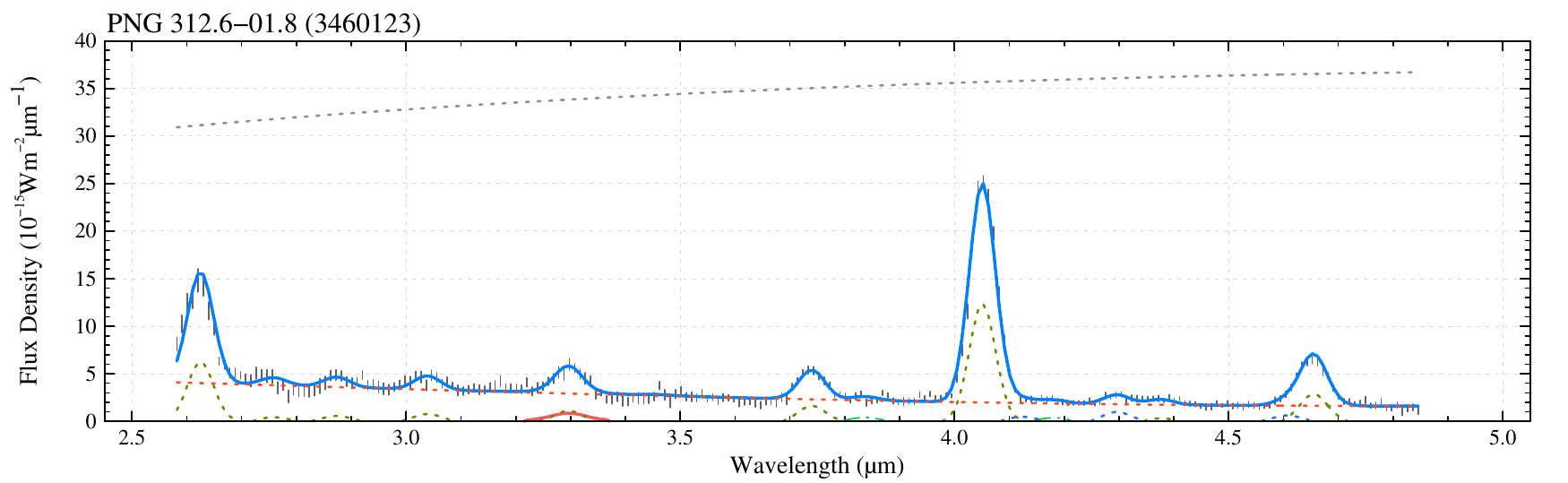}
  \caption{\textit{Cont.}---}
  \label{fig:allspectrum18}
\end{figure*}
\clearpage\addtocounter{figure}{-1}
\begin{figure*}[p]
  \centering
  \plotone{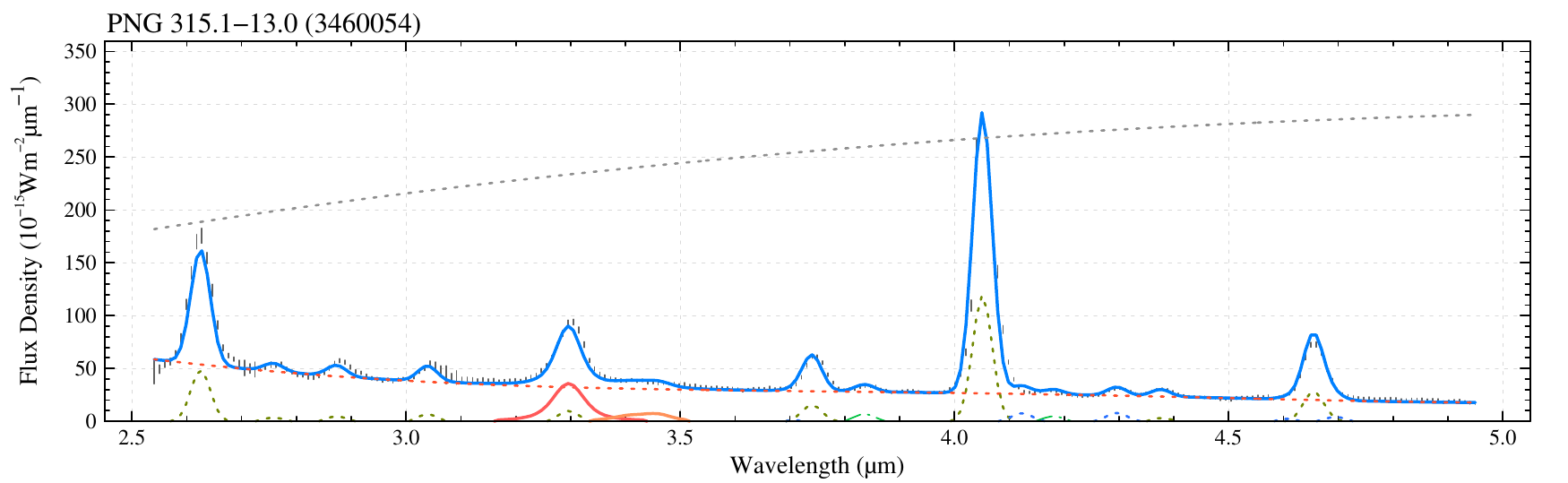}
  \plotone{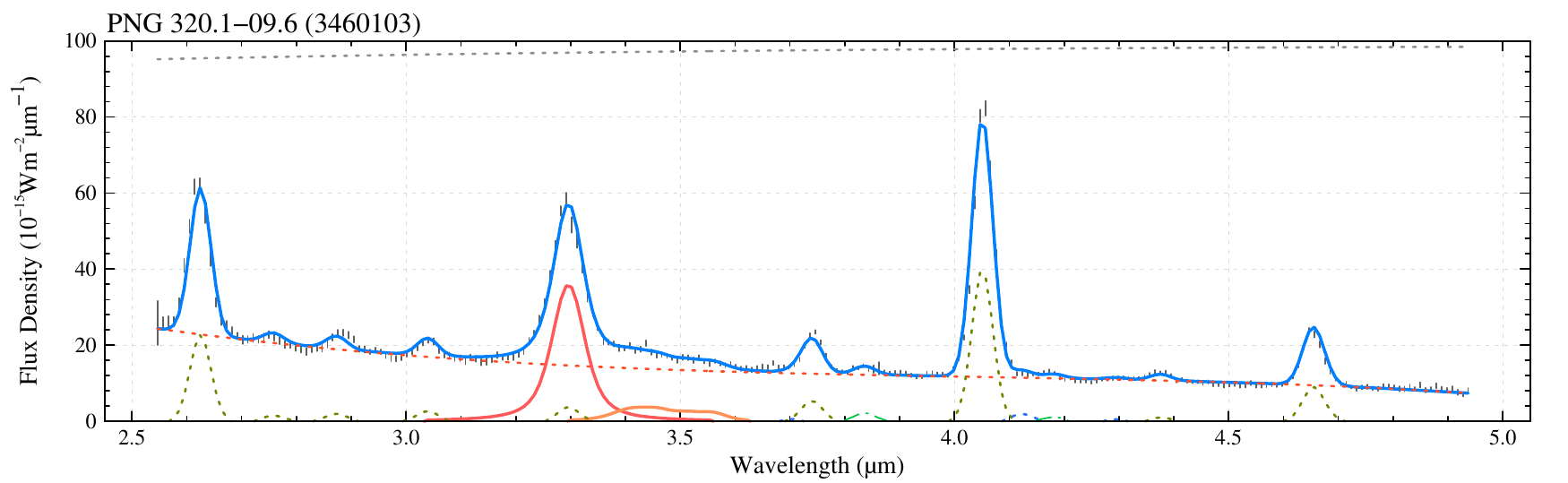}
  \plotone{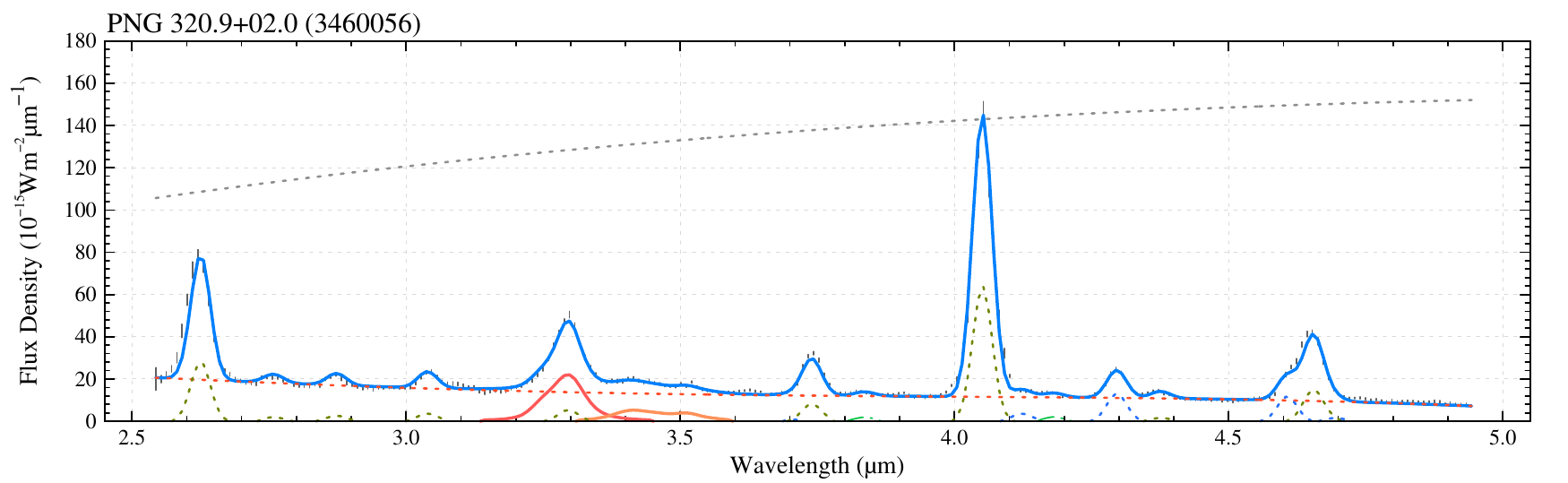}
  \caption{\textit{Cont.}---}
  \label{fig:allspectrum19}
\end{figure*}
\clearpage\addtocounter{figure}{-1}
\begin{figure*}[p]
  \centering
  \plotone{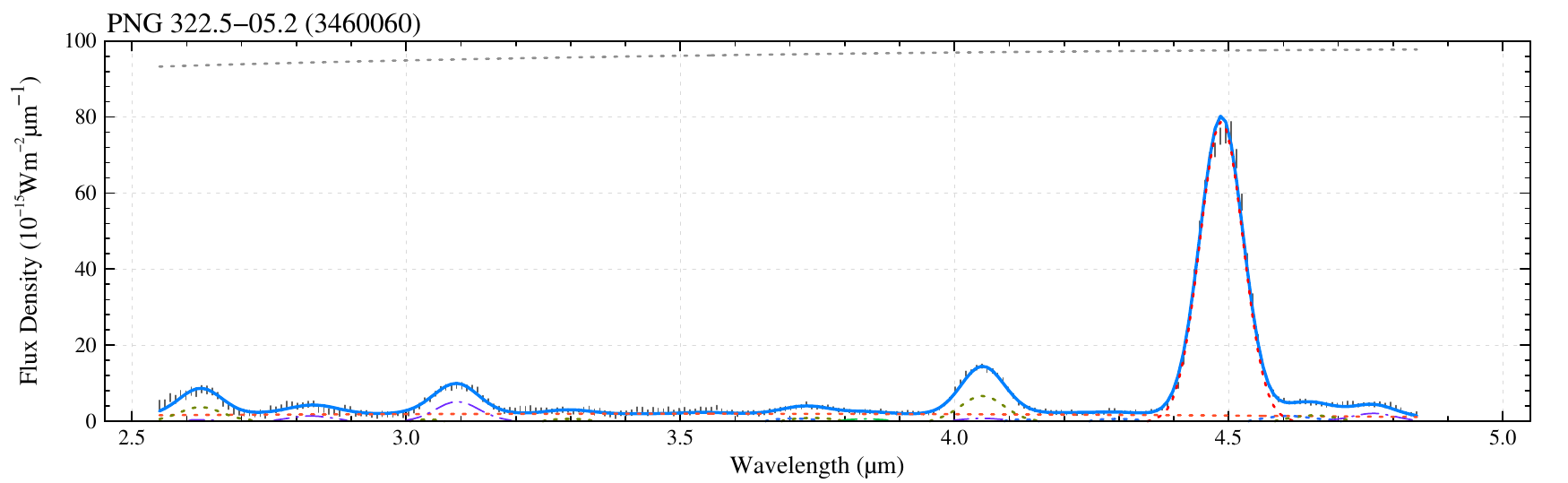}
  \plotone{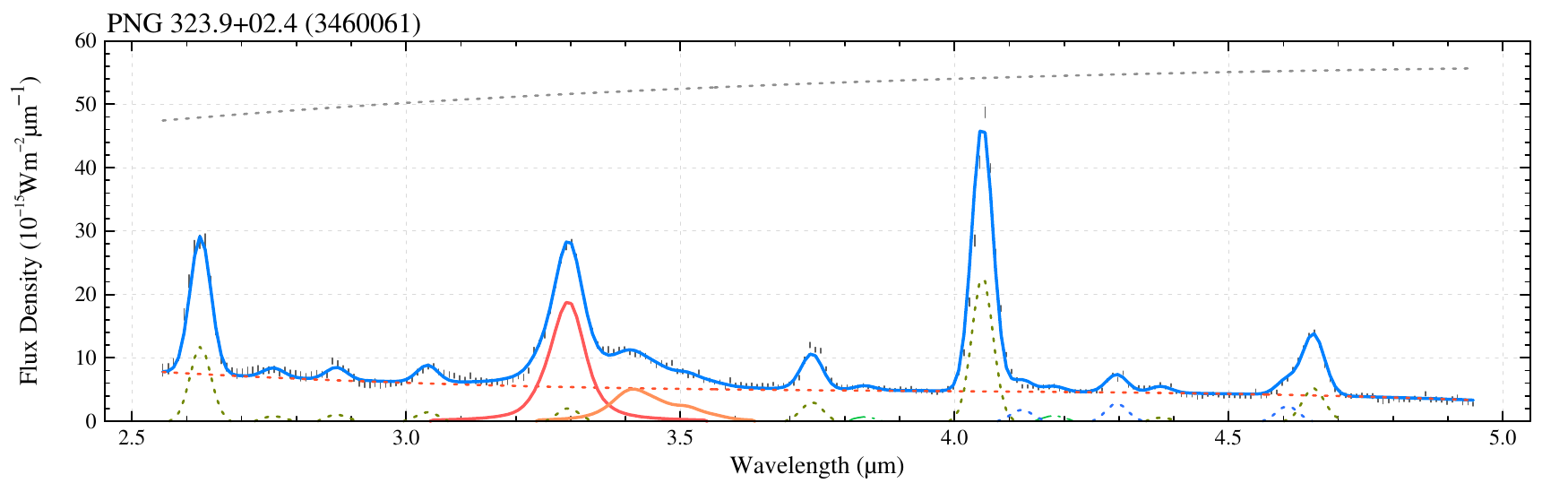}
  \plotone{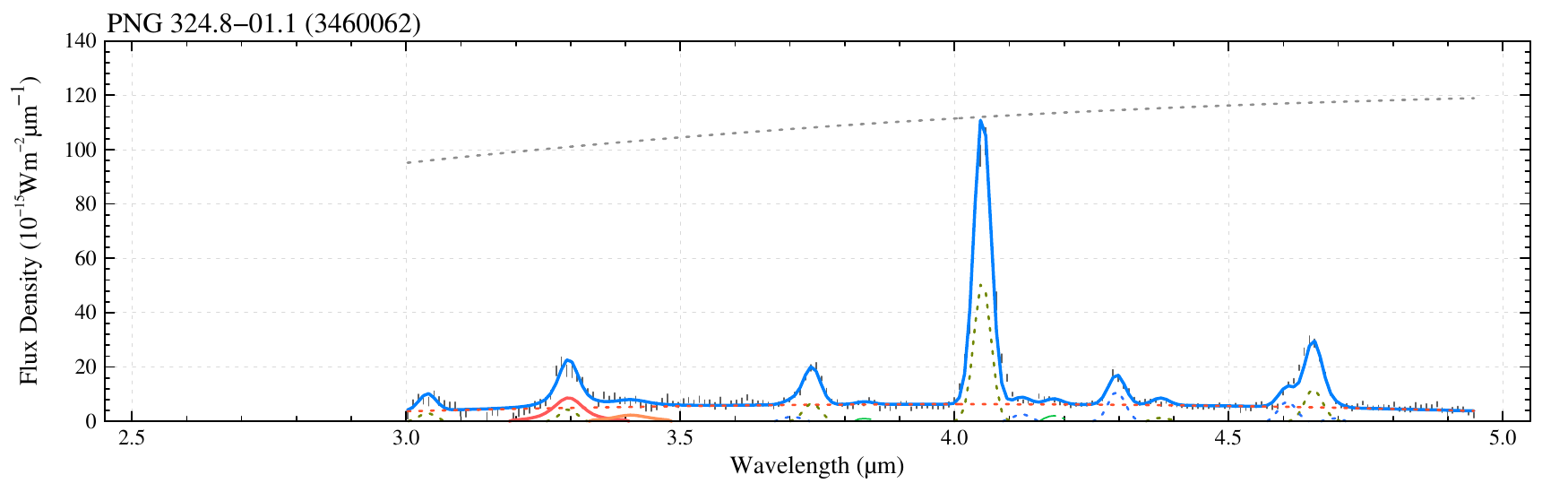}
  \caption{\textit{Cont.}---}
  \label{fig:allspectrum20}
\end{figure*}
\clearpage\addtocounter{figure}{-1}
\begin{figure*}[p]
  \centering
  \plotone{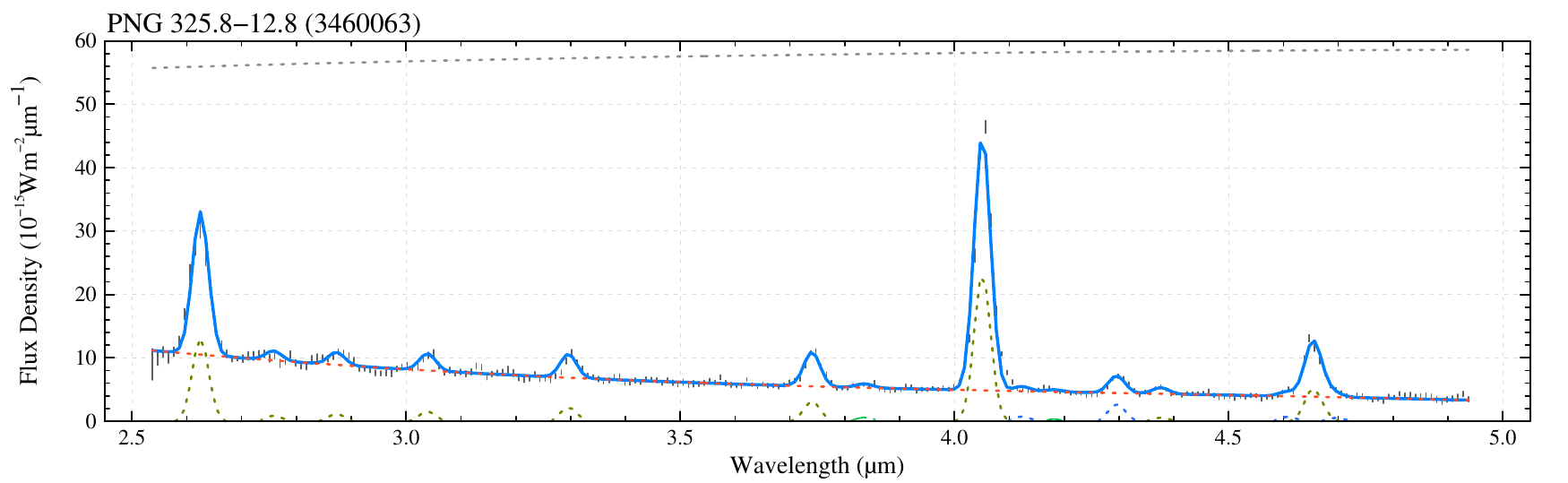}
  \plotone{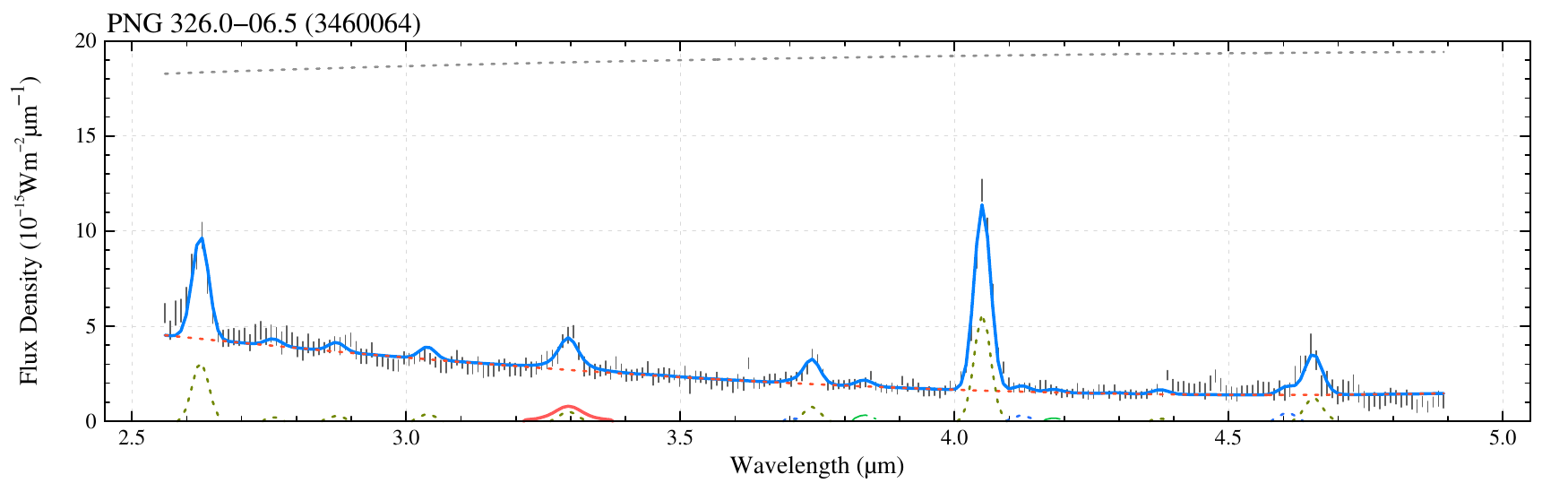}
  \plotone{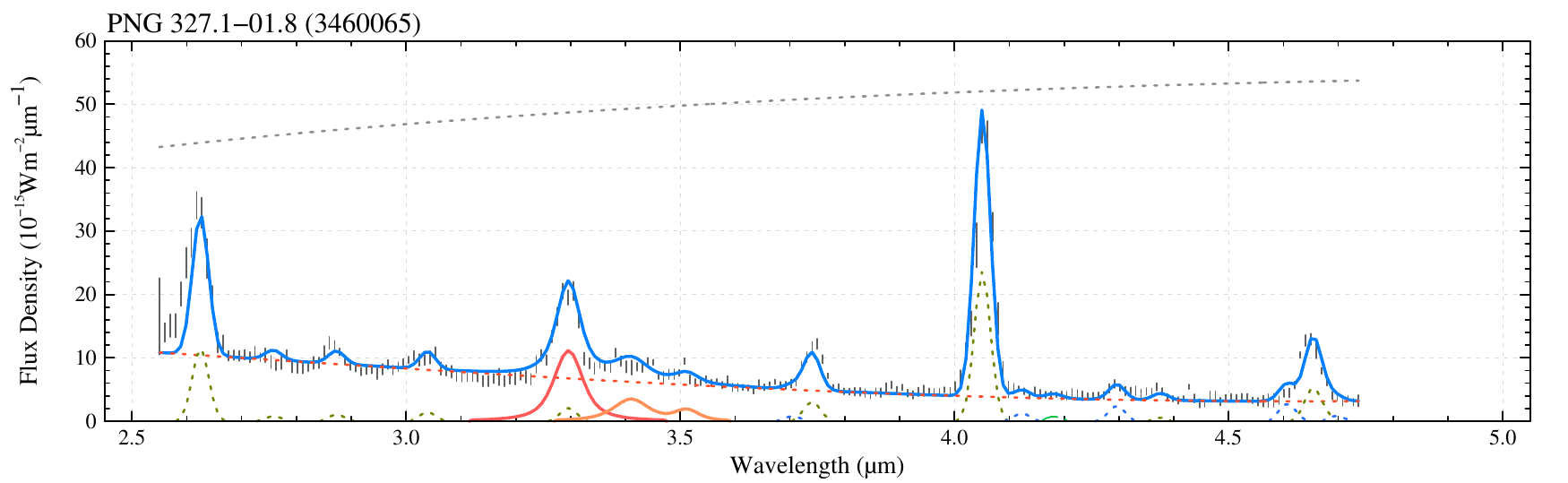}
  \caption{\textit{Cont.}---}
  \label{fig:allspectrum21}
\end{figure*}
\clearpage\addtocounter{figure}{-1}
\begin{figure*}[p]
  \centering
  \plotone{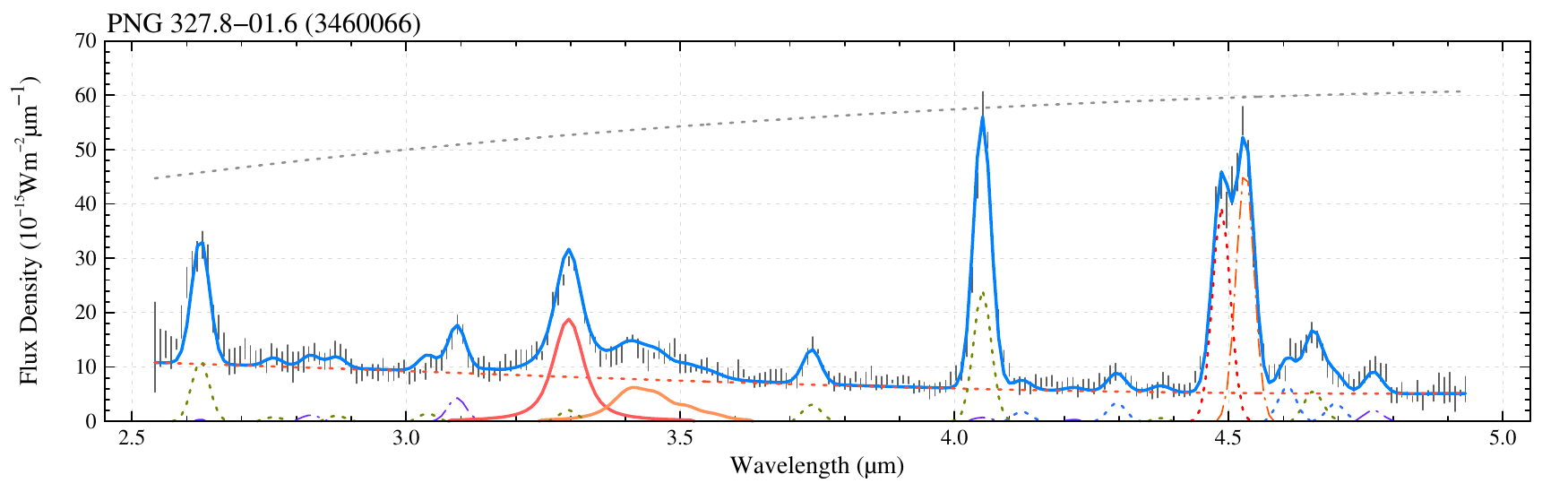}
  \plotone{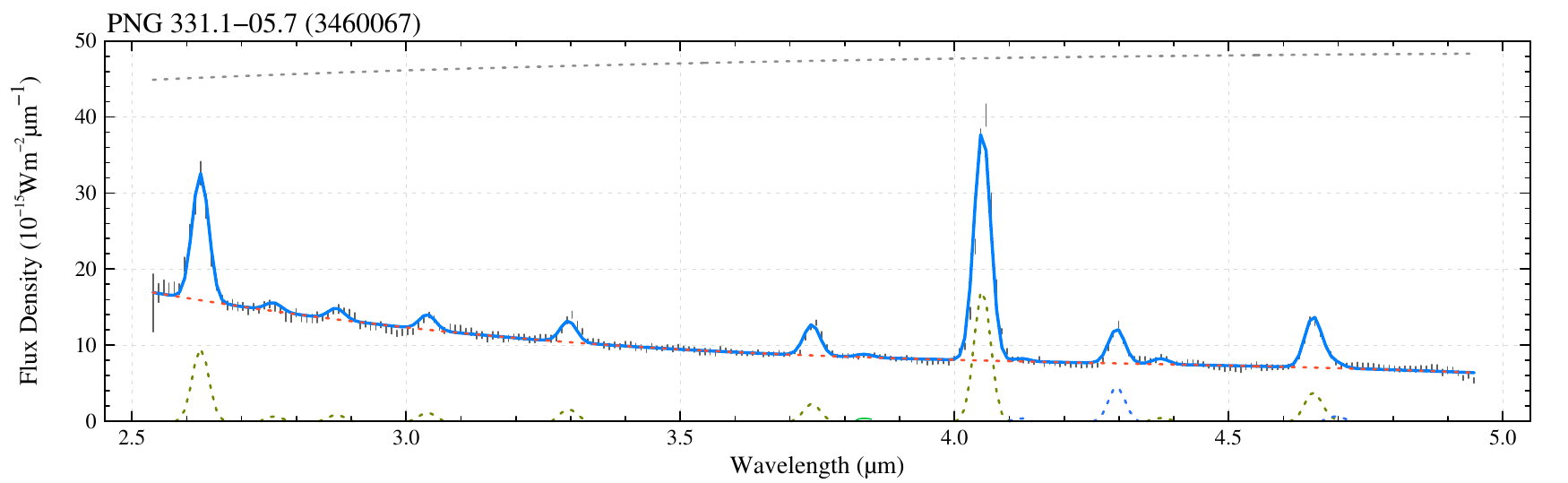}
  \plotone{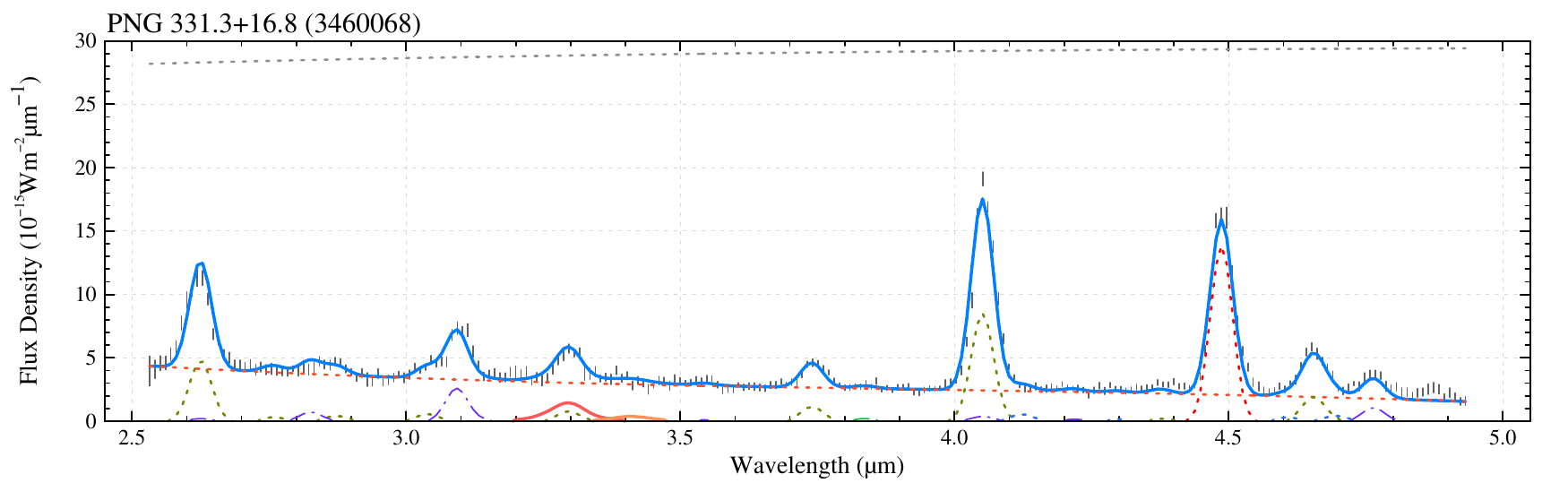}
  \caption{\textit{Cont.}---}
  \label{fig:allspectrum22}
\end{figure*}
\clearpage\addtocounter{figure}{-1}
\begin{figure*}[p]
  \centering
  \plotone{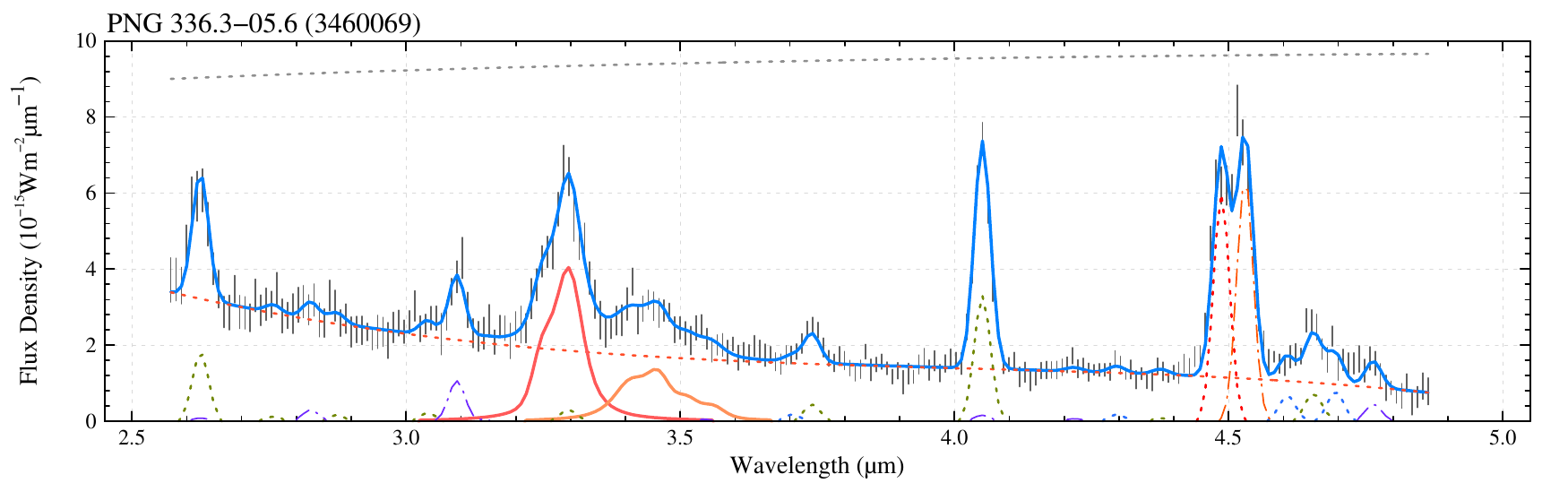}
  \plotone{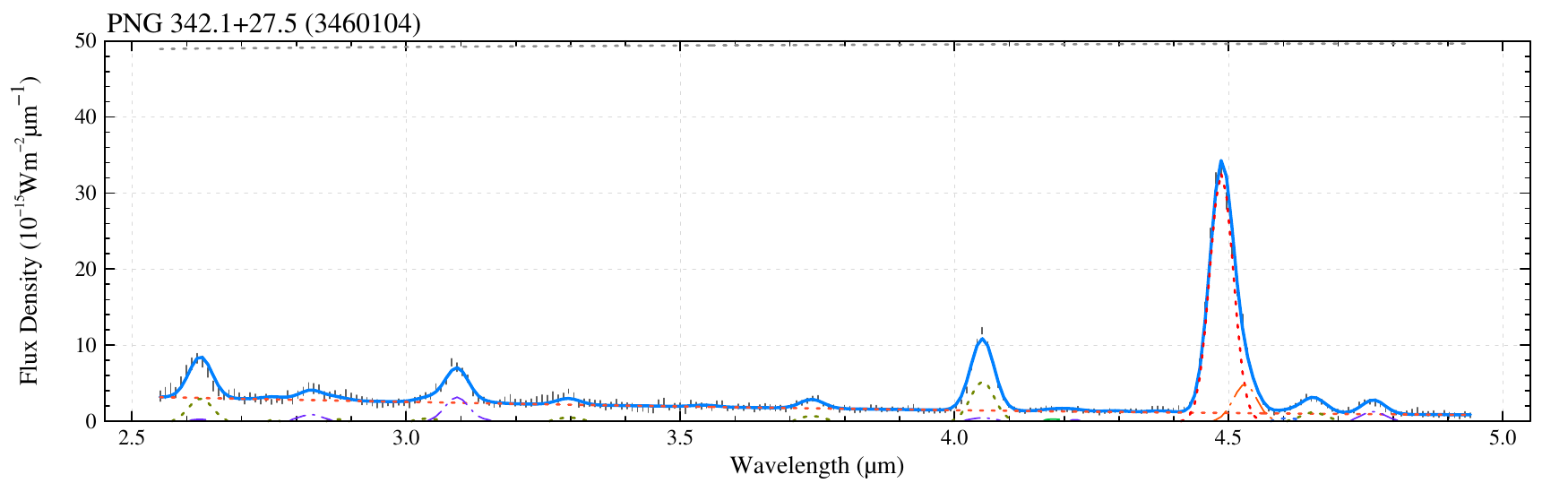}
  \plotone{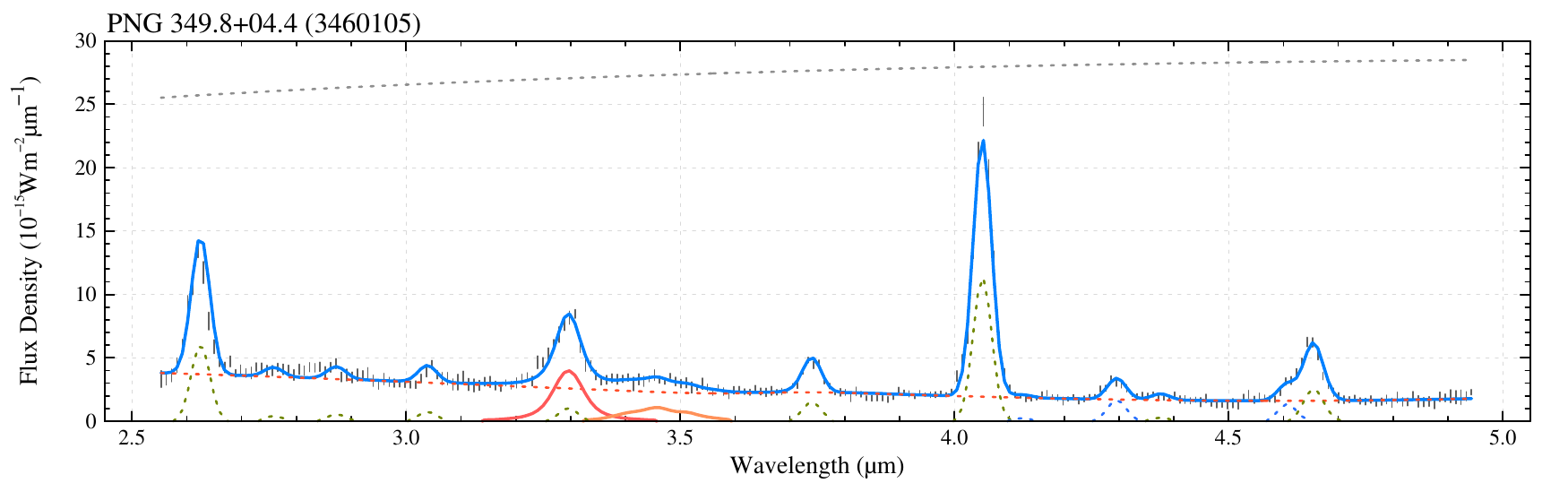}
  \caption{\textit{Cont.}---}
  \label{fig:allspectrum23}
\end{figure*}
\clearpage\addtocounter{figure}{-1}
\begin{figure*}[p]
  \centering
  \plotone{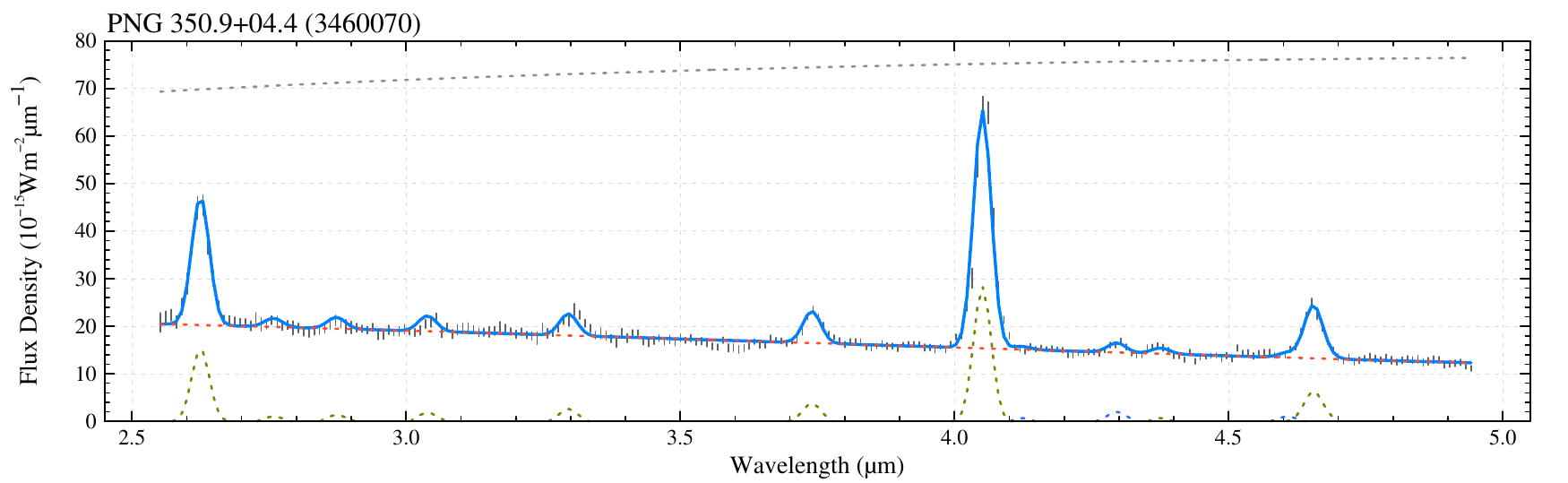}
  \plotone{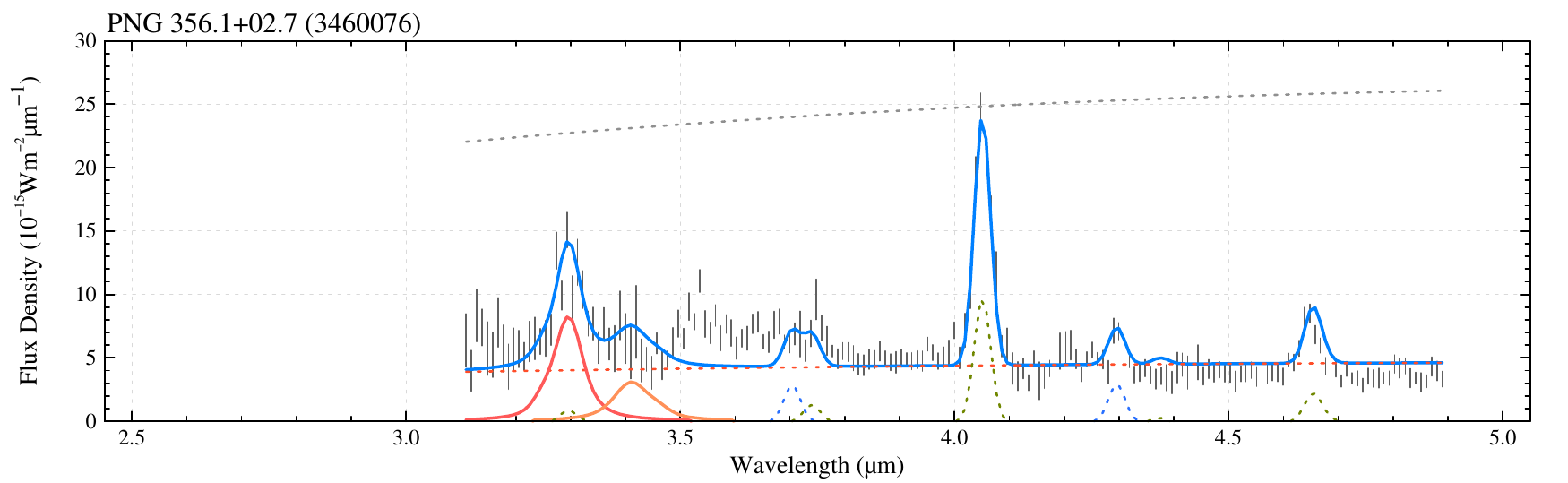}
  \plotone{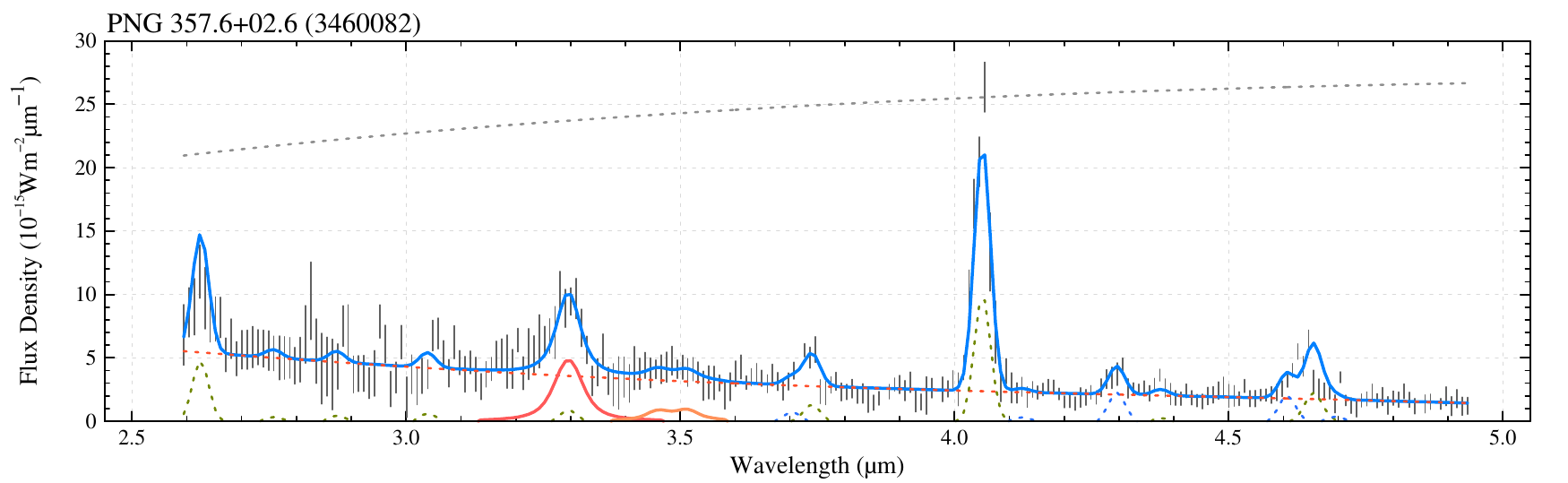}
  \caption{\textit{Cont.}---}
  \label{fig:allspectrum24}
\end{figure*}

\clearpage

\begin{appendix}
  \section{Uncertainty in $A_{V}$}\label{sec:app:uncertainAv}
There are several hydrogen recombination lines in the $2.5$--$5.0\,\mu$m spectrum. Extinction can be estimated based on the intensity ratio of these lines. The accuracy of the estimated extinction heavily depends on the signal-to-noise ratio (${\rm S/N}$) of the intensity ratio. This chapter describes the method to derive the extinction based on the IRC spectrum and shows that a typical uncertainty in ${\rm A}_{V}$ is ${\sim}1\,{\rm mag}$ for the PNSPC samples.

The observed intensity of the line emission $X$ is given by $I_{X}^{\rm obs} = I_{X}{\rm e}^{-\tau_{X}}$, where $I_{X}$ is the intrinsic intensity and $\tau_{X}$ is the extinction at the line $X$. The observed intensity ratio of the line $X$ to $Y$ is
\begin{equation}
  \label{eq:app:ext:lineratio}
  \frac{I_X^{\rm obs}}{I_Y^{\rm obs}} =
  \frac{I_X}{I_Y}\exp\left(-\alpha(X,Y) {\rm A}_V\right),
\end{equation}
where $\alpha(X,Y)$ is defined by $(\tau_X-\tau_Y)/{{\rm A}_V}$, which is a constant value specific to the extinction curve. Thus, the extinction is estimated by
\begin{equation}
  \label{eq:app:ext:aveq}
  {\rm A}_V = \frac{\log x_{\rm int} - \log x_{\rm obs}}{\alpha(X,Y)},
\end{equation}
where $x_{\rm int}$ is $I_X/I_Y$ and $x_{\rm obs}$ is $I_X^{\rm obs}/I_Y^{\rm obs}$. Given that the line $X$ and $Y$ are Brackett-$\alpha$ (Br$\alpha$) and $\beta$ (Br$\beta$), respectively, the intrinsic intensity ratio is about $0.57$ assuming the Case-B condition \citep{baker_physical_1938} with the electron density of $10^4\,$K and the electron density of $10^4\,{\rm cm^{-3}}$. The ratio is less sensitive to either the electron temperature or density. We adopt the extinction curve given by \citet{mathis_interstellar_1990}. The value of $\alpha({\rm Br}\beta,{\rm Br}\alpha)$ is about $0.041$. The extinction estimated from the Brackett-$\beta$ to $\alpha$ ratio is
\begin{equation}
  \label{eq:app:ext:BrB2BrA}
  {\rm A}_{V}({\rm Br}\beta,{\rm Br}\alpha) \simeq {-}24.6\log x_{\rm obs} - 13.7.
\end{equation}

\setcounter{figure}{18}
\begin{figure}[!t]
  \centering
  \includegraphics[width=1.0\linewidth]{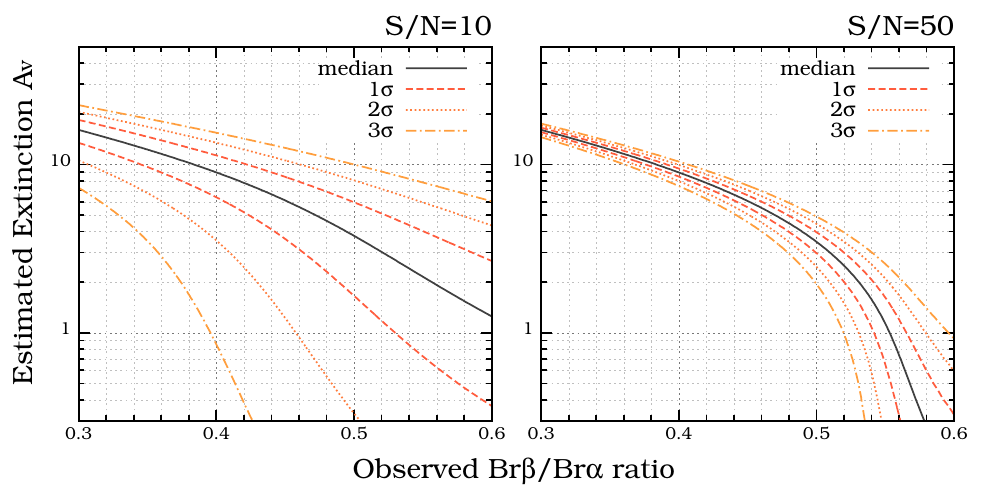}
  \caption[Estimated Extinction and Uncertainty Based on the IRC Spectrum]{Estimated extinction and uncertainty. The dashed, dotted, and dot-dashed contours show the confidence intervals of the 1-, 2-, and 3-$\sigma$, respectively.}
  \label{fig:uncertaintyAv}
\end{figure}

The uncertainty in ${\rm A}_{V}({\rm Br}\beta,{\rm Br}\alpha)$ is investigated. The measured intensity ratio ($x_{\rm obs}$) includes a statistical error. Define the error by $\sigma$. Assuming that the variation in $x_{\rm obs}$ is given by a normal distribution, the posterior probability distribution of the intensity ratio $(\bar{x}_{\rm obs})$ is given by
\begin{equation}
  \label{eq:app:ext:posterior}
  {\rm prob}(\bar{x}_{\rm obs} |\, x_{\rm obs}, \sigma) \propto
  \frac{1}{\sqrt{2\pi}\sigma}\exp\left(-\frac{(x_{\rm obs}-\bar{x}_{\rm obs})^2}{2\sigma^2}\right)
  {\rm prob}(\bar{x}_{\rm obs}),
\end{equation}
where ${\rm prob}(\bar{x}_{\rm obs})$ is the prior distribution of $\bar{x}_{\rm obs}$ assumed to be a uniform distribution from $0$ to $x$. From Equations (\ref{eq:app:ext:BrB2BrA}) and (\ref{eq:app:ext:posterior}), the posterior distribution of ${\rm A}_V({\rm Br}\beta,{\rm Br}\alpha)$ is given by
\begin{equation}
  \label{eq:app:ext:postAv}
  {\rm prob}({\rm A}_V |\, x_{\rm obs}, \sigma) \propto
  {\rm prob}(\bar{x}_{\rm obs} |\, x_{\rm obs}){\rm prob}(\bar{x}_{\rm obs}).
\end{equation}
The results for ${\rm S/N} = 10$ and $50$ are shown in Figure \ref{fig:uncertaintyAv}. The horizontal axis shows the observed intensity ratio ($x_{\rm obs}$), while the vertical axis is the estimated ${\rm A}_V$ value. The gray solid line indicates the median extinction ${\rm A}_V$. The red dashed, orange dotted, and yellow dot-dashed contours show the 1-, 2-, and 3-$\sigma$ confidence intervals, respectively. The typical ${\rm S/N}$ for the PNSPC spectrum is about $50$. Figure~\ref{fig:uncertaintyAv} suggests that the extinction estimated from the Br$\alpha$/Br$\beta$ ratio are quite uncertain when ${\rm A}_V \sim 1$--$3\,$mag. Figure~\ref{fig:exthist} shows that the ${\rm A}_V$ values are typically less than about $2\,{\rm mag}$ for the PNSPC samples. Thus, it is practically impossible to precisely estimate the extinction of the PNSPC samples without using ancillary data such as the intensity of the H$\beta$ emission. The standard errors on ${\rm A}_V$ derived from the  Br$\beta$/Br$\alpha$ and H$\beta$/Br$\alpha$ ratios are estimated to be about $0.5$ and $0.02\,$mag., respectively. The intrinsic line ratio may depend on the electron temperature and density. The impact is much larger for the H$\beta$/Br$\alpha$ ratio, but the variations in them can change ${\rm A}_V$ at most by $0.3\,$mag. Even though these variations are taken into account, the extinction derived from the H$\beta$/Br$\alpha$ ratio is much more reliable than that from the Br$\beta$/Br$\alpha$ ratio.

\end{appendix}

\end{document}